\documentclass[11pt]{article}
\usepackage{amsthm,amsmath,amssymb,natbib}
\usepackage{graphicx,color}
\usepackage[colorlinks]{hyperref}
\definecolor{DarkBlue}{rgb}{0.1,0.1,0.5}
\definecolor{Red}{rgb}{0.9,0.0,0.1}

\theoremstyle{plain}
\newtheorem{thm}{Theorem}[section]
\newtheorem{prop}{Proposition}[section]
\newtheorem{lemma}{Lemma}[section]
\newtheorem{cor}{Corollary}[section]

\def\tr{\text{tr}\,}
\def\bs{\boldsymbol}

\begin{document}


\begin{center}
{\Large{\bf Principal components analysis for sparsely observed
correlated functional data using a kernel smoothing approach}}

\vskip.1in Debashis Paul ~\textit{and}~ Jie Peng

\vskip.1in \textit{University of California, Davis}

\end{center}

\begin{abstract}
In this paper, we consider the problem of estimating the
covariance kernel and its eigenvalues and eigenfunctions from
sparse, irregularly observed, noise corrupted and (possibly)
correlated functional data. We present a method based on
pre-smoothing of individual sample curves through an appropriate
kernel. We show that the naive empirical covariance of the
pre-smoothed sample curves gives highly biased estimator of the
covariance kernel along its diagonal. We attend to this problem by
estimating the diagonal and off-diagonal parts of the covariance
kernel separately. We then present a practical and efficient
method for choosing the bandwidth for the kernel by using an
approximation to the leave-one-curve-out cross validation score.
We prove that under standard regularity conditions on the
covariance kernel and assuming i.i.d. samples, the risk of our
estimator, under $L^2$ loss, achieves the optimal nonparametric
rate when the number of measurements per curve is bounded. We also
show that even when the sample curves are correlated in such a way
that the noiseless data has a \textit{separable} covariance
structure, the proposed method is still consistent and we quantify
the role of this correlation in the risk of the estimator.
\end{abstract}



\section{Introduction}

Noisy functional data arise frequently in various fields, for
example longitudinal data analysis, chemometrics, econometrics, etc
(Ferraty and Vieu, 2006).
Depending on how the measurements are taken, there can be two
different scenarios - (i) individual curves are measured on a
dense, regular grid; (ii) the measurements are observed on a
sparse, and typically irregular set of points in an interval. The
first situation usually arises when the data are recorded by some
automated instrument, e.g. in chemometrics, where the curves
represent the spectra of certain chemical substances. The second
scenario is more typical in longitudinal studies where the
individual curves could represent the level of concentration of
some substance, and the measurements on the subjects may be taken
only at irregular time points.

In these settings, when the goal of analysis is either data
compression, model building or studying covariate effects, one may
want to extract information about the \textit{functional principal
components} (i.e., the eigenvalues and eigenfunctions of the
covariance kernel). The eigenfunctions give a nice basis for
representing the data, and hence are very useful in problems
related to model building and prediction for functional data. For
example, they have been used extensively in functional linear
regression (Cardot, Ferraty and Sarda (1999), Hall and Horowitz (2007),
Cai and Hall (2006)). 
Ramsay and Silverman (2005) and Ferraty and Vieu (2006)
give extensive surveys of the applications of functional principal
components. In the first scenario, i.e., data on a regular grid, as
long as the individual curves are smooth, the measurement noise
level is low, and the grid is dense enough, one can essentially
treat the data to be on a continuum, and employ techniques similar
to the ones used in classical multivariate analysis. However, the
irregular nature of data in the second scenario, and the associated
measurement noise require a different treatment. In this paper, we
propose a kernel smoothing approach to estimate the covariance
surface and its functional principal components based on sparse,
irregularly observed, noise corrupted functional data. This method
is based on the pre-smoothing of individual curves, with suitable
modification of the diagonal, for estimating the covariance kernel.
We prove the consistency and derive the rate of convergence of the
proposed estimator. Also, under many practical circumstances, the
sample curves are correlated, for example, spatio-temporal data
(Hlubinka and Prchal, 2007), 
online auction data (Peng and M\"{u}ller, 2008), 
time course gene expression data (Spellman \textit{et al.}, 1998).
However, in the existing literature, most of the
theoretical study on principal components analysis assume i.i.d.
sample curves. The analysis presented in this paper shows that the
asymptotic consistency of the principal components holds for the
proposed method even under certain types of correlation structures
(as discussed later).




Before we go into the details of the proposed procedure, we first
give an outline of the data model and an overview of different
approaches to this problem.
Suppose that we observe $n$ realizations of an $L^2$-stochastic
process $\{X(t): t \in [0,1]\}$ at a sequence of points on the
interval $[0,1]$ (or, more generally, on an interval $[a,b]$), with
additive measurement noise. That is, the observed data $\{Y_{ij} :
1\leq j \leq m_i; 1 \leq i \leq n\}$ can be modeled as :
\begin{equation}\label{eq:model}
Y_{ij} = X_i(T_{ij}) + \sigma \varepsilon_{ij},
\end{equation}
where $\{\varepsilon_{ij}\}$ are i.i.d. with mean 0 and variance 1.
Since $X(t)$ is an $L^2$ stochastic process, by \textit{Mercer's
Theorem} (Ash, 1972) 
there exists a positive semi-definite kernel
$C(\cdot,\cdot)$ such that $Cov(X(s),X(t)) = C(s,t)$ and each
$X_i(t)$ has the following a.s. representation in terms of the
eigenfunctions of the kernel $C(\cdot,\cdot)$ :
\begin{equation}\label{eq:Karhunen}
X_i(t) = \mu(t) + \sum_{\nu=1}^\infty \sqrt{\lambda_\nu} \psi_\nu(t)
\xi_{i\nu},
\end{equation}
where $\mu(\cdot) = \mathbb{E}(X(\cdot))$ is the mean function;
$\lambda_1 \geq \lambda_2 \geq \ldots \geq 0$ are the eigenvalues
of $C(\cdot,\cdot)$; $\psi_\nu(\cdot)$ are the corresponding
orthonormal eigenfunctions; and the random variables
$\{\xi_{i\nu}:\nu \geq 1\}$, for each $i$, are uncorrelated with
zero mean and unit variance. Furthermore, we assume that for each
pair $(i,j)$ with $1 \leq i\neq j \leq n$, the correlation is
modelled by
\begin{equation*}
\mathbb{E}(\xi_{i\nu}\xi_{j\nu'}) = \delta_{\nu\nu'} \rho_{ij},
\end{equation*}
for $1\leq \nu,\nu' \leq M$, and $\rho_{ij}$ may be nonzero. This
gives rise to a \textit{separable covariance structure} for the
noiseless data. That is, the processes $\{X_i(\cdot)\}_{i=1}^n$
satisfy, $Cov(X_i(s),X_j(t)) = \rho_{ij} C(s,t)$, with $\rho_{ii}
\equiv 1$. This holds, for example when the principal component
scores $\{\xi_{i\nu}\}_{i=1}^n$ for different $\nu$ are i.i.d.
stationary time series. Finally, in the observed data model
(\ref{eq:model}), we assume that $\mathbf{T}_i =
\{T_{ij}:j=1,\ldots,m_i\}$ are randomly sampled from a continuous
distribution.

As an example that is particularly suitable for modeling within the
framework presented above, we consider the data on atmospheric
radiation in Hlubinka and Prchal (2007). 
There, the measurements are taken from balloons from Earth's surface
up to an altitude of 35 km. The data points corresponding to the
$i$-th balloon are of the form $(a_i,z_i)$, where $a$ represents the
altitude and $z$ represents the average number of pulses at altitude
$a$, which is thought to be proportional to the radiation intensity.
Thus, these vertical profiles of atmospheric radiation are
considered as individual realizations of a functional data. That is,
here $a_i$'s are measurement points, $z_i$'s are the measurements
and the subjects are indexed by time. Hence there is a natural
dependence among the sample curves observed over different time
points. Moreover, it is reasonable to assume that the dependence
across time does not change with the vertical distance except
possibly through a long-term trend, i.e., the spatio-temporal
covariance structure is separable.

Below we give a short overview of two existing approaches to the
problem of estimation of functional principal components from sparse
data. Yao, M\"{u}ller and Wang (2005) 
propose a local linear smoothing of the empirical covariances
$\{\widehat C_i(T_{ij},T_{ij'}) : j \neq j'\}_{i=1}^n$:
\begin{equation*}
\widehat C_i(T_{ij},T_{ij'}) = (Y_{ij} - \widehat \mu(T_{ij}))
(Y_{ij'}-\widehat \mu(T_{ij'}))
\end{equation*}
where $\widehat \mu$ is the estimate of the mean function
$\mu(\cdot)$ obtained by local linear smoothing. They prove
asymptotic consistency of this estimator and the estimated
eigenfunctions,
by assuming i.i.d. sample curves.
Hall, M\"{u}ller and Wang (2006) 
prove further that the problem of estimating the covariance kernel
and that of estimating its eigenfunctions are intrinsically
different in that the former is a two-dimensional smoothing problem
while the latter is an one-dimensional problem, which results in
different choices for optimal bandwidth. They also prove that the
proposed local polynomial estimator achieves the optimal
nonparametric convergence rate with the optimal choice of
bandwidths, under the i.i.d. setting, when the number of
measurements per curve is bounded.


Instead of the local polynomial approach, where one imposes
regularization on the estimates by varying the bandwidth of the
kernel, one can impose regularization by restricting the
eigenfunctions in a known basis of smooth functions. This approach
has been used by various researchers including Besse, Cardot and
Ferraty (1997), Cardot (2000), James, Hastie and Sugar (2000) and
Peng and Paul (2007).
Peng and Paul (2007) 
propose to directly maximize the restricted log-likelihood under the
working assumption of Gaussianity, such that the resulting estimator
satisfies the geometry of the parameter space. This method is
implemented through a Newton-Raphson algorithm on the Stiefel
manifold of rectangular matrices with orthonormal columns. The
latter space is the parameter space for the matrix of basis
coefficients for the eigenfunctions. Furthermore, in Paul and Peng
(2007) 
the authors prove that this restricted maximum
likelihood (REML) estimator also achieves the optimal nonparametric
rate when the number of measurements per sample curve is bounded and
the sample curves are i.i.d.


We now give a brief description of the estimation procedure proposed
in this paper. The method is partly motivated by the observation
that the naive sample covariance based on the presmoothed individual
sample curves is a highly bias estimation along the diagonal of the
covariance kernel, when $m_i$, the number of measurements per curve,
is small. As can be seen clearly from (\ref{eq:expect_m_C_tt}) in
Section \ref{subsec:kernel}, this bias does not vanish
asymptotically unless $(\min_{1\leq i \leq n} m_i) h_n \to \infty$
as $n\to \infty$, where $h_n$ is the bandwidth of the kernel
smoother. Under the latter setting, Hall \textit{et al.} (2006) 
discuss the possibility of using a local linear smoother for
individual sample curves and then performing a PCA on the smoothed
curves. Furthermore, when the design points $T_{ij}$ are regularly
spaced and sufficiently dense, they show that using conventional PCA
for functional data (see statements and conditions in Theorem 3 of
that paper for details) one obtains root-$n$ consistent estimates of
the eigenvalues and eigenfunctions so that the problem is
asymptotically equivalent to a parametric problem. It is an
interesting question that whether the naive kernel smoothing
approach can be suitably modified such that it can produce
estimators with good asymptotic risk properties even when the
$m_i$'s are relatively small. Our approach in this paper goes
towards this direction and involves estimating the diagonal and the
off-diagonal portions separately, and then merging them together
using a smooth weight kernel. The estimation of the off-diagonal
portion is based on presmoothing individual sample curves by a
linearized kernel smoother. The estimation of the diagonal part
involves linearized kernel smoothing of the empirical variances. The
task of selecting an appropriate bandwidth, and the number of
nonzero eigenvalues, is addressed through obtaining a
computationally efficient approximation to the leave-one-curve-out
cross validation score. This approximation procedure, as well as the
asymptotical analysis of the estimators, is based on the
perturbation theory of linear operators.


Now we summarize the main contributions of this paper. Our approach
of merging two separate presmoothed linearized kernel estimates of
the diagonal and the off-diagonal parts of the covariance kernel is
new and is computationally very efficient.
We prove that the proposed estimator achieves the optimal
nonparametric rate when the observations are i.i.d. realizations of
a finite dimensional smooth stochastic process, and when the number
of measurements per curve is bounded. This result parallels to the
one obtained by Hall \textit{et al.} (2006) 
for the local polynomial approach. Moreover, we obtain explicit
expressions for the integrated mean squared error of the estimated
eigenfunctions under a regime of separable covariance structure
among the sample curves. The quantification of the role of
correlation in the risk behavior (Theorem \ref{thm2}) is seemingly
new in the literature, under the context of functional data
analysis. We also derive a lower bound on the rate of convergence of
the risk of the first eigenfunction (Theorem \ref{thm3}) which is
sharper than an analogous (but more general) bound obtained in Hall
\textit{et al.} (2006). 
This lower bound and the matching upper bound on the rate of
convergence for the i.i.d. case shows that the proposed estimator
obtains the optimal rate even when $\max_{1\leq i \leq n} m_i \to
\infty$, at least under the restricted setting described in Theorem
\ref{thm3}. Moreover, if the correlation between sample curves is
``weak'' in a suitable sense, then the optimal rate of convergence
for eigenfunctions in the correlated and i.i.d. cases are the same.
Furthermore, we show that our estimation procedure also allows for a
computationally efficient approximation of leave-one-curve-out cross
validation score, which is used for selecting the bandwidth for
estimating the eigenfunctions. This approximation is based on a
perturbation analysis approach that is natural given the form of our
estimator.
In the paper, we also show that the widely used prediction error
loss for cross validation is not correctly scaled under the current
context. Thus we propose to use the empirical Kullback-Leibler loss
for the cross validation criterion.

The rest of the paper is organized as follows. In Section
\ref{sec:merging}, we propose the estimation procedure and contrast
it with the naive kernel smoothing approach. In Section
\ref{sec:CV_approx}, we propose an approximation to the
leave-one-curve-out cross validation score based on the perturbation
theory for linear operators. In Section \ref{sec:asymptotics}, we
state the main results about the consistency and rate of convergence
of the estimators of the covariance kernel and its eigenfunctions.
In Section \ref{sec:proof_thm1and2}, we give an outline of the proof
of the main results (Theorems \ref{thm1} and \ref{thm2}) and discuss
their implications. In Section \ref{sec:discussion}, we give an
overview of various related issues and future research directions.
The proof details are provided in the appendices.

\section{Method}\label{sec:merging}

Throughout this section, we assume that the mean curve has been
estimated separately, and has been subtracted from the data. Thus,
without loss of generality we assume that $\mu = 0$. Also, in the
asymptotic analysis carried out in Section \ref{sec:asymptotics},
we make the same assumption to simplify the exposition. The case
of arbitrary $\mu$ with sufficient degree of smoothness can be
easily handled.

\subsection{Naive kernel smoothing approach}\label{subsec:kernel}

A popular method in nonparametric function estimation is to smooth
the individual sample curves by a kernel averaging of the sample
points. In principle, one can adopt a similar approach in the
current context. This means that first smoothing individual sample
curves, and then computing the covariance of the ``pre-smoothed''
sample curves, followed by an eigen-analysis of this
``pre-smoothed'' empirical covariance. In the following, we first
describe briefly such an approach, and then show that even in the
case of i.i.d. data, the estimator thus obtained has an intrinsic
bias while estimating the diagonal of the covariance kernel, unless
the number of measurements per curve is large.

Let $K(\cdot)$ be a \textit{summability kernel} with an adequate
degree of smoothness, and satisfying the following conditions:
\begin{itemize}
\item[{\bf B1}](i) supp$(K) = [-B_K,B_K]$ for some $B_K
> 0$; (ii) $K$ is symmetric about 0; (iii) $\int K(x)dx = 1$; (iv)
$\int x K(x) dx = 0$; (v) $\int K'(x) dx = 0$; (vi) $\int x K'(x)
dx = 1$.
\end{itemize}
We then define the presmoothed sample curves as follows:
\begin{equation}\label{eq:kernel_smooth}
\widetilde X_i(t) = \frac{1}{m_i} \sum_{j=1}^{m_i} Y_{ij}
K_{h_{n,i}}(t-T_{ij}), ~~~i=1,\ldots,n,
\end{equation}
where $K_h(x) = h^{-1} K(h^{-1} x)$ for $h > 0$ and $h_{n,i}$ is
the bandwidth for the $i$-th curve. Then the empirical covariance
based on the presmoothed curves is simply
\begin{equation}
\label{eqn:empcov} \widetilde{C}(s,t)=\frac{1}{n}\sum_{i=1}^n
\widetilde X_i(t) \widetilde X_i(s).
\end{equation}





In the following, we derive an expression for the expectation of
$\widetilde C(s,t)$ in estimating $C(s,t)$ to  quantify the bias,
when $h_{n,i} = h_n$ for all $i$, under the assumption that
$C(\cdot,\cdot)$ is twice continuously differentiable. Suppose for
simplicity that the density of the design points
$\{T_{ij}\}_{j=1}^{m_i}$, for each subject, is uniform on $[0,1]$.
Define $\overline{C}(t) = C(t,t)$ for $t\in [0,1]$, and
$K_2(\cdot) = \int K(\cdot - u) K(-u) du$. Also, we assume that
$m_i's$ are given. In the following proposition the bounds hold
under $h_n \to 0$.

\begin{prop}\label{prop1} When $s \neq t$,
\begin{eqnarray}\label{eq:expect_m_C_st}
\mathbb{E}[\widetilde X_i(s) \widetilde X_i(t)]  &=& \frac{1}{m_i
h_n} K_2(\frac{s-t}{h_n}) (\overline{C}(t) + \sigma^2) +
\frac{1}{m_i} \overline{C}'(t) \int u K(-u)K(\frac{s-t}{h_n} - u)
du \nonumber\\
&& ~~~~+ (1-\frac{1}{m_i}) C(s,t) + \frac{1}{m_i} O(h_n) + O(h_n^2).
\end{eqnarray}
And,
\begin{equation}\label{eq:expect_m_C_tt}
\mathbb{E}[\widetilde X_i(t)^2] = \frac{1}{m_i h_n} K_2(0)
(\overline{C}(t) + \sigma^2) + (1-\frac{1}{m_i}) \overline{C}(t) +
\frac{1}{m_i} O(h_n) + O(h_n^2)
\end{equation}
The $O(\cdot)$ terms involve $\sup_{t\in[0,1]}|\overline{C}''(t)|$,
$\sup_{s,t \in [0,1]}
\parallel {\cal D}^2 C(s,t)\parallel$ and $\int u^2 K(u) du$,
where ${\cal D}^2$ is the Hessian operator.
\end{prop}

\vskip.1in By Proposition \ref{prop1}, it is easy to see,
$\mathbb{E}[\widetilde X_i(s) \widetilde X_i(t)] =
(1-\frac{1}{m_i}) C(s,t)  + O(h_n^2)$ if $|s-t| > 2 B_K h_n$,
since the first two terms in ~(\ref{eq:expect_m_C_st}) both
vanish, as wells as the $O(h_n)$ term (see the proof in Appendix C
for more details). This shows that $\widetilde C(s,t)$ should be
multiplied by $m_i/(m_i-1)$ to get rid of the trivial bias.
However, (\ref{eq:expect_m_C_st}) and (\ref{eq:expect_m_C_tt})
also show that the empirical covariance $\widetilde C(s,t)$ is a
highly biased estimate of $C(s,t)$ near the diagonal even after
this trivial modification, unless $h_n \min_{1\leq i \leq n} m_i
\to \infty$. This is because the first terms
in(\ref{eq:expect_m_C_st}) and (\ref{eq:expect_m_C_tt}) are always
positive  along the diagonal  (i.e., when $|s-t| < 2 B_K h_n$),
which result in overestimation. In fact the degree of
overestimation gets really big (by a scale factor of $h_n$) as
soon as $|s-t| < 2 B_K h_n$. This demonstrates clearly that the
naive kernel smoothing approach is intrinsically biased and needs
to be appropriately modified.

To understand the reason for this bias, notice that if a pair of
points $(T_{ij},T_{ij'})$, for some $1\leq j\neq j' \leq m_i$, is
randomly sampled from $[0,1]^2$, then it has a probability of the
order $O(h_n^2)$ to be in a neighborhood of length and width $h_n$
of a given point $(s,t)$ (which is away from the diagonal). In
contrast, there is $O(h_n)$ probability of a randomly chosen point
$T_{ij}$ to belong to a neighborhood of length $h_n$ of the point
$(t,t)$ along the diagonal. Therefore, measurements are much denser
along the diagonal and  this explains the difference in rates.


\subsection{Modification to naive kernel
smoothing}\label{subsec:modification}

In this section, we propose a modification to deal with the bias in
the naive kernel smoothing approach described in Section
\ref{subsec:kernel}. We propose to remedy the effect of unequal
scale along the diagonal of the covariance kernel (and the resulting
bias) by estimating the diagonal and the off-diagonal parts
separately. We then use a suitable (smooth) weight kernel to combine
those two estimates together.

Throughout the paper, we assume that the density of the
time-points $\{T_{ij}\}$ is known and is denoted by $g(\cdot)$. In
practice we can estimate $g$ from the data separately. We further
assume that there are constants $0 < c_0 \leq c_1<\infty$ such
that $c_0 \leq g(\cdot) \leq c_1$.

We also propose to use a linearized version of the kernel smoothing
to reduce the bias while controlling the variance. For this purpose,
define $Q(s,t)$ to be a \textit{tensor-product kernel} (that is a
kernel of the form $Q(s,t) = \overline Q(s) \overline Q(t)$ for some
smooth function $\overline Q$) with the following properties,
together referred as condition {\bf B2}:
\begin{itemize}
\item[(i)] $\overline{Q}$ is supported on $[-C_Q,C_Q]$, for some $C_Q >0$, and
$\overline{Q}(\cdot) \geq 0$;
\item[(ii)] $\parallel \overline{Q}\parallel_\infty < \infty$;
\item[(iii)] $\sum_{k \in \mathbb{Z}} \overline{Q}(x-k) = 1$.
\item[(iv)] $\overline{Q}$ is symmetric about 0.
\end{itemize}
Property (iii) can be rephrased as saying that integer translates of
$\overline{Q}$ form a \textit{partition of unity}. As an example,
the B-spline basis functions (Chui, 1987) 
satisfy all four properties. Let $Q_{h}(\cdot,\cdot)$ denote the
kernel $Q(h^{-1} \cdot,h^{-1} \cdot)$.

For estimation of the diagonal $\overline{C}(t)=C(t,t)$, let
$\widehat{\overline{C}}(t)$:= $\widehat C_*(t) - \widehat
\sigma^2$, where $\widehat \sigma^2$ is an estimator of $\sigma^2$
(discussed in Section \ref{subsec:sigma_est}), and $\widehat
C_*(t)$ is the estimate of $\overline{C}(t) + \sigma^2$ obtained
by using a linearized kernel smoothing of the terms
$\{\frac{1}{m_i} Y_{ij}^2 : j=1,\ldots,m_i; i=1,\ldots,n\}$. This
is because, for each pair $(i,j)$, the conditional expectation of
the quantity $Y_{ij}^2$ (conditional on $\mathbf{T}_i$ and $m_i$)
is $C(T_{ij},T_{ij}) + \sigma^2$. Define a grid on $[0,1]$ with
grid spacings $h_n$ and denote the grid points by $\{s_l : l
=1,\ldots,L_n\}$ where $L_n =  \frac{c_L}{h_n}$ for an
appropriately chosen $c_L \approx 1$. Then define,
\begin{equation}\label{eq:diagonal_est}
\widehat C_{*,h_n}(t) = \frac{1}{g(t)} \frac{1}{n} \sum_{i=1}^n
\sum_{l=1}^{L_n} [S_i(s_l) + (t-s_l)
S_i'(s_l)]\overline{Q}_{h_n}(t-s_l),
\end{equation}
with
\begin{equation}\label{eq:S_i_def}
S_i(s) = \frac{1}{m_i} \sum_{j=1}^{m_i} Y_{ij}^2 K_{h_n}(s-T_{ij}).
\end{equation}
Note that, (\ref{eq:diagonal_est}) is a linearized version of the
conventional kernel smoothing, which can be interpreted as a local
linear smoothing of the empirical variances. A similar principle is
applied to construct an estimator of the off-diagonal part (see
(\ref{eq:C_hat_g}) below). The linearization has two advantages: on
one hand, it helps in reducing the bias in the estimate; and on the
other hand it facilitates efficient computation both in terms of
estimation and model selection. The difference of this linearization
approach with the local linear smoothing mainly lies in the fact
that we are using $g(t)$ (or an estimate of $g(t)$) in the
denominator, while in local linear smoothing, the denominator
implicitly is a local estimate of $g$ obtained by averaging the
smoothing kernel in a neighborhood of $t$. Note that, as opposed to
our estimator of $g$, which uses different bandwidth than the one
for estimating the covariance, local linear smoothing essentially
uses the same bandwidth for estimating both $g$ and $C$, and thus it
suffers from instability. More specifically, the local linear
estimator of Yao
\textit{et al.} (2005) 
involves ratios with a denominator consisting of essentially the
number of time points falling in a small interval. Since the time
points are assumed to be randomly distributed and are sparse, in
practice this can cause instability.

Let $\widetilde X_i(t)$ be the $i$-th smoothed sample curve as
defined in (\ref{eq:kernel_smooth}), and $\widetilde X_i'(t)$ be the
derivative of $\widetilde X_i(t)$. Then define the estimate of the
off-diagonal part as (with a slight abuse of notation)
\begin{eqnarray}\label{eq:C_hat_g}
\widetilde C_{h_n}(s,t) &=&
\frac{1}{g(s)g(t)} \frac{1}{n}\sum_{i=1}^n w(m_i)
\sum_{l,l'=1}^{L_n}\left[ (\widetilde X_i(s_l) + (s-s_l) \widetilde
X_i'(s_l))
\right. \nonumber\\
&& ~~\left. \cdot ~ (\widetilde X_i(s_{l'}) + (t-s_{l'}) \widetilde
X_i'(s_{l'}))Q_{h_n}(s-s_l,t-s_{l'})\right].
\end{eqnarray}
Here $w(m_i)=\frac{m_i}{m_i-1}$ is a weight function which is
determined through an asymptotic bias analysis (Proposition
\ref{prop1}). Note that, as long as $|s-t| \geq  A h_n$ for some
constant $A$ depending on $B_K$ and $C_Q$, in the inner sum in
definition (\ref{eq:C_hat_g}), the terms for which $l=l'$ are
absent. Therefore, according to our analysis in the previous
section, they do not contribute anything by way of bias.

Now let $W(\cdot,\cdot)$ be a weight kernel on the domain $[0,1]^2$
defined as
\begin{equation}\label{eq:W_kernel}
W(s,t) := W(s-t) = \begin{cases} 0 & ~\mbox{if}~ |s-t| > \frac{1}{2}\\
                       1 & ~\mbox{if}~ |s-t| \leq \frac{1}{2}\\
\end{cases}
\end{equation}
Define $W_{\widetilde h_n}(s,t) = W((s-t)/\widetilde h_n)$ and
$\overline{W}_{\widetilde h_n}(s,t) = 1 - W_{\widetilde h_n}(s,t)$,
where $\widetilde h_n = A h_n$ for the above $A > 0$. We then smooth
the kernels $W_{\widetilde h_n}$ and $\overline{W}_{\widetilde h_n}$
by convolving them with a Gaussian kernel ${\cal G}_{\tau_n}(\cdot)$
with a small bandwidth $\tau_n$ (in the sense that $\tau_n =
o(h_n)$). And with an abuse of notation, denote the resulting
kernels also by $W_{\widetilde h_n}$ and $\overline{W}_{\widetilde
h_n}$, respectively. Finally, we are ready to define the proposed
combined estimator of $C(s,t)$ as
\begin{eqnarray}\label{eq:C_est_combined}
\widehat C_{c,h_n}(s,t) &=& \overline{W}_{\widetilde h_n}(s,t)
\widetilde C_{h_n}(s,t) + W_{\widetilde h_n}(s,t)
~\max\{\widehat{\overline{C}}_{h_n}(\frac{s+t}{2}),h_n^2\},
\end{eqnarray}
where $\widehat{\overline{C}}_{h_n}(\cdot)$:= $\widehat
C_{*,h_n}(\cdot) - \widehat \sigma^2$. The use of maximum in the
second term is just to guarantee that the estimator of the diagonal
is nonnegative and the bias is $O(h_n^2)$.

We now discuss briefly the computational aspects of the proposed
estimator. A key step is the computation of the functions
$S_i(\cdot)$ and $\widetilde X_i(\cdot)$ and their derivatives at
the grid points ${s_l:l=1,\ldots,L_n}$. Each one of these
computations requires $O(m_i)$ floating point operations (for each
$i=1,\ldots,n$). From these, we obtain $\widetilde C_{h_n}(s,t)$
and $\widehat C_{*,h_n}(t)$ by using (\ref{eq:C_hat_g}) and
(\ref{eq:diagonal_est}), respectively. Both expressions are in the
form of discrete convolutions, and hence can be computed very
rapidly by using the \textit{Fast Fourier Transform}. Thus, the
estimation procedure is computationally very efficient, with
$O(n\overline{m}L_n \log L_n)$ computations on the whole grid,
where $\overline{m}=\max_i m_i$.

\subsection{Estimation of $\sigma^2$}\label{subsec:sigma_est}


Here we briefly outline a method for estimating the error variance
$\sigma^2$. The method is similar to the approach taken in Yao,
M\"{u}ller and Wang (2006), 
and hence we omit the details.



First, for a given bandwidth $h_n$, we estimate the function
$C(s,t)$ for $|s-t| > A h_n$, for some $A$ depending on $B_K$ and
$C_K$, using (\ref{eq:C_hat_g}). Then, as in
Yao \textit{et al.} (2006), 
we estimate the diagonal $\{\overline{C}(t):
t\in [0,1]\}$, using an \textit{oblique} linear interpolation, by
\begin{equation}\label{eq:C_bar_hat_0}
\widehat{\overline{C}}_{0,h_n}(t) = \int_{A_1}^{A_2}
\frac{1}{2}(\widetilde C_{h_n}(t-uh_n,t+uh_n) + \widetilde C_{h_n}
(t+uh_n,t-uh_n)) d\widetilde G(u),
\end{equation}
for some probability distribution function $\widetilde G$ supported
on $[A_1,A_2]$ where $A_1 > A$. On the other hand, we estimate the
curve $\{\overline{C}(t)+\sigma^2 : t\in[0,1]\}$ by $\widehat
C_{*,h_n}(t)$ defined in (\ref{eq:diagonal_est}).
Now, we estimate $\sigma^2$ by
\begin{equation}\label{eq:sigma_hat}
\widehat \sigma^2 = \frac{1}{T_1 - T_0} \int_{T_0}^{T_1} (\widehat
C_{*,h_n}(t) - \widehat{\overline{C}}_{0,h_n}(t)) dt,
\end{equation}
where $0 < T_0 < T_1 < 1$. It can be shown that (Corollary
\ref{cor1} in Section \ref{sec:asymptotics}) the estimator
$\widehat \sigma^2$ thus obtained is consistent for an appropriate
choice of $h_n$.

\section{Bandwidth selection}\label{sec:CV_approx}


The choice of optimal bandwidth for the kernel is a key step in any
kernel-based estimation procedure. Yao \textit{et al.} (2005)
use a leave-one-curve-out cross validation score based on the
prediction error for selecting the bandwidth of the smoother, and an
AIC approach for selecting the number of non-zero eigenvalues.
However, leave-one-curve-out cross validation is computationally
very expensive. Also, as shown below, the \textit{prediction error
loss} is not an appropriate criterion for cross validation under the
current context. Therefore, in this paper we address the issue of
model selection by producing an approximation to the
leave-one-curve-out cross validation score based on the
\textit{empirical Kullback-Leibler loss}. The approximation is based
on the idea that the estimator obtained by dropping any single curve
is a small perturbation of the estimator based on the whole data
(Peng and Paul, 2007). 
In particular, we use perturbation theory of linear operators to
quantify this perturbation and produce a first order approximation
to the CV score that is computationally efficient. It also enables
us to select the bandwidth and the dimension of the process
simultaneously.


We first discuss the choice of the loss function, which is very
important for a cross-validation scheme. We want to point out
that, the prediction problem is intrinsically different from the
estimation of the covariance kernel. We find out that the
criterion based on prediction error loss is not correctly scaled,
as opposed to the one based on empirical Kullback-Leibler loss. To
make this point clear, we examine these two cross validation
criteria in details.

Define $\mathbf{Y}_i = (Y_{ij})_{j=1}^{m_i}$, $\boldsymbol{\mu}_i =
(\mu(T_{ij}))_{j=1}^{m_i}$, $\boldsymbol{\psi}_{i\nu} =
(\psi_\nu(T_{ij}))_{j=1}^{m_i}$.
We assume that the covariance
kernel can be represented using $K$ orthonormal eigenfunctions for
some $K \geq 1$. Then the leave-one-curve-out cross validation score
based on the prediction error loss is given by
\begin{equation}\label{eq:CV}
CV(K,h_n) =
\sum_{i=1}^n \sum_{j=1}^{m_i} (Y_{ij} - \widehat
Y_i^{(-i)}(T_{ij}))^2.
\end{equation}
Here $\widehat Y_i^{(-i)}(t) = \widehat\mu^{(-i)}(t) +
\sum_{\nu=1}^K \widehat \xi_{i\nu}^{(-i)} \widehat
\psi_\nu^{(-i)}(t)$, where $\widehat \mu^{(-i)}(t)$ and $\widehat
\psi_\nu^{(-i)}(t)$ are the estimates of $\mu(t)$ and $\psi_\nu(t)$
computed from observations $\{\mathbf{Y}_{i'}\}_{i' \neq i}^n$.
Also, $\widehat \xi_{i\nu}^{(-i)}$ is the estimated principal
component score based on observations $\{\mathbf{Y}_{i'}\}_{i' \neq
i}^n$. Note that, the estimated principal components scores
$\xi_{i\nu}^{(-i)}$ can be obtained through the procedure described
in Yao \textit{et al.} (2005), 
even though it will not be necessary for the model selection
procedure we shall adopt.

On the other hand, the CV score based on the empirical
Kullback-Leibler loss is given by
\begin{equation}\label{eq:CV_KL}
CV_*(K,h_n) = \sum_{i=1}^n \ell_i(\mathbf{Y}_i;\widehat
{\boldsymbol{\mu}}_i^{(-i)}, \widehat \Sigma_{i,K}^{(-i)}),
\end{equation}
where
$$
\widehat \Sigma_{i,K}^{(-i)} = \sum_{\nu=1}^K \widehat
\lambda_\nu^{(-i)} \widehat{\bs{\psi}}_{i\nu}^{(-i)} (\widehat
{\bs{\psi}}_{i\nu}^{(-i)})^T + \widehat \sigma_{(-i)}^2 I_{m_i},
$$
and $\ell_i$ is (up to an additive constant) the negative
log-likelihood of the $i$-th observation under the working
assumption of Gaussianity,  which is
$$
\ell_i(\mathbf{Y}_i;\boldsymbol{\mu}_i,\Sigma_{i}) = \frac{1}{2}
\log |\Sigma_{i}| + \frac{1}{2} \tr(\Sigma_{i}^{-1} (\mathbf{Y}_i -
\boldsymbol{\mu}_i) (\mathbf{Y}_i - \boldsymbol{\mu}_i)^T).
$$



To gain an understanding of what these CV scores are approximating,
we assume that we have two independent samples, each with $n$ i.i.d.
sample curves. Furthermore, to simplify exposition, we assume that
$\mu \equiv 0$. Suppose that the estimates $\widehat \Psi =
\{\widehat \psi_\nu\}_{\nu=1}^K$, $\widehat \Lambda = \{\widehat
\lambda_\nu\}_{\nu=1}^K$ are obtained from the first sample. Then a
leave-one-curve-out CV score can be reasonably approximated by
substituting these estimates in the corresponding empirical loss
function based on the second sample, and with an abuse of notation
we also denote this quantity by $CV$. If $\ell_i(\Psi,\Lambda)$
denotes the loss function corresponding to the $i$-th observation in
the second sample, then the CV score is given by
$\frac{1}{n} \sum_{i=1}^n \ell_i(\widehat \Psi,\widehat \Lambda)$.
For simplicity, we assume that there is a true model
$(\Psi_*,\Lambda_*)$ within the class of models we are considering.
A first order expansion of the difference between the CV scores
under the true and estimated parameters for the empirical
Kullback-Leibler loss shows that, with high probability,
\begin{eqnarray}\label{eq:CV_difference_KL_approx}
&& \frac{1}{n} \sum_{i=1}^n \ell_i(\widehat \Psi,\widehat \Lambda) -
\frac{1}{n} \sum_{i=1}^n \ell_i(\Psi_*,\Lambda_*) \nonumber\\
&=&
\frac{1}{4n} \sum_{i=1}^n \parallel \widehat
\Sigma_i^{-1/2}(\Sigma_{*i} - \widehat \Sigma_i) \widehat
\Sigma_i^{-1/2}\parallel_F^2
(1+o(1)) \nonumber\\
&& + O\left(\sqrt{\frac{\log n}{n}}\left[\frac{1}{n}\sum_{i=1}^n
\parallel \Sigma_{*i}^{1/2}(\widehat\Sigma_i^{-1} -
\Sigma_{*i}^{-1})\Sigma_{*i}^{1/2}\parallel_F^2\right]^{1/2}\right),
\end{eqnarray}
where $\parallel \cdot \parallel_F$ is the Frobenius norm, and
$\Sigma_{*i}$ and $\widehat\Sigma_i$ are the covariance matrices of
the observations $\mathbf{Y}_i = (Y_{i1},\ldots,Y_{im_i})^T$,
corresponding to the true parameter $(\Psi_*,\Lambda_*)$ and
estimates $(\widehat \Psi, \widehat \Lambda)$, respectively.  Since
we can essentially ignore the $O(\cdot)$ term in
(\ref{eq:CV_difference_KL_approx}) as long as $\frac{1}{n}
\sum_{i=1}^n
\parallel \widehat \Sigma_i^{-1/2}(\Sigma_{*i} - \widehat \Sigma_i)
\widehat \Sigma_i^{-1/2}\parallel_F^2$ is not too small,
(\ref{eq:CV_difference_KL_approx}) gives a quadratic approximation
to the CV score. Notice that, in each term within the summation of
this quadratic approximation, directions with high variability are
down-weighted by the multiplicative factors $\widehat
\Sigma_i^{-1/2}$. Therefore this CV score based on the empirical
Kullback-Leibler loss is properly scaled. Moreover, note that
approximation (\ref{eq:CV_difference_KL_approx}) does not really
depend on Gaussianity but only on the tail of the distributions
involved.

On the other hand, it can be shown by simple algebra that, up to a
multiplicative factor, the CV score based on the prediction error
loss is $CV = \frac{1}{n} \sum_{i=1}^n \widetilde
\ell_i(\Psi,\Lambda)$ where $\widetilde \ell_i(\Psi,\Lambda) =
\tr(\widehat\Sigma_i^{-2} S_i)$, where, $S_i = (\mathbf{Y}_i -
\widehat{\bs{\mu}}_i)(\mathbf{Y}_i - \widehat{\bs{\mu}}_i)^T$ is the
empirical covariance matrix corresponding the $i$ observation
vector. The corresponding difference of the CV scores between
estimated and true parameters becomes (ignoring the multiplicative
constant),
\begin{eqnarray}\label{eq:CV_difference_pred}
&& \frac{1}{n} \sum_{i=1}^n \widetilde \ell_i(\widehat \Psi,\widehat
\Lambda) - \frac{1}{n} \sum_{i=1}^n \widetilde
\ell_i(\Psi_*,\Lambda_*) \nonumber\\
&=& \frac{1}{n} \sum_{i=1}^n \tr[(\widehat \Sigma_i^{-2}
-\Sigma_{*i}^{-2})\Sigma_{*i}] + \frac{1}{n} \sum_{i=1}^n
\tr[(\widehat \Sigma_i^{-2} -\Sigma_{*i}^{-2})(S_i -
\Sigma_{*i})] \nonumber\\
&=& \frac{1}{n} \sum_{i=1}^n \tr[\Sigma_{*i}^{-1/2}(A_i^2 - I_{m_i})
\Sigma_{*i}^{-1/2}] \nonumber\\
&& ~~~~ + O\left(\sqrt{\frac{\log
n}{n}}\left[\frac{1}{n}\sum_{i=1}^n
\parallel \Sigma_{*i}^{-1/2}(A_i^2 - I_{m_i})
\Sigma_{*i}^{-1/2}\parallel_F^2\right]^{1/2}\right)
\end{eqnarray}
with high probability. Here $A_i = \Sigma_{*i}^{1/2}
\widehat\Sigma_i^{-1}\Sigma_{*i}^{1/2}$ which is already properly
scaled. Therefore, from (\ref{eq:CV_difference_pred}) it is clear
that this CV score itself is not correctly scaled. Also, the
expression $\frac{1}{n} \sum_{i=1}^n \tr[\Sigma_{*i}^{-1/2}(A_i^2
- I_{m_i}) \Sigma_{*i}^{-1/2}]$ appearing in
(\ref{eq:CV_difference_pred}) is not necessarily nonnegative. This
means that the prediction error loss does not enjoy the pleasing
property of the Kullback-Leibler loss that the minimum of the
expected loss occurs at the true parameter. Hence the use of the
prediction error loss is not recommended for the current problem.


\subsection{First order approximation}

Direct computation of the criterion $CV_*(K,h_n)$  (equation
(\ref{eq:CV_KL})) is a laborious process since we need to compute
$\widehat C_c^{(-i)}(s,t)$ and perform its eigen-analysis for every
$i=1,\ldots,n$.
Therefore, we propose to approximate $CV_*(K,h_n)$ by using a
first order approximation to the quantities
$\widehat{\boldsymbol{\mu}}_i^{(-i)}$, $\widehat
\psi_\nu^{(-i)}(\cdot)$ and $\widehat \lambda_\nu^{(-i)}$ around
the estimates $\widehat{\boldsymbol{\mu}}_i$, $\widehat
\psi_\nu(\cdot)$ and $\widehat \lambda_\nu$, respectively. The
approximations of the eigenfunctions and eigenvalues is based on a
perturbation analysis approach. The key idea is that the
leave-one-curve-out estimator $\widehat C_c^{(-i)}$ of the
covariance can be viewed as a perturbation of the linear operator
$\widehat C_c$. The key component is Proposition \ref{prop2} which
uses a result on perturbation of eigenfunctions of a linear
operator (Lemma \ref{lemmaA.1} in Appendix A). Note that, our
approximation scheme can also be applied to CV scores based on
some other loss functions, such as $CV(K,h_n)$.

Using Lemma \ref{lemmaA.1}, we can get a first order approximation
to the quantities $\widehat{\boldsymbol{\psi}}_{i \nu}^{(-i)}$ and
$\widehat \lambda_\nu^{(-i)}$ that depends on the observations
through a term that is linear in $\Delta_i(s,t) = \widehat C_c(s,t)
- \widehat C_c^{(-i)}(s,t)$ (for convenience we omit $h_n$ in the
notation). Since the latter quantity has a rather simple expression
which involves \textit{essentially} only the $i$-th observation,
this step substantially reduces the computational burden of the
cross-validation procedure.


\begin{prop}\label{prop2}
For the proposed estimator $\widehat C_c$ given by
(\ref{eq:C_est_combined}), we have,
\begin{itemize}
\item[(i)]
\begin{equation}\label{eq:psi_diff_approx}
\widehat{\boldsymbol\psi}_{i\nu}^{(-i)}
 - \widehat{\boldsymbol\psi}_{i\nu} = (\widehat\psi_\nu^{(-i)}(T_{ij})
-\widehat\psi_\nu(T_{ij}))_{j=1}^{m_i} \approx ((\widehat H_\nu
\Delta_i \widehat \psi_\nu)(T_{ij}))_{j=1}^{m_i};
\end{equation}
\item[(ii)]
\begin{equation}\label{eq:lambda_diff_approx}
\widehat \lambda_\nu^{(-i)} - \widehat \lambda_\nu \approx
-\tr(\widehat P_\nu \Delta_i);
\end{equation}
\end{itemize}
where
\begin{itemize}
\item[(a)]
$\widehat P_\nu = \widehat \psi_\nu \otimes \widehat \psi_\nu$
where, for $f,g \in L^2([0,1])$, $f \otimes g$ denotes the integral
operator with kernel $f(x)g(y)$ and acts on any $w \in L^2([0,1])$
as $(f \otimes g)(w)(x) = (\int_0^1 g(y)w(y) dy) f(x)$;
\item[(b)]
\begin{eqnarray}\label{eq:H_nu_expand}
\widehat H_\nu(t,u) &=& \sum_{k\neq \nu}^K \frac{1}{\widehat
\lambda_k - \widehat \lambda_\nu} \widehat \psi_k(t) \widehat
\psi_k(u) - \frac{1}{\widehat \lambda_\nu}\left(\delta(t-u) -
\sum_{k=1}^K
\widehat \psi_k(t)  \widehat \psi_k(u)\right) \nonumber\\
&=& \sum_{k\neq \nu}^K \frac{\widehat \lambda_k}{\widehat
\lambda_\nu (\widehat \lambda_k - \widehat \lambda_\nu)} \widehat
\psi_k(t) \widehat \psi_k(u) + \frac{1}{\widehat \lambda_\nu}
\widehat\psi_\nu(t) \widehat \psi_\nu(u) - \frac{1}{\widehat
\lambda_\nu}\delta(t-u),
\end{eqnarray}
with $\delta$ being the Dirac $\delta$- function, i.e., $\int
\delta(t-u)w(u) du = w(t)$ for any smooth $w \in L^2([0,1])$. Here
$\tr(\widehat P_\nu \Delta_i)$ and $(\widehat H_\nu \Delta_i
\widehat \psi_\nu)(t)$ are defined as follows:
\item[(a')]
\begin{equation}\label{eq:trace_P_Delta}
\tr(\widehat P_\nu \Delta_i) = \int \widehat \psi_\nu(u)
\Delta_i(u,v) \widehat \psi_\nu(v) du dv;
\end{equation}
\item[(b')]
\begin{equation}\label{eq:H_Delta_psi} (\widehat H_\nu \Delta_i
\widehat \psi_\nu)(t) = \int \int \widehat H_\nu(t,u)
\Delta_i(u,v) \widehat \psi_\nu(v) du dv.
\end{equation}
\end{itemize}
Also,
\begin{itemize}
\item[(iii)]
\begin{equation}\label{eq:mu_hat_minus_i}
\widehat \mu^{(-i)}(t)  - \widehat \mu(t) = \frac{1}{n-1} \widehat
\mu(t) - \frac{1}{n-1} \frac{1}{m_i} \sum_{j=1}^{m_i} Y_{ij}
K_{h_\mu}(t-T_{ij}),
\end{equation}
\end{itemize}
where $\widehat \mu(t) = \frac{1}{n} \sum_{i=1}^n
\frac{1}{m_i}\sum_{j=1}^{m_i} Y_{ij} K_{h_\mu}(t-T_{ij})$, with
$h_\mu$ being the bandwidth for estimating $\mu$ (chosen
separately).
\end{prop}

\vskip.1in After we obtain the approximations for
$\widehat{\boldsymbol{\psi}}_{i \nu}^{(-i)}$ and $\widehat
\lambda_\nu^{(-i)}$ from Proposition \ref{prop2}, we plug them
back in equation (\ref{eq:CV_KL}) for $CV_*(K,h_n)$ to obtain the
final approximation of the CV score, denoted by
$\widetilde{CV}_*(K,h_n)$:
\begin{equation*}
\widetilde{CV}_*(K,h_n) = \frac{1}{2n} \sum_{i=1}^n \log
|\widetilde\Sigma_{i}| + \frac{1}{2n} \sum_{i=1}^n
\tr(\widetilde\Sigma_{i}^{-1} (\mathbf{Y}_i -
\widehat{\boldsymbol{\mu}}_i^{(-i)}) (\mathbf{Y}_i -
\widehat{\boldsymbol{\mu}}_i^{(-i)})^T),
\end{equation*}
where $\widetilde \Sigma_i = \sum_{\nu=1}^K \widetilde
\lambda_{i\nu} \widetilde{\boldsymbol{\psi}}_{i\nu}
\widetilde{\boldsymbol{\psi}}_{i\nu}^T + \widehat \sigma_{(-i)}^2
I_{m_i}$, with
\begin{equation*}
\widetilde \lambda_{i\nu} = \widehat \lambda_\nu - \tr(\widehat
P_\nu \Delta_i) \qquad \mbox{and} \qquad
\widetilde{\boldsymbol{\psi}}_{i\nu} =
\widehat{\boldsymbol{\psi}}_{i\nu} + ((\widehat H_\nu \Delta_i
\widehat \psi_\nu)(T_{ij}))_{j=1}^{m_i},
\end{equation*}
and $\widehat{\boldsymbol{\mu}}_i^{(-i)} =
(\widehat\mu^{(-i)}(T_{ij}))_{j=1}^{m_i}$, with $\widehat
\mu^{(-i)}$ given by (\ref{eq:mu_hat_minus_i}). An expression for
$\widehat \sigma_{(-i)}^2 - \frac{n}{n-1}\widehat \sigma^2$ is
easily obtained by using (\ref{eq:diagonal_est}),
(\ref{eq:C_bar_hat_0}) and (\ref{eq:sigma_hat}). Note that this step
does not require any extra computation beyond that for computing
$\widehat\sigma^2$.

Observe that our objective of minimizing  the criterion
$\widetilde{CV}_*(K,h_n)$ is to estimate the number of nonzero
eigenvalues and to select an appropriate bandwidth for estimating
the eigenfunctions. If instead the objective is to select an
appropriate bandwidth for estimating the covariance kernel, we can
do so by replacing the term $\sum_{\nu=1}^K \widetilde
\lambda_{i\nu} \widetilde{\boldsymbol{\psi}}_{i\nu}
\widetilde{\boldsymbol{\psi}}_{i\nu}^T $ in the definition of
$\widetilde \Sigma_i$ with the leave-one-curve-out estimate of
covariance kernel, viz. $\widehat C_{c,h_n}^{(-i)}$ evaluated at the
design points, and minimizing the corresponding CV criterion. This
distinction is important since the theoretical results (Theorems
\ref{thm1} and \ref{thm2}) show that the optimal rates for the
bandwidth $h_n$ are different for estimating the covariance kernel
and its eigenfunctions.


\subsection{Representation of $\widehat H_\nu \Delta_i \widehat \psi_\nu$ and
$\tr(\widehat P_\nu \Delta_i)$}

In order to obtain the approximate CV score
$\widetilde{CV}_*(K,h_n)$ efficiently, we need to compute the
quantities $(\widehat H_\nu \Delta_i \widehat \psi_\nu)(t)$ and
$\tr(\widehat P_\nu \Delta_i)$ in an efficient manner.
Thus we have the following further approximation based on Lemma
\ref{lemmaA.1}.


\begin{prop}\label{prop3}
We have
\begin{itemize}
\item[(i)]
\begin{eqnarray}\label{eq:first_order_psi}
&&(\widehat H_\nu \Delta_i \widehat \psi_\nu)(t) \nonumber\\
&\approx& \frac{w(m_i)}{n-1} \sum_{k\neq \nu}^K \frac{\widehat
\lambda_k}{\widehat \lambda_\nu (\widehat \lambda_k - \widehat
\lambda_\nu)} \gamma_{k,h_n}(i) \gamma_{\nu,h_n}(i) \widehat
\psi_k(t) \nonumber\\
&&+ \frac{w(m_i)}{n-1} \frac{1}{\widehat \lambda_\nu}
(\gamma_{\nu,h_n}(i))^2 \widehat \psi_\nu(t)\nonumber\\
&& - \frac{w(m_i)}{n-1} \frac{1}{\widehat \lambda_\nu}\frac{1}{g(t)}
\sum_{l=1}^{L_n}(\widetilde X_i(s_l) +(t-s_l)\widetilde
X_i'(s_l))\overline{Q}_{h_n}(s_l - t)\widetilde
\gamma_{\nu,h_n}(i,t) \nonumber\\
&& - \frac{w(m_i)}{n-1} \left[\sum_{k\neq \nu}^K \frac{\widehat
\lambda_k}{\widehat \lambda_\nu (\widehat \lambda_k - \widehat
\lambda_\nu)} \overline{\gamma}_{k,\nu,h_n}(i)\widehat\psi_k(t) +
\frac{1}{\widehat \lambda_\nu}
\overline{\gamma}_{\nu,\nu,h_n}(i)\widehat\psi_\nu(t)\right]
\nonumber\\
&& + \frac{1}{n-1} \sum_{k\neq \nu}^K \frac{\widehat
\lambda_k}{\widehat \lambda_\nu (\widehat \lambda_k - \widehat
\lambda_\nu)} \widehat \psi_k(t) \sum_{l=1}^{L_n} \int
\frac{\widehat \psi_k(u)\widehat\psi_\nu (u)}{g(u)} (S_i(s_l)
\beta_{1,h}(u,s_l) +
S_i'(s_l) \beta_{2,h}(u,s_l)) du \nonumber\\
&& + \frac{1}{n-1} \frac{1}{\widehat\lambda_\nu} \widehat
\psi_\nu(t) \sum_{l=1}^{L_n} \int \frac{(\widehat
\psi_\nu(u))^2}{g(u)} (S_i(s_l) \beta_{1,h}(u,s_l) + S_i'(s_l)
\beta_{2,h}(u,s_l)) du \nonumber\\
&& ~~~~~  -  \frac{1}{n-1} \frac{1}{\widehat\lambda_\nu}
\sum_{l=1}^{L_n} (S_i(s_l) \beta_{1,h}(t,s_l) + S_i'(s_l)
\beta_{2,h}(t,s_l) )
\frac{\widehat \psi_\nu(t)}{g(t)}\nonumber\\
&& +~ (\widehat \sigma_{(-i)}^2 - \frac{n}{n-1} \widehat\sigma^2)
(\int \widehat H_\nu(t,u) du) (\int \widehat \psi_\nu(u)du);
\end{eqnarray}
\item[(ii)]
\begin{eqnarray}\label{eq:first_order_lambda}
\tr(\widehat P_\nu \Delta_i) &\approx& -\frac{\widehat
\lambda_\nu}{n-1} + \frac{w(m_i)}{n-1} \left[(\gamma_{\nu,h_n}(i))^2
- \overline{\gamma}_{\nu,\nu,h_n}(i)\right]\nonumber\\
&& + \frac{1}{n-1}  \int \frac{(\widehat \psi_\nu(u))^2}{g(u)}
(S_i(u) \beta_{1,h}(u) + S_i'(u) \beta_{2,h}(u)) du,\nonumber \\
&& + (\widehat \sigma_{(-i)}^2 - \frac{n}{n-1}\widehat\sigma^2)
(\int \widehat \psi_\nu(u) du)^2
\end{eqnarray}
\end{itemize}
where
\begin{itemize}
\item[(a)]
\begin{equation}\label{eq:gamma}
\gamma_{k,h_n}(i) = \sum_{l=1}^{L_n} \widetilde X_i(s_l)
G_0\left(\frac{\widehat \psi_k}{g},\overline{Q}_{h_n}\right) (s_l) -
\sum_{l=1}^{L_n} \widetilde X_i'(s_l) G_1 \left(\frac{\widehat
\psi_k}{g},\overline{Q}_{h_n}\right)(s_l),
\end{equation}
where, for any two functions $f_1$ and $f_2$ defined on $[0,1]$,
\begin{eqnarray*}
G_0(f_1,f_2)(s) &=& (f_1 * f_2)(s) = \int f_1(x)f_2(s-x) dx, \\
G_1(f_1,f_2)(s) &=& (f_1 * (x f_2))(s) = \int f_1(x) (s-x) f_2(s-x)
dx;
\end{eqnarray*}
\item[(b)]
\begin{equation}\label{eq:gamma_tilde}
\widetilde \gamma_{k,h_n}(i,t) = \sum_{l=1}^{L_n}\int
(1-W_{\widetilde h_n}(t,v)) [(\widetilde X_i(s_l) +
(v-s_l)\widetilde X_i'(s_l)) \overline{Q}_{h_n}(s_l-v)]
\frac{\widehat\psi_k(v)}{g(v)} dv;
\end{equation}
\item[(c)]
\begin{eqnarray}\label{eq:gamma_bar}
\overline{\gamma}_{k,k',h_n}(i)  &=&  \sum_{j,j'=0}^1
\sum_{l,l'=1}^{L_n}
 X_i^{(j)}(s_l) X_i^{(j')}(s_{l'}) \int W_{\widetilde h_n}(u,v) \frac{\widehat
\psi_k(u)}{g(u)} \frac{\widehat \psi_{k'}(v)}{g(v)}  \nonumber\\
&& ~~~~~~\cdot (u-s_l)^j (v-s_{l'})^{j'}\overline{Q}_{h_n}(s_l - u)
\overline{Q}_{h_n}(s_{l'} - v) du dv;
\end{eqnarray}
\item[(d)]
\begin{eqnarray}\label{eq:beta_def}
\beta_{1,h}(u,s) &=& \int_{u-\frac{A}{2}h}^{u+\frac{A}{2}h}
\overline{Q}_h(\frac{u+v}{2}
- s) dv \\
\beta_{2,h}(u,s) &=& \int_{u-\frac{A}{2}h}^{u+\frac{A}{2}h}
(\frac{u+v}{2} - s)\overline{Q}_h(\frac{u+v}{2} - s) dv.
\end{eqnarray}
\end{itemize}
\end{prop}

In the above, the computation of $\gamma_{k,h_n}(i)$ can be easily
done by using fast fourier transformation. Also, $\widetilde
\gamma_{k,h_n}(i,t) \approx \gamma_{k,h_n}(i)$ for all $t \in
[0,1]$. However, the computation of
$\overline{\gamma}_{k,k',h_n}(i)$ involves a double integration.
Thus we need to do some approximations to simplify the
computation. A computationally efficient approximation to
$\overline{\gamma}_{k,k',h_n}(i)$ is described in
Appendix B. Computation of $\beta_{1,h}(u)$ and $\beta_{2,h}(u)$ can
be done in closed form whenever $\overline{Q}_h(\cdot)$ has a
``nice'' functional form (e.g. a B-spline). From Propositions
\ref{prop2} and \ref{prop3} it is clear that most of the components
have already been computed in constructing the estimator, and
convolutions can be performed in a fast manner by using FFT. Thus,
the key advantage afforded by Proposition \ref{prop3} is to replace
the expensive computation of double integrals to a much cheaper
computation of single integrals and convolutions. See
Appendix F for details of some of these steps.

\section{Asymptotic properties}\label{sec:asymptotics}

In this section, we present the theoretical properties of the
proposed estimators through a large sample analysis. Our main
interest is in the estimation accuracy of the covariance kernel
and its eigenfunctions. The statements of the results and the
associated regularity conditions are given below.


We first state the following assumptions on $g$, the density of
the design points; $C$, the covariance kernel; and
$\{\psi_k\}_{k=1}^M$, the eigenfunctions.
\begin{itemize}
\item[{\bf A1}] $g$ is twice continuously differentiable and the
second derivative is H\"{o}lder$(\alpha)$, for some $\alpha \in
(0,1)$. Also, the same holds for the covariance kernel $C$.
\item[{\bf A2}] $\max_k \{\parallel
\psi_k\parallel_\infty,\parallel \psi_k'\parallel_\infty,
\parallel \psi_k''\parallel_\infty\}$ is bounded. \item[{\bf A3}]
There are constants $0 < c_0 \leq c_1 < \infty$ such that $c_0
\leq g(\cdot) \leq c_1$.
\end{itemize}
We also assume that the kernels $K(\cdot)$ and
$\overline{Q}(\cdot)$ satisfy conditions {\bf B1} and {\bf B2},
respectively. We need to make further assumptions about the
covariance kernel $C$ and the correlations among the sample
curves. Let $\mathbf{R}$ denote an $n \times n$ matrix with
$(i,j)$-th entry $\rho_{ij}$. Assume:
\begin{itemize}
\item[{\bf C1}] $\lambda_1 > \lambda_2 > \cdots > \lambda_M > 0$
and $\lambda_{M+1} = \cdots = 0$. That is, the nonzero eigenvalues
are all distinct and the covariance kernel is of finite dimension.
\item[{\bf C2}] $\max_{1\leq \nu \leq M} (\lambda_\nu -
\lambda_{\nu+1})^{-1}$ is bounded above. \item[{\bf C3}]
$\frac{1}{n^2} \tr[(\mathbf{R}-I_n)^2] \to 0$ as $n \to \infty$,
and $\parallel \mathbf{R} \parallel \leq \kappa_n$ for $\kappa_n >
0$.
\end{itemize}
Note that, the first part of {\bf C3} quantifies the total
contribution of the correlations among the sample curves in the
variance of the estimated covariance kernel (see Theorem
\ref{thm1}). The second part of {\bf C3} imposes a stability
condition on the correlation matrix ${\bf R}$. In other words, the
sample curves are ``weakly correlated'' as $\parallel \mathbf{R}
\parallel$ is bounded by $\kappa_n$. Define $\underline{m} =
\min_{1\leq i \leq n} m_i$ and $\overline{m} = \max_{1\leq i \leq n}
m_i$. We further assume that
\begin{itemize}
\item[{\bf C4}] $\overline{m}/\underline{m}$ is bounded above as
$n\to \infty$.
\end{itemize}
We now give the bias and variance of the proposed combined
estimator.


\begin{thm}\label{thm1}
Suppose that conditions {\bf A1}-{\bf A3}, {\bf B1}-{\bf B2}  and
{\bf C3}-{\bf C4} hold. Assume further that $\sigma^2$ is known
and $\widehat{\overline{C}}(\cdot) = \widehat C_*(\cdot)
-\sigma^2$ where $\widehat C_*(\cdot)$ is defined through
(\ref{eq:diagonal_est}). Suppose further that in the definition
(\ref{eq:C_est_combined}), $\widetilde h_n = Ah_n$ for some
constant $A \geq 4(B_K + C_Q)$. Then, with $h_n = o(1)$ and $n
h_n^2 \to \infty$, the estimator $\widehat C_c$ satisfies:
\begin{eqnarray}
\mathbb{E}[\widehat C_c (s,t)] &=& C(s,t) + O(h_n^2),
\label{eq:C_hat_bias}\\
Var[\widehat C_c (s,t)] &=& O\left(\frac{1}{n}\right) +
O\left(\max\{\frac{1}{nh_n^2\underline{m}^2},\frac{1}{n h_n
\underline{m}}\} \right)  + \left(\frac{1}{n^2} \sum_{i\neq j}^n
\rho_{ij}^2\right) O(1), \label{eq:C_hat_variance}
\end{eqnarray}
where the $O(\cdot)$ terms are uniform in $s,t \in [0,1]$.
\end{thm}

\vskip.1in\noindent One implication of Theorem \ref{thm1} is that
it gives the rate of convergence of the estimator
$\widehat\sigma^2$ defined in (\ref{eq:sigma_hat}) as illustrated
in the following corollary.

\begin{cor}\label{cor1}
Suppose that conditions {\bf A1}-{\bf A3}, {\bf B1}-{\bf B2} and
{\bf C3}-{\bf C4} hold, and in the definition
(\ref{eq:C_est_combined}), $\widetilde h_n = Ah_n$ for some constant
$A \geq 4(B_K + C_Q)$. Then, with $h_n = o(1)$ and $n h_n^2 \to
\infty$,
\begin{equation}\label{eq:sigma_risk}
\mathbb{E}(\widehat \sigma^2 - \sigma^2)^2 =
O\left(\frac{1}{n}\right) +
O\left(\max\{\frac{1}{nh_n^2\underline{m}^2},\frac{1}{n h_n
\underline{m}}\} \right)  + \left(\frac{1}{n^2} \sum_{i\neq j}^n
\rho_{ij}^2\right) O(1) + O(h_n^4),
\end{equation}
where the $O(\cdot)$ terms are uniform in $s,t \in [0,1]$.
\end{cor}

\vskip.1in\noindent Using Corollary \ref{cor1} and Theorem
\ref{thm1}, we get a bound on the variance of the proposed
estimator of the covariance kernel when $\sigma^2$ is estimated by
$\widehat\sigma^2$ defined in (\ref{eq:sigma_hat}).

\begin{cor}\label{cor2}
Suppose that conditions {\bf A1}-{\bf A3}, {\bf B1}-{\bf B2} and
{\bf C3}-{\bf C4} hold, and in the definition
(\ref{eq:C_est_combined}), $\widetilde h_n = Ah_n$ for some constant
$A \geq 4(B_K + C_Q)$. Then, with $h_n = o(1)$ and $n h_n^2 \to
\infty$,
\begin{equation}\label{eq:C_hat_variance_sigma}
Var[\widehat C_c (s,t)] = O\left(\frac{1}{n}\right) +
O\left(\max\{\frac{1}{nh_n^2\underline{m}^2},\frac{1}{n h_n
\underline{m}}\} \right) + \left(\frac{1}{n^2} \sum_{i\neq j}^n
\rho_{ij}^2\right) O(1) + O(h_n^4),
\end{equation}
where the $O(\cdot)$ terms are uniform in $s,t \in [0,1]$.
\end{cor}

\vskip.1in Next we state the result about the asymptotic behavior
of the estimated eigenfunctions. Let the loss function for
$\psi_\nu$ be the \textit{modified $L^2$-loss} given by
\begin{equation}\label{eq:psi_loss}
L(\widehat \psi_\nu,\psi_\nu) = \parallel \widehat\psi_\nu -
\mbox{sign}(\langle \widehat
\psi_\nu,\psi_\nu\rangle)\psi_\nu\parallel_2^2,
\end{equation}
where $\parallel \cdot \parallel_2$ denotes the $L^2$ norm, and
$\langle \widehat \psi_\nu,\psi_\nu\rangle = \int_0^1
\widehat\psi_\nu(x)\psi_\nu(x)dx$. For the statement of Theorem
\ref{thm2}, we only need to assume that the estimator $\widehat
\sigma^2$ of $\sigma^2$ satisfies $\mathbb{E}(\widehat \sigma^2 -
\sigma^2)^2 = o(1)$.

\begin{thm}\label{thm2}
Suppose that conditions {\bf A1}-{\bf A3}, {\bf B1}-{\bf B2} and
{\bf C1}-{\bf C4} hold. Suppose further that in the definition
(\ref{eq:C_est_combined}), $\widetilde h_n = Ah_n$ for some constant
$A > 4(B_K + C_Q)$. If ~$\overline{m} h_n = o(1)$, $n h_n^2 \to
\infty$ and $\kappa_n \overline{m} h_n^{-1} n^{-1/2+\epsilon'} \to
0$ for some $\epsilon'
>0$, then the estimator $\widehat \psi_\nu$, which is the eigenfunction
corresponding to the $\nu$-th largest eigenvalue of $\widehat C_c$,
satisfies: for any arbitrary but fixed $\epsilon > 0$,
\begin{eqnarray}\label{eq:psi_hat_risk}
\sup_{(C,g) \in \Theta} \mathbb{E}L(\widehat\psi_\nu,\psi_\nu)
&\leq& (1+\epsilon) \frac{1}{n}\left(\sum_{1\leq k\neq \nu\leq M}
\frac{\lambda_k\lambda_\nu}{(\lambda_k-\lambda_\nu)^2}\right)\nonumber \\
&& + (1+\epsilon) \left(\frac{1}{n^2}\sum_{i\neq
j}^{n}\rho_{ij}^2\right)\left(\sum_{1\leq k\neq \nu\leq M}
\frac{\lambda_k\lambda_\nu}{(\lambda_k-\lambda_\nu)^2} +
O(h_n)\right) \nonumber\\
&& ~~~~~~+ O(h_n^4)  + O\left(\frac{1}{nh_n\underline{m}}\right),
\end{eqnarray}
where $\Theta$ denotes the class of covariance-density pairs $(C,g)$
satisfying the conditions {\bf A1}-{\bf A3}, {\bf B1}-{\bf B2} and
{\bf C1}-{\bf C4}.
\end{thm}



\vskip.1in One important implication of Theorems \ref{thm1} and
\ref{thm2} is that, if the correlation between sample curves is
``weak'' in a suitable sense, then the best rate of convergence for
the correlated and i.i.d. cases are the same. Comparing with the
i.i.d. case, we immediately see that, in order for this to hold,
under the conditions of Theorem \ref{thm2}, we need
\begin{equation}\label{eq:rho_optimal}
\frac{1}{n^2} \sum_{j\neq i} \rho_{ij}^2 = o\left(\frac{1}{n h_{n,*}
\underline{m}}\right),
\end{equation}
where $h_{n,*}$ is the optimal bandwidth choice (at the level of
rates of convergence) for the i.i.d. case.
Also, in order to ensure the optimal rate for the estimate of the
covariance kernel in the correlated case is the same as that in the
i.i.d. case, it is sufficient that (by Corollary \ref{cor2})
\begin{equation}\label{eq:rho_optimal_covar}
\frac{1}{n^2} \sum_{j\neq i} \rho_{ij}^2 = o\left(\max\{\frac{1}{n
h_{n,*} \underline{m}},\frac{1}{n h_{n,*}^2
\underline{m}^2}\}\right),
\end{equation}
where $h_{n,*}$ is the optimal bandwidth choice for the covariance
estimator (at the level of rates of convergence) for the i.i.d.
case. Specifically, for estimating the covariance, $h_{n,*} = {(n
\underline{m}^2)}^{-1/6}$ (by Theorem \ref{thm1}, and under the
setting where $\overline{m}h_{n,*} = o(1)$), and for estimating the
eigenfunctions, $h_{n,*} = {(n \underline{m})}^{-1/5}$ (by Theorem
\ref{thm2}). Thus, one notices that the optimal bandwidth for
estimating the covariance and its eigenfunctions are different, at
least in the case where $\overline{m}$ can only grow rather slowly
with $n$. Combining the lower bound given by (Theorem 2) in Hall
\textit{et al.} (2006) 
and the upper bound from Theorem
\ref{thm2}, it follows that when $\overline{m}$ is bounded, the rate
of convergence of $L^2$-risk is optimal if (\ref{eq:rho_optimal})
holds. Thus, under this setting the proposed estimator of the
eigenfunctions is optimal even in the situation when the sample
curves are \textit{weakly correlated}. Similarly, under the setting
of Theorem \ref{thm1}, if (\ref{eq:rho_optimal_covar}) holds, then
the $L^2$-risk of the proposed estimator of covariance also has the
optimal rate under an appropriate choice of bandwidth.

\vskip.1in Another important point is that the conditions in Theorem
\ref{thm2}, specifically that $\overline{m}h_n = o(1)$, $n h_n^{2}
\to \infty$, and $\overline{m} \kappa_n h_n^{-1} n^{-1/2 +
\epsilon'} = o(1)$, which imply that $\overline{m} = o(n^{1/4})$,
are not the most general conditions. We conjecture that
(\ref{eq:psi_hat_risk}) hold under weaker conditions.
Indeed, in the i.i.d. case, (\ref{eq:psi_hat_risk}) holds (without
the second term on the RHS) under a much wider range of possible
values of $\overline{m}$ as indicated by the following result. The
following result gives a lower bound on the rate of convergence of
the first eigenfunction when $\overline{m} \to \infty$ under the
i.i.d. setting. This bound is a refinement over an analogous result
(Theorem 2) in Hall \textit{et al.} (2006), 
even though the latter holds for all eigenfunctions. Notice that
this lower bound, together with the upper bound elucidated in the
paragraph following Theorem \ref{thm2}, implies that at least for
the first eigenfunction, the best rate of convergence for
eigenfunctions, viz. $O((n\underline{m})^{-4/5})$ is optimal when
$\overline{m}\to \infty$ at a faster rate and if
(\ref{eq:rho_optimal}) holds.


\begin{thm}\label{thm3}
Let ${\cal C}$ denote the class of covariance kernels
$\overline{\Sigma}(\cdot,\cdot)$ on $[0,1]^2$ with rank $\geq 1$,
and nonzero eigenvalues $\{\lambda_j\}_{j \geq 1}$ satisfying $C_0
\geq \lambda_1 > \lambda_2 \geq 0$ with $\lambda_1 - \lambda_2
\geq C_1$, and the first eigenfunction $\psi_1$ being twice
differentiable and satisfying $\parallel
\psi_1^{\prime\prime}\parallel_\infty \leq C_2$, for some
constants $C_0, C_1, C_2 > 0$. Also, let ${\cal G}$ denote the
class of continuous densities $g$ on $[0,1]$ such that $c_1 \leq g
\leq c_2$ for some $0 < c_1 \leq 1 \leq c_2 < \infty$. Suppose
that we observe data according to models (\ref{eq:model}) where
$X_i(\cdot)$ are i.i.d. Gaussian processes with mean 0 and
covariance kernel $\overline{\Sigma}$. Also suppose that the
number of measurements $m_i$'s satisfy $\underline{m} \leq m_i
\leq \overline{m}$, for $\overline{m} \geq \underline{m} \geq 4$,
such that $\overline{m}/\underline{m} \leq C_3$ for some $C_3 <
\infty$, and $\overline{m} = o(n^{2/3})$. Let ${\cal D}$ denote
the space of such designs $D=\{m_i\}_{i=1}^n$. Then for
sufficiently large $n$, for any estimator $\widehat \psi_1$ with
$l_2$ norm one, the following holds:
\begin{equation}
\sup_{D \in {\cal D}} \sup_{g \in {\cal G}} \sup_{\overline{\Sigma}
\in {\cal C}} \mathbb{E}\parallel \widehat \psi_1 - \psi_1
\parallel_2^2 \geq C_4 (n \underline{m})^{-4/5}.
\end{equation}
\end{thm}

\vskip.1in The proof of Theorem \ref{thm3} is given in Appendix G.

\section{Outline of the Proof of Theorems \ref{thm1} and \ref{thm2}}\label{sec:proof_thm1and2}

In this section, we briefly describe the main ideas leading to the
proof of Theorems \ref{thm1} and \ref{thm2}. The technical
arguments are given in the appendices. The proof of Theorem
\ref{thm1} uses direct computation (Appendices C and D). The basic
idea in the computation of the moments is to treat the diagonal
and the off-diagonal parts of $\widehat C_c(\cdot,\cdot)$
separately. The proof of Theorem \ref{thm2} heavily relies on an
application of Lemma \ref{lemmaA.1}. In view of this, the key
quantity in the derivation of asymptotic risk is the computation
of $\mathbb{E}\parallel H_\nu \widehat C_c \phi_\nu\parallel_2^2$,
where $\parallel f
\parallel_2^2$ denotes $\int_0^1 f^2(x) dx$ for a function $f \in
L^2([0,1])$. Once we obtain an expression for this (as given in
Section \ref{subsec:risk_psi}), we use a probabilistic bound on
the operator norm of the difference between estimated and true
covariance kernels, to complete the proof. Proofs of Theorems
\ref{thm1} and \ref{thm2} require repeated computation of mixed
moments of correlated Gaussian random variables. The details of
all these computations are given in the appendices.

\subsection{Asymptotic risk for estimating
$\psi_\nu$}\label{subsec:risk_psi}


The key result in this section is the following proposition.

\begin{prop}\label{prop4}
Under the assumptions of Theorem \ref{thm2}, we have
\begin{eqnarray}\label{eq:expect_H_C_psi_L_2}
\mathbb{E}\parallel H_\nu \widehat C_c\psi_\nu
\parallel_2^2
&\leq& (1+\epsilon)\frac{1}{n}\left(\sum_{1\leq k\neq \nu\leq M}
\frac{\lambda_k\lambda_\nu}{(\lambda_k-\lambda_\nu)^2}\right)
\nonumber\\
&& + (1+\epsilon) \left(\frac{1}{n^2}\sum_{i_1\neq
i_2}^{n}\rho_{i_1i_2}^2\right)\left(\sum_{1\leq k\neq \nu\leq M}
\frac{\lambda_k\lambda_\nu}{(\lambda_k-\lambda_\nu)^2} +
O(h_n)\right) \nonumber\\
&& ~~~~~~+ O(h_n^4)  + O\left(\frac{1}{nh_n\underline{m}}\right)
\end{eqnarray}
for any arbitrary but fixed $\epsilon > 0$.
\end{prop}

\vskip.1in Here we briefly describe the main idea of the proof. For
convenience of exposition, throughout we replace
$\max\{\widehat{\overline{C}}(\frac{s+t}{2}),h_n^2\}$ in the
definition (\ref{eq:C_est_combined}) by
$\widehat{\overline{C}}(\frac{s+t}{2})$. Using appropriate
exponential inequalities for $\widehat C_*(t)$, it can be shown
that, asymptotically this does not make any difference as long as
$\min_{t \in [0,1]} C(t,t) > c_3$ for some $c_3 > 0$. Also, for
computational purposes, it is helpful to consider the unsmoothed
version (\ref{eq:W_kernel}) of the kernel $W$, and take $\widetilde
h_n = A h_n$, where $A \geq 4(B_K + C_Q)$. The advantage of this is
in being able to deal with the contributions from the diagonal and
off-diagonal parts of the estimator separately. Since the definition
of $H_\nu$ involves the Dirac-$\delta$ operator, we need to account
for the contribution of terms involving $\delta$ carefully. The
estimation error in $\sigma^2$ also plays a role, and is taken into
account separately. The main decompositions that facilitate the
computations are given through (\ref{eq:H_nu_repr}) -
(\ref{eq:sigma_contr_bound}) in  Appendix D. The last bound reduces
the task of bounding $\mathbb{E}\parallel H_\nu \widehat C_c
\psi_\nu\parallel_2^2$ to that of bounding $\mathbb{E}\parallel
H_\nu \widetilde C_c \psi_\nu\parallel_2^2$, with $\widetilde
C_c(\cdot,\cdot)$ as described in
Appendix D. Note also that, if $\sigma^2$ is assumed to be known,
then the decomposition (\ref{eq:sigma_contr_bound}) is not required,
and we can get rid of the multiplicative factor $(1+\epsilon)$ in
the expression (\ref{eq:psi_hat_risk}) for the risk in Theorem
\ref{thm2}.

\subsection{Norm bound on $\widehat C_c - \mathbb{E}\widehat C_c$}\label{subsec:norm_bound}

To complete the proof of the theorems, we need to find a
probabilistic bound for $\parallel \widehat C_c -
\mathbb{E}\widehat C_c\parallel$, where $\parallel \cdot
\parallel$ denotes the operator norm. We shall first find a bound
on the sup norm of $\widehat C_c - \mathbb{E}\widehat C_c$, and
then we can bound the operator norm $\parallel \widehat C_c -
\mathbb{E}\widehat C_c\parallel$ via the inequality $\parallel
\widehat C_c - \mathbb{E}\widehat C_c\parallel \leq
\parallel C_c - \mathbb{E}\widehat C_c\parallel_F$, where
$\parallel \cdot\parallel_F$ denotes the Hilbert-Schmidt norm.
This is in turn due to the inequality,
\begin{equation*}
\parallel \widehat C_c - \mathbb{E} \widehat C_c \parallel_F \leq
\sup_{x,y \in [0,1]} |\widehat C_c(x,y) - \mathbb{E}\widehat
C_c(x,y)| =: \parallel \widehat C_c - \mathbb{E}\widehat
C_c\parallel_\infty.
\end{equation*}
Note that, by piecewise differentiability of the estimate $\widehat
C_c$, in order to provide exponential bounds for the deviations of
$\parallel \widehat C_c - \mathbb{E}\widehat C_c\parallel_\infty$,
it is enough to provide exponential bounds for the fluctuations of
$|\widehat C_c(s,t) - \mathbb{E}[\widehat C_c(s,t)]|$ for a finite
(but polynomially growing with $n$) number of points $(s,t) \in
[0,1]$. Thus, we fix an arbitrary $(s,t)\in [0,1]$ and derive an
exponential inequality for the deviation of estimate at this point.
For simplifying the computations, without loss of generality, we
assume that $g$ is the density of the Uniform(0,1) distribution.
Then we have the following proposition.

\begin{prop}\label{prop5}
Under the conditions of Theorem \ref{thm2}, given $\eta > 0$, there
is a $c_\eta
> 0$ such that, for every fixed $s,t \in [0,1]$,
\begin{equation}\label{eq:hat_C_c_sup_norm_bound}
\mathbb{P}\left(|\widehat C_c(s,t) - \mathbb{E}(\widehat C_c(s,t))|
> c_\eta \overline{m} \kappa_n \sqrt{\frac{\log n}{nh_n^2}}\right)
\leq n^{-\eta}.
\end{equation}
\end{prop}


The proof of Theorem \ref{thm2} then follows by noticing first that
by Lemma \ref{lemmaA.1} and the fact that $\parallel
\widehat\psi_\nu
\parallel_2 = \parallel \psi_\nu \parallel_2 = 1$,
\begin{equation*}
\mathbb{E}L(\widehat \psi_\nu,\psi_\nu) \leq \mathbb{E}\parallel
H_\nu \widehat C_c\psi_\nu\parallel_2^2 (1+\delta_{n,\eta}) + 2
\mathbb{P}\left(\parallel \widehat C_c - \mathbb{E}(\widehat
C_c)\parallel > c_\eta' \overline{m} \kappa_n \sqrt{\frac{\log
n}{nh_n^2}}\right)
\end{equation*}
for some $\eta > 0$, $c_\eta' > 0$ and $\delta_{n,\eta} \to 0$
appropriately chosen, and then using Propositions \ref{prop4} and
\ref{prop5}.

\subsection{Connection to parametric rate for ``purely functional''
data}

It is instructive to compare the optimal rate for our procedure with
that obtained by Hall \textit{et al.} (2006). 
We can regard the first line on the right hand side  of
(\ref{eq:expect_H_C_psi_L_2}), as the parametric component of the
risk and the second line as the nonparametric component. If we take
$h= O(n^{-1/5})$, then for bounded $\overline{m}$ we get the optimal
nonparametric rate. For consistency of $\widehat \psi_\nu$ in $L^2$
sense, we clearly need $\frac{1}{n^2}\sum_{i_1\neq
i_2}\rho_{i_1i_2}^2 = o(1)$ (used in Theorem \ref{thm2}). If
$\underline{m}$ increases with increasing sample size, then the rate
also improves. But there is no result about optimality.

When the observations are i.i.d., it can be checked by using a
modification to the proof of Proposition \ref{prop5} that, if
$\underline{m}\to \infty$, $h \to 0$, such that
$(\underline{m}h)^{-1} = o(1)$ and $h = o(n^{-1/4})$, we obtain the
parametric rate for the $L^2$-risk of $\widehat\psi_\nu$ (as
indicated in Hall \textit{et al.}, 2006). 
In other words, under that setting there is asymptotically no
difference between the risk of estimating the eigenfunctions from
data obtained with observational noise and measured at randomly
distributed points, and that from data measured on the continuum
without noise. Indeed, such a scenario is possible if
$\underline{m}^{-1} = o(n^{-1/4-\epsilon})$, for an $\epsilon
>0$. Then, by taking $h_n = o(n^{-1/4})$, and assuming that
either $\sigma^2$ is known, or an estimator $\widehat\sigma^2$
satisfying $|\widehat\sigma^2 - \sigma^2| = O_P(h_n^2)$ is
available, we attain the conditions mentioned above. We conjecture
that the same result holds even when the observations are ``weakly"
correlated.




\section{Discussion}\label{sec:discussion}

In this paper, we presented a procedure for estimating the
covariance kernel and its eigenfunctions from sparsely observed,
noise corrupted and correlated functional data. The estimator for
the covariance kernel is based on merging two separate estimators:
(i) the estimator of the off-diagonal part based on computing
linearized empirical covariances of the smoothed version of
individual sample curves; (ii) the estimator of the diagonal part
based on linearized kernel smoothing of the empirical variances.
The importance of this modification to the naive kernel smoothing
approach, especially in the scenario when the number of design
points per curve is small, is demonstrated through an asymptotic
bias analysis. The linearized version of the kernel smoothing
helps in reducing bias, while controlling the variance, and is
computationally appealing. Asymptotic risk behavior of the
proposed estimators is studied under the assumption that the
sample curves have a ``separable covariance'' structure and are
``weakly" correlated. Exact quantification of the asymptotic risk
for the eigenfunctions is obtained under the Gaussian setting
(Theorem \ref{thm2}). It is also shown that the $L^2$-risk for the
eigenfunctions achieves the optimal rate, under an appropriate
choice of the bandwidth, when the number of measurements per curve
is bounded. Also, in the i.i.d. case, we obtain a lower bound on
the rate of convergence for estimating the first eigenfunction
that is sharper than bounds in the existing literature, which
proves the rate-optimality of our estimator in a wider regime.
Finally, we propose a computationally tractable model selection
procedure based on minimizing an approximation to the
leave-one-curve-out cross validation score that uses the empirical
Kullback-Leibler loss. We also show that in the context of
estimating the covariance kernel or its eigenfunctions, it has
clear advantages over the commonly used prediction error loss.

The proposed procedure for estimation and model selection is easily
implementable and computationally more tractable as compared to some
of the existing methods. Moreover, due to the linear structure of
the pre-smoothing of individual curves, our estimator is stable.
Furthermore, the linear structure of the proposed estimator also
allows for a simple approximation to the cross validation score.
Finally, even though the results are proved under Gaussianity of the
noise process, it can be shown that at the level of rates of
convergence, the upper bounds hold under sufficient moment
conditions on the noise, and hence the estimator is expected to be
robust to distributional assumptions.

There are a few aspects of the estimation procedure that need
further exploration. In the asymptotic analysis, we assumed that
$g$, the density function of the design points, is known. In
practice it has to be estimated from the data. Additional
computations are needed to show that the results derived here hold
under that setting as well. It will be useful also to study its
impact on the estimation procedure through simulation studies, and
in real data applications when the assumption of exact randomness
of the design points may be violated.

A natural generalization of the framework studied in this paper will
be when the principal component scores jointly form a stationary
vector autoregressive process. Under such a setting, we would like
to extend the estimation and model selection procedures described
here to exploit the special structures of such processes. This is
likely to summarize the statistical properties of some real-life
phenomena and also help in model building and prediction, for
example in spatio-temporal models when the covariance is not
separable.


\section{Appendix}

\subsection*{Appendix A}

\subsubsection*{Perturbation of
eigen-structure}\label{subsec:perturb_eigen}

The following lemma is a modified version of a similar result in
Paul and Johnstone (2007). 
Several variants of this lemma appear in the literature (see, e.g.,
Kneip and Utikal (2001), Cai and Hall (2006)),
and most of them implicitly use the approach taken in Kato (1980).
In the following we use $\parallel A
\parallel$ to denote the operator norm of an operator $A$, i.e., the
largest singular value of $A$.

\begin{lemma}\label{lemmaA.1}
Let $A$ and $B$ be two symmetric Hilbert-Schmidt operators acting
on $L^2([0,1])$. Let the eigenvalues of the operator $A$ be
denoted by $\lambda_1(A), \lambda_2(A),\cdots$. Set $\lambda_0(A)
= \infty$ and $\lambda_\infty(A) = -\infty$. For any $r \geq 1$,
if $\lambda_r(A)$ is a unique eigenvalue of $A$, i.e., if
$\lambda_r(A)$ is of multiplicity 1, then denoting by
$\mathbf{p}_r$ the eigenfunction associated with the $r$-th
eigenvalue. Then
\begin{equation*}
\mathbf{p}_r(A+B) - \mbox{sign}\langle \mathbf{p}_r(A+B),
\mathbf{p}_r(A)\rangle \mathbf{p}_r(A) = - H_r(A) B \mathbf{p}_r(A)
+ R_r
\end{equation*}
where $H_r(A) := \sum_{s \neq r} \frac{1}{\lambda_s(A) -
\lambda_r(A)} P_{{\cal E}_s}(A)$ and $P_{{\cal E}_s}(A)$ denotes the
orthogonal projection operator onto the eigen-subspace ${\cal E}_s$
corresponding to eigenvalue $\lambda_s(A)$ (possibly
multi-dimensional). Define $\delta_r$ and  $\overline{\delta}_r$ as
\begin{eqnarray*}
\delta_r &:=& \frac{1}{2} \left[\parallel H_r(A) B \parallel +
|\lambda_r(A+B)
- \lambda_r(A)| \parallel H_r(A) \parallel\right] \\
\overline{\delta}_r &=& \frac{\parallel B\parallel} {\min_{1\leq j
\neq r \leq \infty} |\lambda_j(A) - \lambda_r(A)|}~.
\end{eqnarray*}
Then, the residual term $R_r$ can be bounded as
\begin{equation*}
\parallel R_r \parallel \leq \min\left( 10\overline{\delta}_r^2, ~\parallel H_r(A)
B \mathbf{p}_r(A)
\parallel \left[\frac{2\delta_r(1+2\delta_r)}{1-2\delta_r(1+2\delta_r)} +
\frac{\parallel H_r(A) B \mathbf{p}_r(A)\parallel}
{(1-2\delta_r(1+2\delta_r))^2}\right]\right)
\end{equation*}
where the second bound holds only if $\delta_r <
\frac{\sqrt{5}-1}{4}$.

In addition, if $1\leq r_1 \leq r_2$ are such that
$\lambda_{r_1}(A)
> \lambda_{r_1+1}(A) = \cdots = \lambda_{r_2}(A) >
\lambda_{r_2+1}(A)$, then
\begin{equation*}
\sum_{k=r_1}^{r_2} \left(\lambda_k(A+B)-\lambda_k(A)\right) =
\tr(P_{{\cal E}_{r_1}}(A) B) + \overline{R}_{r_1,r_2},
\end{equation*}
where $P_{{\cal E}_{r_1}}(A)$ is the orthogonal projection operator
of $A$ corresponding to the eigenvalues
$\lambda_{r_1}(A),\ldots,\lambda_{r_2}(A)$, and the residual
$\overline{R}_{r_1,r_2}$ satisfies
\begin{equation*}
|\overline{R}_{r_1,r_2}| \leq (r_2-r_1+1) \frac{6
\parallel B\parallel^2} {\min_{1\leq j \neq r \leq \infty} |\lambda_j(A)
- \lambda_r(A)|}.
\end{equation*}
\end{lemma}

\subsubsection*{Large deviations of quadratic
forms}\label{subsec:large_dev}

The following lemmas are from Paul (2004). 
Suppose that $\Phi : {\cal X} \rightarrow \mathbb{R}^{n\times n}$ is
a measurable function. Let $Z$ be a random variable taking values in
${\cal X}$.

\begin{lemma}\label{lemmaA.2}
Suppose that $X$ and $Y$ are i.i.d. $N_n(0,I)$ and are independent
of $Z$. Then for every $L
> 0$ and $0 <\delta < 1$, for all $0 < t < \frac{\delta}{1-\delta}
L$,
\begin{equation*}
\mathbb{P}(\frac{1}{n}|X^T \Phi(Z) Y| > t, \parallel \Phi(Z)
\parallel \leq L) \leq 2
\exp\left(-\frac{(1-\delta)nt^2}{2L^2}\right).
\end{equation*}
\end{lemma}

\begin{lemma}\label{lemmaA.3}
Suppose that $X$ is distributed as $N_n(0,I)$ and is independent
of $Z$. Also assume that $\Phi(z) = \Phi^T(z)$ for all $z \in
{\cal X}$. Then for every $L > 0$ and $0 < \delta < 1$, for all $0
< t < \frac{2\delta}{1-\delta} L$,
\begin{equation*}
\mathbb{P}(\frac{1}{n}|X^T \Phi(Z) X - Tr(\Phi(Z))|
> t, \parallel \Phi(Z) \parallel \leq
L) \leq 2 \exp\left(-\frac{(1-\delta)nt^2}{4L^2}\right).
\end{equation*}
\end{lemma}

\subsubsection*{Computation of conditional mixed moments}
\label{subsec:cond_moment}

In order to calculate the bias and variance of the proposed
estimator, we need to compute the conditional expectations
$\mathbb{E}(Y_{i_1 j_1} Y_{i_1j_1'}Y_{i_2j_2}Y_{i_2j_2'}
|\mathbf{T}_{i_1},\mathbf{T}_{i_2})$ for various choices of
$i_1,i_2$, $j_1,j_1',j_2,j_2'$. We shall use the following
well-known result, which is a special case of \textit{Wick formula}
(Nica and Speicher, 2006, p. 129) 
for computation of mixed moments of a Gaussian random vector.

\begin{lemma}\label{lemmaA.4}
If $W_1$,$W_2$,$W_3$ and $W_4$ are jointly Gaussian with mean zero
and covariance matrix $\Sigma$, then
\begin{equation}\label{eq:Wick}
\mathbb{E}(W_1W_2W_3W_4) = \Sigma_{12}\Sigma_{34} +
\Sigma_{13}\Sigma_{24} + \Sigma_{14}\Sigma_{23}.
\end{equation}
\end{lemma}

\vskip.1in We shall use the formula to compute the above  mixed
moments with the observation that
Cov$(X_{i_1j_1},X_{i_2j_2}|\mathbf{T}_{i_1},\mathbf{T}_{i_2}) =
\rho_{i_1 i_2} C(T_{i_1j_1},T_{i_2j_2})$. The details of this
computation in various generic cases are given in Appendix F.


\subsection*{Appendix B}

In Appendix B and the following appendices, we shall often write $h$
and $\widetilde h$ to denote $h_n$ and $\widetilde h_n$,
respectively, and we shall drop the subscript $h_n$ from the
covariance estimates. For example, $\widehat C_{c}$ will be used to
denote $\widetilde C_{c,h_n}$.

\subsubsection*{Proof of Proposition \ref{prop2}}

This is a straightforward application of Lemma \ref{lemmaA.1}, by
taking the estimated covariance kernel $\widehat C_c$ as operator
$A$ and $-\Delta_i = \widehat C_c^{(-i)} - \widehat C_c$ as operator
$B$. Note that in (\ref{eq:H_nu_expand}) the last term corresponds
to the zero eigenvalues of $\widehat C_c$.

\subsubsection*{Proof of Proposition \ref{prop3}}

We can express $\Delta_i(u,v)$ as $\widehat \Delta_i(u,v) + R_i(u,v)
+ (\widehat\sigma_{(-i)}^2 - \frac{n}{n-1}\sigma^2) - \frac{1}{n-1}
\widehat C_c(u,v)$, where
\begin{eqnarray}\label{eq:Delta_tilde}
&& \widehat \Delta_i(u,v) \nonumber\\
&=& (1-W_{\tilde h_n}(u,v)) \frac{w(m_i)}{n-1} \frac{1}{g(u)g(v)} \nonumber\\
&& ~\cdot~ \sum_{l,l'=1}^{L_n} (\widetilde X_i(s_l) + (u-s_l)
\widetilde X_i'(s_l)) (\widetilde X_i(s_{l'}) + (v-s_l) \widetilde
X_i'(s_{l'})) \overline{Q}_{h_n} (u-s_l)
\overline{Q}_{h_n}(v-s_{l'}) \nonumber\\
&& + W_{\tilde
h_n}(u,v)\frac{1}{n-1}\frac{1}{g(\frac{u+v}{2})}\sum_{l=1}^{L_n}
\left[S_i(\frac{u+v}{2})+ (\frac{u+v}{2}-s_l)S_i'(\frac{u+v}{2})
\right] \overline{Q}_{h_n}(\frac{u+v}{2}-s_l)
\end{eqnarray}
and $R_i(u,v)$ equals (with $z$ denoting $\frac{u+v}{2}$)
\begin{eqnarray*}
&&(1-W_{\tilde h_n}(u,v)) \frac{\sigma}{n-1}\sum_{j\neq i}^n w(m_j)
\left[(\widehat \mu_{*,j}^{(-i)}(u) - \widehat \mu_{*,j}(u))\widehat
\varepsilon_{j}(v) + \widehat \varepsilon_{j}(u)(\widehat
\mu_{*,j}^{(-i)}(v) -\widehat \mu_{*,j}(v))\right] \\
&& + (1-W_{\tilde h_n}(u,v))\sum_{j\neq i}
\frac{w(m_j)}{n-1}\left[(\widehat \mu_{*,j}^{(-i)}(u) - \widehat
\mu_{*,j}(u))(\mu_{*,j}(v) - \widehat \mu_{*,j}(v)) \right. \nonumber\\
&& ~~  \left.  + (\widehat \mu_{*,j}^{(-i)}(v) - \widehat
\mu_{*,j}(v))(\mu_{*,j}(u) - \widehat \mu_{*,j}(u)) -(\widehat
\mu_{*,j}^{(-i)}(u) - \widehat
\mu_{*,j}(u))(\widehat \mu_{*,j}^{(-i)}(v) - \widehat \mu_{*,j}(v))\right] \nonumber\\
&& + W_{\tilde h_n}(u,v) \frac{2\sigma}{(n-1)g(z)}\sum_{j\neq i}^n
\frac{1}{m_j} \sum_{k=1}^{m_j}(\widehat \mu(T_{jk}) - \widehat
\mu^{(-i)}(T_{jk}))\varepsilon_{jk} \sum_{l=1}^{L_n} \widetilde
K_{z,l}(T_{jk}) \overline{Q}_{h_n}(z-s_l)
\nonumber\\
&& + W_{\tilde h_n}(u,v) \frac{2}{(n-1)g(z)}\sum_{j\neq i}^n
\frac{1}{m_j} \sum_{k=1}^{m_j}(\widehat \mu^{(-i)}(T_{jk}) -
\widehat \mu_(T_{jk}))(\mu(T_{jk}) - \widehat
\mu(T_{jk}))\sum_{l=1}^{L_n}
\widetilde K_{z,l}(T_{jk}) \overline{Q}_{h_n}(z-s_l) \nonumber\\
&& - W_{\tilde h_n}(u,v) \frac{1}{(n-1)g(z)}\sum_{j\neq i}^n
\frac{1}{m_j} \sum_{k=1}^{m_j}(\widehat \mu^{(-i)}(T_{jk}) -
\widehat \mu(T_{jk}))^2\sum_{l=1}^{L_n} \widetilde K_{z,l}(T_{jk})
\overline{Q}_{h_n}(z-s_l)
\end{eqnarray*}
where, the kernel $\widetilde K_{s,l}(\cdot) \equiv \widetilde
K_{s,l,h_n}(\cdot)$ is defined as
\begin{equation}\label{eq:K_tilde_l}
\widetilde K_{s,l}(u) =  \frac{1}{h_n}\left[K(\frac{s_{l}-u}{h_n}) +
\left(\frac{s - s_{l}}{h_n}\right)K'(\frac{s_{l}-u}{h_n})\right],
\end{equation}
for $s \in [0,1]$ and $l=1,\ldots,L_n$; and for any function $f$,
\begin{equation*}
f_{*,j}(x) = (g(x))^{-1} \sum_{l=1}^{L_n} (\widetilde f_j(s_l) +
(x-s_l) \widetilde f_j'(s_l)) \overline{Q}_{h_n}(x-s_l)
\end{equation*}
with $\widetilde f_j(s) := \frac{1}{m_j} \sum_{k=1}^{m_j}
f(T_{jk})K_{h_n}(s-T_{jk})$; and
\begin{equation*}
\widehat \varepsilon_{j}(x) = (g(x))^{-1}\sum_{l=1}^{L_n}
[\widetilde \varepsilon_{j}(s_l) + (x-s_l) \widetilde
\varepsilon_{j}'(s_l)]\overline{Q}_{h_n}(x-s_l)
\end{equation*}
with $\widetilde \varepsilon_{j}(s) = \frac{1}{m_j}\sum_{k=1}^{m_j}
\varepsilon_{jk} K_{h_n}(s-T_{jk})$.

Since $\widehat H_\nu \widehat C_c \widehat \psi_\nu = \widehat
\lambda_\nu \widehat H_\nu \widehat \psi_\nu = 0$, it follows from
the representation of $\Delta_i$ that
\begin{equation*}
\widehat H_\nu \Delta_i \widehat \psi_\nu = \widehat H_\nu (\widehat
\Delta_i + R_i)\widehat \psi_\nu + (\widehat \sigma_{(-i)}^2 -
\frac{n}{n-1}\widehat \sigma^2) (\widehat H_\nu \mathbf{1} \widehat
\psi_\nu),
\end{equation*}
where $\mathbf{1}(u,v) = \mathbf{1}_{\{0 \leq u,v \leq 1\}}$. It is
easy to see from the expression for $R_i(u,v)$ and
(\ref{eq:mu_hat_minus_i}) that for reasonable choices of $h_\mu$,
the contribution of $R_i(u,v)$ can be ignored, since it is of a
smaller asymptotic order (in fact can be shown to be $o_P(n^{-1})$).
Hence, we end up with the approximation $\widehat H_\nu \Delta_i
\widehat \psi_\nu \approx \widehat H_\nu \widehat \Delta_i \widehat
\psi_\nu + (\widehat \sigma_{(-i)}^2 - \frac{n}{n-1}\widehat
\sigma^2) (\widehat H_\nu \mathbf{1} \widehat \psi_\nu)$.

Thus, we can separate out the first term on the RHS of
(\ref{eq:Delta_tilde}) into two parts - one with multiplier 1, and
the other with multiplier $W_{\tilde h_n}(u,v)$. Then using
(\ref{eq:H_Delta_psi}) and the second representation of $\widehat
H_\nu$ in (\ref{eq:H_nu_expand}), we obtain the expressions in the
first four lines on the RHS of (\ref{eq:first_order_psi}). Next,
using the fact that $W_{\widetilde h_n}(u,v) \approx
\mathbf{1}_{\{|u-v| \leq \frac{A}{2}h_n\}}$, and using the
approximations $\widetilde \psi_\nu(v) \approx \widetilde
\psi_\nu(u)$ and $g(\frac{u+v}{2}) \approx g(u)$ on the interval
$[u-\frac{A}{2}h_n, u+\frac{A}{2}h_n]$, we obtain the last three
terms on the RHS of (\ref{eq:first_order_psi}). Now, using
(\ref{eq:trace_P_Delta}), noting that $\tr(\widehat P_\nu \widehat
C_c) = \widehat \lambda_\nu$, and following similar arguments, we
have (\ref{eq:first_order_lambda}).

\subsubsection*{Approximation of
$\overline{\gamma}_{k,k',h_n}(i)$}\label{subsec:gamma_bar}


First, to fix notation, suppose that $\widetilde h_n = A h_n$ for
some constant $A>0$. Then, by definition of $W_{\widetilde h_n}$,
and the symmetry of $\overline{Q}$, the integral appearing in
(\ref{eq:gamma_bar}) can be expressed as (ignoring the boundaries)
\begin{equation}\label{eq:d_ll_prime}
d_{ll',kk';h_n}^{jj'}  := \int_0^1 \frac{\widehat \psi_k(u)}{g(u)}
(u-s_l)^j \overline{Q}_{h_n}(s_l-u)
\int_{u-\frac{A}{2}h_n}^{u+\frac{A}{2}h_n} \frac{\widehat
\psi_{k'}(v)}{g(v)} (v-s_{l'})^{j'} \overline{Q}_{h_n}(v-s_{l'}) dv
du.
\end{equation}
Noticing that, on $[u-\frac{A}{2}h_n,u+\frac{A}{2}h_n]$,
$\frac{\widehat \psi_{k'}(v)}{g(v)}$ can be approximated as
$\frac{\widehat \psi_{k'}(u)}{g(u)}$, we can approximate the inner
integral (with respect to $v$) by
\begin{eqnarray*}
&& \frac{\widehat \psi_{k'}(u)}{g(u)}
\int_{u-\frac{A}{2}h_n}^{u+\frac{A}{2}h_n} (v-s_{l'})^{j'}
\overline{Q}_{h_n}(v-s_{l'}) dv\\
&=& h_n^{j'+1} \frac{\widehat \psi_{k'}(u)}{g(u)}
\int_{\frac{u-s_{l'}}{h_n} - \frac{A}{2}}^{\frac{u-s_{l'}}{h_n} +
\frac{A}{2}} w^{j'} \overline{Q}(w) dw, ~~(\mbox{setting}~ w =
\frac{v-s_{l'}}{h_n})\\
&=:& h_n^{j'+1} \frac{\widehat \psi_{k'}(u)}{g(u)}
G_{j'}^Q\left(\frac{u-s_{l'}}{h_n}\right) ~=:~ h_n^{j'+1}
\frac{\widehat \psi_{k'}(u)}{g(u)} G_{j',l';h_n}^Q(u).
\end{eqnarray*}
Substituting this in (\ref{eq:d_ll_prime}), we have the
approximation
\begin{eqnarray}\label{eq:d_ll_prime_approx}
d_{ll',kk';h_n}^{jj'} &\approx& (-1)^j h_n^{j'+1}  \int_0^{1}
\frac{\widehat\psi_k(u)\widehat\psi_{k'}(u)}{g^2(u)}
G_{j',l';h_n}^Q(u) (s_l-u)^j
\overline{Q}_{h_n}(s_l - u) du \nonumber \\
&=&  (-1)^j h_n^{j'+1} G_j\left(\left(\frac{\widehat\psi_k
\widehat\psi_{k'}}{g^2}\right) G_{j',l';h_n}^Q ,
\overline{Q}_{h_n}\right)(s_l) ~=:~
\overline{d}_{ll',kk';h_n}^{jj'},
\end{eqnarray}
by definition of $G_j(f_1,f_2)(\cdot)$, $j=0,1$.
Since
\begin{equation*}
\left[\frac{u-s_{l'}}{h_n} - \frac{A}{2},\frac{u-s_{l'}}{h_n} +
\frac{A}{2}\right] \cap [-C_Q,C_Q] = \phi \Leftrightarrow |u-s_{l'}|
> (C_Q+\frac{A}{2})h_n,
\end{equation*}
then $G_{j'}^Q\left(\frac{u-s_{l'}}{h_n}\right) \equiv 0$ if
$|u-s_{l'}|
> (C_Q+\frac{A}{2})h_n$. Furthermore, $\overline{Q}_{h_n}(u-s_l)
\equiv 0$ if $|u - s_l| > C_Q h_n$. This means that, if either
$|u-s_l| > C_Q h_n$, or $|u-s_{l'}| > (C_Q +\frac{A}{2})h_n$, then
the integrand in the first step of (\ref{eq:d_ll_prime_approx}) is
zero. So the domain of integration is, effectively, $[s_l - C_Q
h_n, s_l + C_Q h_n]\cap [s_{l'} - (C_Q +\frac{A}{2}) h_n, s_{l'} +
(C_Q +\frac{A}{2}) h_n]$. This implies that if $|s_l - s_{l'}| >
(2C_Q + \frac{A}{2}) h_n$, then the effective domain of
integration is empty, meaning that
\begin{equation*}
\overline{d}_{ll',kk';h_n}^{jj'} = 0 ~~~\mbox{if}~~~ |s_l - s_{l'}|
> (2C_Q + \frac{A}{2})h_n.
\end{equation*}
If $\overline{Q}(\cdot)$ is chosen to be a centered cubic B-spline
(so that $C_Q = 2$), we can compute $G_{j'}^Q(\cdot)$ explicitly,
without having to perform a numerical integration (Appendix F).

\subsection*{Appendix C}

In the following, we often drop the subscript $n$ from $h_n$ for
simplicity and sometimes we even drop the subscript $h$ from the
notation.

\subsubsection*{Proof of Proposition \ref{prop1}}

By elementary calculations, and supposing that $m_i \geq 2$ for each
$1\leq i \leq n$, we have
\begin{eqnarray}\label{eq:expect_m_C}
&& \mathbb{E}[\widetilde X_i(s) \widetilde X_i(t)] \nonumber\\ &=&
\frac{1}{m_i^2} \sum_{j,j'=1}^{m_i} \mathbb{E} [Y_{ij}Y_{ij'}
\frac{1}{h_n^2} K(\frac{s-T_{ij}}{h_n})
K(\frac{s-T_{ij'}}{h})]\nonumber\\
&=& \frac{m_i}{m_i^2} \frac{1}{h_n^2} \int (C(u,u) +
\sigma^2)K(\frac{s-u}{h_n}) K(\frac{t-u}{h_n}) du + \nonumber\\
&& ~~ \frac{m_i(m_i-1)}{m_i^2}\frac{1}{h_n^2} \int\int
C(u,v)K(\frac{s-u}{h_n})
K(\frac{t-u}{h_n}) du dv \nonumber\\
&=& \frac{1}{m_i} \frac{1}{h_n} \int (C(t+h_n u,t+h_n u) + \sigma^2)
K(-u)K(\frac{s-t}{h_n}-u) du \nonumber\\
&&+ \frac{m_i-1}{m_i} \int \int C(s+h_n u,t+h_n v) K(-u)K(-v) du dv
\nonumber\\
&=& \frac{1}{m_i h_n} \left[(\overline{C}(t) + \sigma^2) \int  K(-u)
K(\frac{s-t}{h_n}-u) du  + h_n \overline{C}'(t) \int u K(-u)
K(\frac{s-t}{h_n}-u) du + O(h_n^2)\right] \nonumber\\
&& + \left(1-\frac{1}{m_i}\right) C(s,t) \int\int K(-u)K(-v) du dv
\nonumber\\
&& + \left(1-\frac{1}{m_i}\right) h_n \int\int [C_s(s,t) u +
C_t(s,t) v] K(-u)K(-v) du dv + O(h_n^2),
\end{eqnarray}
where the last step is by Taylor series expansions. Now, noticing
that $K$ is symmetric about 0, $\int K(x) dx = 1$ and $\int x K(x)
dx = 0$, (\ref{eq:expect_m_C_st}) and (\ref{eq:expect_m_C_tt})
follow from (\ref{eq:expect_m_C}) after simplifications.

\subsubsection*{Asymptotic pointwise bias (\ref{eq:C_hat_bias})}\label{subsec:bias}

We first compute the expected value of the estimate described by
(\ref{eq:C_est_combined}). For simplicity of notations, we express
$\widetilde X_i(s_l) + (s-s_l)\widetilde X_i'(s_l)$ by $\widetilde
X_{i,l}(s)$. Observe that
\begin{equation*}
\widetilde X_{i,l}(s) = \frac{1}{m_i} \sum_{j=1}^{m_i} Y_{ij}
\frac{1}{h}\left[ K(\frac{s_l - T_{ij}}{h}) + \frac{s-s_l}{h}
K'(\frac{s_l - T_{ij}}{h})\right]
\end{equation*}
Let the support of kernel $K(\cdot)$ be denoted by $[-B_K,B_K]$.
Then, for each fixed $j=1,\ldots,m_i$, and $i=1,\ldots,n$,
\begin{eqnarray}\label{eq:Y_ij_square_expect}
&& \mathbb{E}\left[Y_{ij}^2\left((K(\frac{s_l - T_{ij}}{h}) +
\frac{s-s_l}{h} K'(\frac{s_l - T_{ij}}{h})) (K(\frac{s_{l'} -
T_{ij}}{h}) + \frac{s-s_{l'}}{h} K'(\frac{s_{l'} -
T_{ij}}{h}))\right)\right]\nonumber\\
&=& \int [C(u,u)+\sigma^2] g^2(u)\nonumber\\
&& ~~\cdot~\left((K(\frac{s_l-u}{h}) + \frac{s-s_l}{h} K'(\frac{s_l
- u}{h}))(K(\frac{s_{l'} - v}{h}) + \frac{t-s_{l'}}{h}
K'(\frac{s_{l'} - v}{h}))\right)du,
\end{eqnarray}
which is 0 if $|s_l - s_{l'}| > 2B_Kh$, since this implies that
$K(\frac{s_l -u}{h}) K(\frac{s_{l'} -u}{h}) = 0$ for all $u \in
\mathbb{R}$. If $|s_l - s_{l'}|\leq 2B_Kh$, there is nonzero
contribution of the term (\ref{eq:Y_ij_square_expect}) in
$\mathbb{E}[\widetilde X_{i,l}(s) \widetilde X_{i,l'}(t)]
\overline{Q}_h(s-s_l)\overline{Q}_h(t-s_{l'})$ only if $|s-t| \leq
2(B_K+C_Q)h$, where $supp(\overline{Q}) = [-C_Q,C_Q]$. Thus, if $A >
4(B_K+C_Q)$, then for $|s-t|
> \frac{1}{2} Ah$,  we have
\begin{eqnarray}\label{eq:mean_pointwise_TS}
&& w(m_i) \mathbb{E}(\widetilde X_{i,l}(s) \widetilde X_{i,l'}(t)) \nonumber\\
&=& \int\int C(u,v) g(u)g(v) \frac{1}{h^2} \left[(K(\frac{s_l-u}{h})
+ \frac{s-s_l}{h} K'(\frac{s_l - u}{h}))(K(\frac{s_{l'} - v}{h}) +
\frac{t-s_{l'}}{h} K'(\frac{s_{l'} - v}{h}))\right]du dv \nonumber\\
&=& \int\int C(s_l + x h, s_{l'} + y h) g(s_l + x h)g(s_{l'} + y h)
\nonumber\\
&& ~~~~~~~\cdot~\left[(K(x) + \frac{s-s_l}{h} K'(-x))(K(y) +
\frac{t-s_{l'}}{h} K'(-y))\right] dx dy.
\end{eqnarray}

We assume that the conditions in Section \ref{sec:asymptotics} hold.
Then using the representation (\ref{eq:mean_pointwise_TS}), and the
calculations done in 
Appendix F, we get an expression for the asymptotic bias in
estimating $C(\cdot,\cdot)$ as a function of the bandwidth $h \equiv
h_n$. These results are summarized in the following lemmas, where
$C_s$, $C_{ss}$ and $C_t$, $C_{tt}$ denote the first and second
partial derivatives of $C(s,t)$ with respect to $s$ and $t$,
respectively.


\begin{lemma}\label{lemmaC.1}
({\bf Expectation of $\widetilde C(s,t)$}): Let $K_2 = \int x^2 K(x)
dx$,
\begin{equation*}
\mathbf{Q}_h(s) = \sum_{l=1}^{L_n} \overline{Q}_h(s-s_l),
~~~~and~~~~ \mathbf{Q}_h^{(2)}(s) =
\sum_{l=1}^{L_n}(\frac{s-s_l}{h})^2 \overline{Q}_h(s-s_l).
\end{equation*}
Then, for $|s-t| > 2Ah_n$,
\begin{eqnarray}\label{eq:expect_C_tilde}
\mathbb{E}\widetilde C(s,t)
&=& C(s,t) \mathbf{Q}_h(s)\mathbf{Q}_h(t) \nonumber\\
&& + \frac{h^2}{2} C(s,t) \left[\frac{g''(s)}{g(s)} (K_2
\mathbf{Q}_h(s) - \mathbf{Q}_h^{(2)}(s)) \mathbf{Q}_h(t) +
\frac{g''(t)}{g(t)} (K_2 \mathbf{Q}_h(t) - \mathbf{Q}_h^{(2)}(t))
\mathbf{Q}_h(s)\right]\nonumber\\
&& + h^2 C_s \frac{g'(s)}{g(s)}(K_2 \mathbf{Q}_h(s) -
\mathbf{Q}_h^{(2)}(s)) \mathbf{Q}_h(t) + h^2C_t
\frac{g'(t)}{g(t)}(K_2 \mathbf{Q}_h(t) -
\mathbf{Q}_h^{(2)}(t))\mathbf{Q}_h(s)\nonumber\\
&& + \frac{h^2}{2} \frac{1}{g(s)g(t)} \left[C_{ss}(K_2
\mathbf{Q}_h(s) - \mathbf{Q}_h^{(2)}(s)) \mathbf{Q}_h(t) +
C_{tt}(K_2 \mathbf{Q}_h(t) - \mathbf{Q}_h^{(2)}(t))
\mathbf{Q}_h(s)\right]\nonumber\\
&& + O(h^{2+\alpha}).
\end{eqnarray}
\end{lemma}

Note that because of property (iii) of the kernel $\overline{Q}$,
and the fact that $s_l = (l+a)h$ for $l=1,\ldots,L_n$, for some
constant $a \in [-3,3]$, we have for $s \in (c,1-c)$, for some $c
\in (0,1)$,
\begin{equation*}
\mathbf{Q}_h(s) = \sum_{l=1}^{L_n} \overline{Q}(\frac{s}{h} - a - l)
= 1.
\end{equation*}
Therefore, we can choose $L_n$ and the sequence of points
$\{s_l\}_{l=1}^{L_n}$ so that $L_n \approx h_n^{-1}$, and
$\mathbf{Q}_h(s) \equiv 1$ for all $s \in [0,1]$. That is, from
Lemma \ref{lemmaC.1}, we have $\mathbb{E}\widetilde C(s,t) = C(s,t)
+ O(h^2)$.


\begin{lemma}\label{lemmaC.2}
({\bf Expectation of $\widehat{C}_*(t)$}): Let $\overline{C}'(t)$
and $\overline{C}''(t)$ denote the first and second derivative of
the function $\overline{C}(t):=C(t,t)$. Then, uniformly in $t$,
\begin{eqnarray}\label{eq:expect_C_star}
\mathbb{E}\widehat C_*(t) &=& (\overline{C}(t) +
\sigma^2)\mathbf{Q}_h(t) + \frac{h^2}{2} (\overline{C}(t) +
\sigma^2)\left(\frac{g''(t)}{g(t)}\right) (K_2 \mathbf{Q}_h(t) -
\mathbf{Q}_h^{(2)}(t)) \nonumber\\
&& + h^2\overline{C}'(t) \left(\frac{g'(t)}{g(t)}\right)(K_2
\mathbf{Q}_h(t) - \mathbf{Q}_h^{(2)}(t)) + \frac{h^2}{2}
\frac{\overline{C}''(t)}{g(t)}(K_2 \mathbf{Q}_h(t) -
\mathbf{Q}_h^{(2)}(t)) \nonumber\\
&& + O(h^{2+\alpha}).
\end{eqnarray}
\end{lemma}

Proof of Lemma \ref{lemmaC.2} follows along the lines of Lemma
\ref{lemmaC.1}. Furthermore, if an estimator $\widehat \sigma^2$ is
such that $\mathbb{E}\widehat\sigma^2 = \sigma^2 + O(h^2)$, then it
follows from Lemma \ref{lemmaC.2} that the estimator
$\widehat{\overline{C}}(t) := \widehat C_*(t) - \widehat \sigma^2$
satisfies
\begin{equation}\label{eq:diag_C_hat_bias}
\mathbb{E}\widehat{\overline{C}}(t) = C(t) + O(h^2),
\end{equation}
uniformly on $t \in [0,1]$, since $\mathbf{Q}_h(t) \equiv 1$ on $t
\in [0,1]$. Next, since $C(s,t) = C(t,s)$ and $C(\cdot,\cdot)$ is
smooth, it follows that $C_s - C_t \equiv 0$. Consequently, using a
Taylor series expansion, it follows that, for any $A > 0$,
\begin{equation}\label{eq:diag_C_approx}
C(s,t) = \overline{C}\left(\frac{s+t}{2}\right) +
O(h^2),~~~\mbox{for}~~ |s-t| \leq \frac{A}{2}h.
\end{equation}
Combining (\ref{eq:diag_C_hat_bias}) and (\ref{eq:diag_C_approx}) we
get,
\begin{equation}\label{eq:C_hat_bar_bias}
\mathbb{E}\widehat{\overline{C}}\left(\frac{s+t}{2}\right) = C(s,t)
+ O(h^2), ~~~\mbox{for}~~|s-t| \leq \frac{A}{2}h, ~s,t \in [0,1].
\end{equation}

\subsection*{Appendix D}

\subsubsection*{Proof of Proposition
\ref{prop4}}\label{subsec:proof_prop4}

We shall extensively use the following representation
\begin{eqnarray}\label{eq:H_nu_repr}
H_\nu(x,y) &=& \overline{H}_\nu(x,y) -
\frac{1}{\lambda_\nu}\delta(x-y), \\
~\mbox{where}~~~~ \overline{H}_\nu(x,y) &=& \sum_{1\leq k \neq \nu
\leq M} \frac{\lambda_k}{\lambda_k - \lambda_\nu} \psi_k(x)\psi_k(y)
+ \frac{1}{\lambda_\nu} \psi_\nu(x)\psi_\nu(y). \nonumber
\end{eqnarray}
The first step is to express $\widehat C_c(s,t)$ as $\widetilde
C_c(s,t) - W_{\widetilde h_n}(s,t) (\widehat\sigma^2 - \sigma^2)$,
where
\begin{equation}\label{eq:C_est_modified}
\widetilde C_c(s,t) = \overline{W}_{\widetilde h_n}(s,t) \widetilde
C(s,t) + W_{\widetilde h_n}(s,t) (\widehat
C_*(\frac{s+t}{2})-\sigma^2).
\end{equation}
Therefore, in order to separate the effect of estimating $\sigma^2$,
use the fact that for any fixed $\epsilon
> 0$,
\begin{eqnarray}\label{eq:sigma_contr_bound}
\parallel H_\nu \widehat C_c\psi_\nu \parallel_2^2
&\leq&
(1+\epsilon)\parallel H_\nu \widetilde C_c\psi_\nu
\parallel_2^2 + \left(1+\frac{1}{\epsilon}\right)  (\widehat
\sigma^2 -\sigma^2)^2 \parallel H_\nu W_{\widetilde h_n}
\psi_\nu\parallel_2^2 \nonumber\\
&=& (1+\epsilon)\parallel H_\nu \widetilde C_c\psi_\nu
\parallel_2^2 + \left(1+\frac{1}{\epsilon}\right)
(\widehat \sigma^2 -\sigma^2)^2 O(h_n^4).
\end{eqnarray}
The equality follows since using $H_\nu \psi_\nu =0$, the
definition of $W_{\widetilde h_n}$, and the \textit{Mean Value
Theorem}, we have
\begin{eqnarray*}
|(H_\nu W_{\widetilde h_n} \psi_\nu) (x)|
&=& \left|\int H_\nu(x,s) \int_{(s+\frac{Ah_n}{2})\vee
0}^{(s-\frac{Ah_n}{2})\wedge 1} (\psi_\nu(t)-\psi_\nu(s)) dt
ds \right| \nonumber\\
&\leq& \frac{A^2h_n^2}{2}\parallel \psi_\nu'\parallel_\infty\left[
\int |\overline{H}_\nu(x,s)|ds + \frac{1}{\lambda_\nu}\right].
\end{eqnarray*}
Since $\mathbb{E}(\widehat \sigma^2 -\sigma^2)^2 = o(1)$, it is
enough to show that $\mathbb{E}\parallel H_\nu \widetilde C_c
\psi_\nu\parallel_2^2$ has the bound given by the RHS of
(\ref{eq:expect_H_C_psi_L_2}), without the multiplicative factor
$(1+\epsilon)$. With a slight abuse of notation, we write
$\widehat{\overline{C}}(s)$ to indicate $\widehat
C_*(s)-\sigma^2$. Then, since
\begin{equation*}
(H_\nu \widetilde C_c \psi_\nu)(x) = \int\int H_\nu(x,s)
\overline{W}_{\widetilde h_n}(s,t) \widetilde C(s,t) \psi_\nu(t) ds
dt + \int\int H_\nu(x,s) W_{\widetilde h_n}(s,t)
\widehat{\overline{C}}(\frac{s+t}{2}) \psi_\nu(t) ds dt,
\end{equation*}
it follows that, $\parallel H_\nu \widetilde C_c
\psi_\nu\parallel_2^2$ equals
\begin{eqnarray}\label{eq:H_C_psi_L_2}
&&  \int\int\int\int\int
H_\nu(x,s_1)H_\nu(x,s_2)\overline{W}_{\widetilde
h_n}(s_1,t_1)\overline{W}_{\widetilde h_n}(s_2,t_2)
\nonumber\\
&& ~~~~~~~~~~~\cdot~ \widetilde C(s_1,t_1)\widetilde C(s_2,t_2)
\psi_\nu(t_1)\psi_\nu(t_2)
ds_1ds_2dt_1dt_2 dx\nonumber\\
&& + \int\int\int\int\int H_\nu(x,s_1)H_\nu(x,s_2)W_{\widetilde
h_n}(s_1,t_1)W_{\widetilde h_n}(s_2,t_2) \nonumber\\
&& ~~~~~~~~~~~\cdot~ \widehat{\overline{C}}(\frac{s_1+t_1}{2})
\widehat{\overline{C}}(\frac{s_2+t_2}{2}) \psi_\nu(t_1)\psi_\nu(t_2)
ds_1ds_2dt_1dt_2 dx \nonumber\\
&& + \int\int\int\int\int H_\nu(x,s_1)H_\nu(x,s_2)W_{\widetilde
h_n}(s_1,t_1)\overline{W}_{\widetilde h_n}(s_2,t_2)
\nonumber\\
&& ~~~~~~~~~~~\cdot~ \widetilde C(s_1,t_1)
\widehat{\overline{C}}(\frac{s_2+t_2}{2}) \psi_\nu(t_1)\psi_\nu(t_2)
ds_1ds_2dt_1dt_2 dx.
\end{eqnarray}
Thus, in order to obtain $\mathbb{E}\parallel H_\nu \widetilde C_c
\psi_\nu\parallel_2^2$, we need to evaluate the quantities
$\mathbb{E}[\widetilde C(s_1,t_1)\widetilde C(s_2,t_2)]$,
$\mathbb{E}[\widehat{\overline{C}}(\frac{s_1+t_1}{2})
\widehat{\overline{C}}(\frac{s_2+t_2}{2})]$, and
$\mathbb{E}[\widetilde
C(s_1,t_1)\widehat{\overline{C}}(\frac{s_2+t_2}{2})]$.

Let
\begin{equation}\label{eq:U_i_def}
U_i(s,t) = \sum_{l,l'=1}^{L_n} \frac{1}{m_i^2} \sum_{j,j'=1}^{m_i}
Y_{ij}Y_{ij'}\widetilde K_{s,l}(T_{ij})\widetilde
K_{s,l}(T_{ij'})\overline{Q}_h(s-s_l)\overline{Q}_h(t-s_{l'}),
\end{equation}
where $\widetilde K_{s,l}(\cdot)$ is as in (\ref{eq:K_tilde_l}).
Then we can express the expectation of the first term on the RHS of
(\ref{eq:H_C_psi_L_2}) as
\begin{eqnarray}\label{eq:H_C_psi_offdiag}
&&~~~~~~~\frac{1}{n^2}\sum_{i=1}^n w^2(m_i)\int\int\int\int\int
H_\nu(x,s_1)H_\nu(x,s_2)\overline{W}_{\widetilde
h_n}(s_1,t_1)\overline{W}_{\widetilde
h_n}(s_2,t_2) \nonumber\\
&& ~\cdot [g(s_1)g(s_2)g(t_1)g(t_2)]^{-1}
\mathbb{E}[U_i(s_1,t_1)U_i(s_2,t_2)]
\psi_\nu(t_1)\psi_\nu(t_2) ds_1ds_2dt_1dt_2 dx\nonumber\\
&& + \frac{1}{n^2}\sum_{i_1\neq i_2}^n
w(m_{i_1})w(m_{i_2})\int\int\int\int\int
H_\nu(x,s_1)H_\nu(x,s_2)\overline{W}_{\widetilde
h_n}(s_1,t_1)\overline{W}_{\widetilde
h_n}(s_2,t_2) \nonumber\\
&& ~~\cdot [g(s_1)g(s_2)g(t_1)g(t_2)]^{-1}
\mathbb{E}[U_{i_1}(s_1,t_1)U_{i_2}(s_2,t_2)]
\psi_\nu(t_1)\psi_\nu(t_2) ds_1ds_2dt_1dt_2 dx.
\end{eqnarray}
The following proposition is the key to get a simplified bound on
(\ref{eq:H_C_psi_offdiag}). It is proved using a lengthy, but fairly
straightforward calculation. The details are given in Appendix F.

\begin{prop}\label{propC.1}
Suppose that $A > 4(B_K + C_Q)$. Then for $|s_k - t_k| >
\frac{1}{2} Ah_n$ ($k=1,2$), we have
\begin{eqnarray}\label{eq:sum_E_U_ii}
&& \frac{1}{n^2} \sum_{i=1}^n w^2(m_i) \frac{\mathbb{E}[U_i(s_1,t_1)
U_i(s_2,t_2)]}{g(s_1)g(s_2)g(t_1)g(t_2)} \nonumber\\
&=& \frac{1}{n}\sum_{i=1}^n \frac{(m_i-2)(m_i-3)}{m_i(m_i-1)}
\left[(C(s_1,t_1) + O(h^2))(C(s_2,t_2) + O(h_n^2)) \right.
\nonumber\\
&& \left. + (C(s_1,s_2)+O(h_n^2))(C(t_1,t_2) + O(h_n^2)) +
(C(s_1,t_2)+O(h_n^2))(C(s_2,t_1) + O(h_n^2))\right] \nonumber\\
&& + Z_1 + Z_2 + Z_3 + Z_4 + Z_5 + Z_6,
\end{eqnarray}
where the quantities $Z_j := Z_j(s_1,s_2,t_1,t_2)$, $j=1,\ldots,6$
where $Z_1,\ldots,Z_4$ are \textit{asymptotically equivalent} to
$Z(s_1,s_2)$, $Z(s_1,t_2)$, $Z(t_1,s_2)$ and $Z(t_1,t_2)$,
respectively; and $Z_5, Z_6$ are \textit{asymptotically equivalent}
to $\widetilde Z(s_1,s_2,t_1,t_2)$ and $\widetilde
Z(s_1,t_2,t_1,s_2)$, respectively, where
\begin{equation*}
Z(s,t) = \begin{cases} O(\frac{1}{nh_n\underline{m}})
& ~\mbox{if}~ |s-t|\leq \frac{Ah_n}{2}\\
0 & ~\mbox{otherwise};
\end{cases}
\end{equation*}
and
\begin{equation*}
\widetilde Z(s_1,s_2,t_1,t_2) =
\begin{cases}
O(\frac{1}{nh_n^2\underline{m}^2}) & ~\mbox{if}~
\max\{|s_1-s_2|,|t_1-t_2|\}\leq\frac{Ah_n}{2}\\
O(\frac{1}{nh_n\underline{m}^2}) & ~\mbox{if}~|s_1-s_2|
\leq\frac{Ah_n}{2} ~\mbox{and}~ |t_1-t_2| > \frac{Ah_n}{2}\\
O(\frac{1}{nh_n\underline{m}^2}) & ~\mbox{if}~|s_1-s_2|
> \frac{Ah_n}{2}~\mbox{and}~ |t_1-t_2| \leq \frac{Ah_n}{2}\\
0 & ~\mbox{otherwise}.
\end{cases}
\end{equation*}
Also,
\begin{eqnarray}\label{eq:sum_E_U_i1i2}
&& \frac{1}{n^2} \sum_{i_1 \neq i_2}^n w(m_{i_1})w(m_{i_2})
\frac{\mathbb{E}[U_{i_1}(s_1,t_1)
U_{i_2}(s_2,t_2)]}{g(s_1)g(s_2)g(t_1)g(t_2)} \nonumber\\
&=&
\frac{n-1}{n}(C(s_1,t_1) + O(h_n^2))(C(s_2,t_2) + O(h_n^2))
\nonumber\\
&& + \frac{1}{n^2}(\sum_{i_1\neq i_2}^n
\rho_{i_1i_2}^2)\left[(C(s_1,s_2) + O(h_n^2))(C(t_1,t_2) + O(h_n^2))
\right. \nonumber\\
&& ~~~~~~~~~~~~~~~~~~\left.  + (C(s_1,t_2)+O(h_n^2))(C(s_2,t_1) +
O(h_n^2))\right].
\end{eqnarray}
In all of above the $O(\cdot)$ terms are uniform in
$s_1,s_2,t_1,t_2$ in their respective domains.
\end{prop}

Now we deal with the last two terms on the RHS of
(\ref{eq:H_C_psi_L_2}). Let
\begin{equation}\label{eq:V_i_def}
V_i(s) = \sum_{l=1}^{L_n} \frac{1}{m_i} \sum_{j=1}^{m_i} Y_{ij}^2
\widetilde K_{s,l}(T_{ij}).
\end{equation}
Then,
\begin{equation*}
\widehat C_*(s) = \frac{1}{n}\sum_{i=1}^n
[g(s)]^{-1}V_i(s)\overline{Q}_{h_n}(s-s_l).
\end{equation*}
For convenience, in the rest of this subsection we shall use $z_k$
to denote $(s_k+t_k)/2$, for $k=1,2$. Then the following proposition
describes the contribution of the quantities of the type
$\mathbb{E}[V_{i_1}(z_1)V_{i_2}(z_2)]$ and
$\mathbb{E}[U_{i_1}(s_1,t_1) V_{i_2}(z_2)]$.

\begin{prop}\label{propC.2}
Suppose that $A > 4(B_K + C_Q)$. Then for (i) $|s_k - t_k| \leq
\frac{Ah_n}{2}$, $k=1,2$,
\begin{eqnarray}\label{eq:sum_E_V_i1i2}
&& \frac{1}{n^2} \sum_{i=1}^n
\frac{\mathbb{E}(V_i(z_1)V_i(z_2))}{g(z_1)g(z_2)} + \frac{1}{n^2}
\sum_{i_1 \neq i_2}^n
\frac{\mathbb{E}(V_{i_1}(z_1)V_{i_2}(z_2))}{g(z_1)g(z_2)} - \sigma^2
[\mathbb{E}(\widehat C_*(z_1)) + \mathbb{E}(\widehat C_*(z_2))] +
\sigma^4\nonumber\\
&=& C(s_1,t_1)C(s_2,t_2)- \left(\frac{1}{n^2}\sum_{i=1}^n
\frac{1}{m_i}\right) (C(s_1,t_1)+\sigma^2)(C(s_2,t_2)+\sigma^2)  + O(h_n^2)\nonumber\\
&& + \left(\frac{1}{n}(1-\frac{1}{n}\sum_{i=1}^n \frac{1}{m_i}) +
\frac{1}{n^2}\sum_{i_1\neq i_2}^n \rho_{i_1i_2}^2\right)
(C(s_1,s_2)C(t_1,t_2) + C(s_1,t_2)C(s_2,t_1)
+ O(h_n))\nonumber\\
&& + Z_7,
\end{eqnarray}
where $Z_7 := Z_7(z_1,z_2)$ is asymptotically equivalent to
$Z(z_1,z_2)$.
Next, if (ii) $|s_1 - t_1| > \frac{Ah_n}{2}$ and $|s_2 - t_2| \leq
\frac{Ah_n}{2}$, then
\begin{eqnarray}\label{eq:sum_E_U_i1_V_i2}
&& \frac{1}{n^2} \sum_{i=1}^n w(m_i) \mathbb{E}(U_i(s_1,t_1)
V_i(z_2)) + \frac{1}{n^2} \sum_{i_1\neq i_2} w(m_{i_1})
\mathbb{E}(U_{i_1}(s_1,t_1) V_{i_2}(z_2)) - \sigma^2
\mathbb{E}\widetilde C(s_1,t_1)\nonumber\\
&=& (C(s_1,t_1) + O(h_n^2))(C(s_2,t_2) + O(h_n^2))\nonumber\\
&& - \left(\frac{1}{n^2}\sum_{i=1}^n \frac{2}{m_i}\right)(C(s_1,t_1)
+ O(h_n^2))(C(s_2,t_2) + \sigma^2 + O(h_n^2))\nonumber\\
&& + \left(\frac{1}{n}(1-\frac{1}{n}\sum_{i=1}^n \frac{2}{m_i}) +
\frac{1}{n^2}\sum_{i_1\neq i_2}^n \rho_{i_1i_2}^2\right)
(C(s_1,s_2)C(t_1,t_2) + C(s_1,t_2)C(s_2,t_1) + O(h_n)) \nonumber\\
&& + Z_8 + Z_9,
\end{eqnarray}
where the $O(h_n^2)$ terms within brackets in the first term on the
RHS depend on $(s_1,t_1)$ and $(s_2,t_2)$ respectively, and $Z_j :=
Z_j(s_1,t_1,z_2)$, $j=8,9$ satisfy
\begin{eqnarray*}
Z_8 &=& \begin{cases} O(\frac{1}{nh_n^2\underline{m}}) &~\mbox{if}~ |s_1-s_2|\leq \frac{Ah_n}{2}\\
0 & ~\mbox{otherwise};
\end{cases}\\
Z_9 &=& \begin{cases} O(\frac{1}{nh_n^2\underline{m}}) &~\mbox{if}~ |t_1-s_2|\leq \frac{Ah_n}{2}\\
0 & ~\mbox{otherwise}.
\end{cases}
\end{eqnarray*}
\end{prop}

The proof of Proposition \ref{prop4} is now completed by using the
definitions of $\mathbb{E}[\widetilde C(s_1,t_1)\widetilde
C(s_2,t_2)]$,
$\mathbb{E}[\widehat{\overline{C}}(\frac{s_1+t_1}{2})
\widehat{\overline{C}}(\frac{s_2+t_2}{2})]$, and
$\mathbb{E}[\widetilde
C(s_1,t_1)\widehat{\overline{C}}(\frac{s_2+t_2}{2})]$; using the
properties of the kernel $H_\nu(x,y)$; and the bounds in
Propositions \ref{propC.1} and \ref{propC.2} and plugging
everything back into the expectation of (\ref{eq:H_C_psi_L_2}).
The details can be found in Appendix F.


\subsection*{Appendix E}

\subsubsection*{Asymptotic pointwise variance (\ref{eq:C_hat_variance})}\label{subsec:variance}


In this section, we prove (\ref{eq:C_hat_variance}),
(\ref{eq:sigma_risk}) and (\ref{eq:C_hat_variance_sigma}). Most of
the derivations are similar to that of Proposition \ref{prop4}.
Thus we simply give a brief outline.

First, using the fact that $\overline{W}_{\widetilde
h_n}(s,t)W_{\widetilde h_n}(s,t) = 0$, we obtain
\begin{eqnarray*}\label{eq:C_est_var}
\mbox{Var}(\widehat C_c(s,t)) &=& \overline{W}_{\widetilde h_n}(s,t)
\mbox{Var}(\widetilde C(s,t)) + W_{\widetilde h_n}(s,t)
\mbox{Var}(\widehat C_*(\frac{s+t}{2}) - \widehat
\sigma^2)\nonumber\\
&\leq& \overline{W}_{\widetilde h_n}(s,t)\mbox{Var}(\widetilde
C(s,t)) + 2W_{\widetilde h_n}(s,t) \left[\mbox{Var}(\widehat
C_*(\frac{s+t}{2})) + \mbox{Var}(\widehat\sigma^2)\right].
\end{eqnarray*}
Since $\mathbb{E}(\widehat\sigma^2-\sigma^2)^2$ has the rate given
by (\ref{eq:sigma_risk}) (Corollary \ref{cor1}), we only need to
provide bounds for $\overline{W}_{\widetilde
h_n}(s,t)\mbox{Var}(\widetilde C(s,t))$ and $W_{\widetilde
h_n}(s,t)\mbox{Var}(\widehat C_*(\frac{s+t}{2}))$. We state these
in the following propositions.

\begin{prop}\label{propC.3}
\begin{equation}\label{eq:var_C_tilde_offdiag}
\overline{W}_{\widetilde h_n}(s,t)\mbox{Var}(\widetilde C(s,t)) =
O\left(\frac{1}{n}\right) + \left(\frac{1}{n^2} \sum_{i_1\neq i_2}^n
\rho_{i_1i_2}^2\right) O(1) +
O\left(\max\{\frac{1}{nh_n^2\underline{m}^2},\frac{1}{nh_n
\underline{m}}\}\right).
\end{equation}
\end{prop}

\begin{prop}\label{propC.4}
\begin{equation}\label{eq:var_C_star_diag}
W_{\widetilde h_n}(s,t)\mbox{Var}(\widehat C_*(\frac{s+t}{2})) =
O\left(\frac{1}{n}\right) + \left(\frac{1}{n^2} \sum_{i_1\neq i_2}^n
\rho_{i_1i_2}^2\right) O(1) +
O\left(\frac{1}{nh_n\underline{m}}\right).
\end{equation}
\end{prop}

\vskip.1in\noindent The proof of (\ref{eq:C_hat_variance_sigma})
is finished by combining Propositions \ref{propC.3} and
\ref{propC.4} and Corollary \ref{cor1}.

\subsubsection*{Proof of Corollary \ref{cor1}}

First observe that,
\begin{eqnarray}\label{eq:sigma_risk_expand}
\mathbb{E}(\widehat \sigma^2 - \sigma^2)^2  &=&
\frac{1}{(T_1-T_0)^2} \int_{T_0}^{T_1} \int_{T_0}^{T_1}
\mathbb{E}[(\widehat C_*(t) - \sigma^2 -
\widehat{\overline{C}}_0(t))(\widehat C_*(s) - \sigma^2 -
\widehat{\overline{C}}_0(s))] ds dt \nonumber\\
&\leq& \sup_{t \in [T_0, T_1]} \mathbb{E}(\widehat C_*(t) - \sigma^2
- \widehat{\overline{C}}_0(t))^2 ~~~~~~~(\mbox{by Cauchy-Schwarz
inequality}) \nonumber\\
&\leq& 2 \sup_{t \in [T_0, T_1]} \mbox{Var}(\widehat C_*(t)) + 2
\sup_{t \in [T_0, T_1]} \mbox{Var}(\widehat{\overline{C}}_0(t))
\nonumber\\
&& + \sup_{t \in [T_0, T_1]} \left(\mathbb{E}(\widehat C_*(t)) -
\sigma^2 - \mathbb{E}(\widehat{\overline{C}}_0(t))\right)^2.
\end{eqnarray}
By Propositions \ref{propC.3} and \ref{propC.4}, and the definition
(\ref{eq:C_bar_hat_0}) of $\widehat{\overline{C}}_0$, the sum of the
first two term on the RHS on (\ref{eq:sigma_risk_expand}) is bounded
by
\begin{equation*}
O\left(\frac{1}{n}\right) + \left(\frac{1}{n^2} \sum_{i_1\neq i_2}^n
\rho_{i_1i_2}^2\right) O(1) +
O\left(\max\{\frac{1}{nh_n^2\underline{m}^2},\frac{1}{nh_n\underline{m}}\}\right).
\end{equation*}
On the other hand, since for any bounded $u \in [A_1,A_2]$,
\begin{equation*}
\left|\frac{1}{2}(C(t-h_nu,t+h_nu)+C(t+h_nu,t-h_nu)) - C(t,t)\right|
= O(h_n^2),
\end{equation*}
uniformly in $t \in [T_0,T_1]$, it follows from Lemmas
\ref{lemmaC.1} and \ref{lemmaC.2} (Appendix C) that the last term
on the RHS of (\ref{eq:sigma_risk_expand}) is $O(h_n^4)$.

\subsubsection*{Proof of Proposition \ref{prop5}}

Without loss of generality we assume $g$ to be uniform density on
$[0,1]$.  We need to consider two cases separately : (i) $|s-t|
> \frac{Ah}{2}$ and (ii) $|s-t|\leq \frac{Ah}{2}$.
\begin{itemize}
\item [(i)] $|s-t| > \frac{Ah}{2}$: In this case, we have
$\widehat C_c(s,t) - \mathbb{E}[\widehat C_c(s,t)] =
\overline{W}_{Ah}(s,t)(\widetilde C(s,t) - \mathbb{E}[\widetilde
C(s,t)])$. Let
\begin{equation*}
B_i(s,T_{ij}) = \sum_{l=1}^{L_n} \widetilde K_{s,l}(T_{ij})
\overline{Q}_h(s-s_l),~~~1\leq j\leq m_i, 1\leq i \leq n.
\end{equation*}
Since $|\widetilde K_{s,l}(T_{ij})| = O(h^{-1})$ and the summands
are nonzero for finitely many $l$, there exists a constant $C_3
> 0$ such that
\begin{equation}\label{eq:B_i_bound}
\sup_{s\in [0,1]} \max_{1\leq i \leq n}\max_{1\leq j \leq m_i}
|B_i(s,T_{ij})| \leq C_3 h^{-1}.
\end{equation}
Note further that $B_i(s,T_{ij}) = 0$ if $|s-T_{ij}| >
2(B_K+C_Q)h$. Next,
\begin{eqnarray*}\label{eq:X_i_tilde_expand}
\sum_{l=1}^{L_n}\widetilde X_{i,l}(s) &=& \frac{1}{m_i}
\sum_{j=1}^{m_i} (X_i(T_{ij}) +
\sigma \varepsilon_{ij}) B_i(s,T_{ij})\nonumber\\
&=& \sum_{k=1}^M \sqrt{\lambda_k} \xi_{ik} \left(\frac{1}{m_i}
\sum_{j=1}^{m_i} \psi_k(T_{ij}) B_i(s,T_{ij})\right) + \sigma
\frac{1}{m_i} \sum_{j=1}^{m_i} \varepsilon_{ij}
B_i(s,T_{ij})\nonumber\\
&=& \sum_{k=1}^M  \sqrt{\lambda_k} \xi_{ik} B_{1i,k}(s) + \sigma
\frac{1}{m_i}\sum_{j=1}^{m_i} \varepsilon_{ij} B_i(s,T_{ij}),
\end{eqnarray*}
where $B_{1i,k}(s) := \frac{1}{m_i} \sum_{j=1}^{m_i} \psi_k(T_{ij})
B_i(s,T_{ij})$. By (\ref{eq:B_i_bound}), there exists $C_4 > 0$ such
that
\begin{equation}\label{eq:B_12_i_bound}
\sup_{s \in [0,1]}\max_{1\leq k \leq M} \max_{1\leq i \leq n}
|B_{1i,k}(s)| \leq C_4 h^{-1}.
\end{equation}
Also, since $A > 4(B_K+C_Q)h$ and since $|s-t| \geq \frac{Ah}{2}$,
it follows that $B_i(s,T_{ij})B_i(t,T_{ij})=0$. Moreover,
$B_i(s,T_{ij})B_i(t,T_{ij'}) \neq 0$ only if $1\leq j\neq j'\leq
m_i$ are such that $|s-T_{ij}|\leq 2(B_K+C_Q)h$ and
$|t-T_{ij'}|\leq 2(B_K+C_Q)h$. This implies that
\begin{equation*}\label{eq:B_ik_prob_bound}
\mathbb{P}_g\left(B_{1i,k}(s)B_{1i,k}(t) \neq 0\right) \leq C_5
m_i(m_i-1) h^2 ~~~\mbox{for some}~~~C_5 := C_5(A) > 0.
\end{equation*}
Furthermore, for each $k=1,\ldots, M$, $\{B_{1i,k}(s)\}_{i=1}^n$ are
independent, and these random variables are independent of
$\{\xi_{ik}:1\leq k\leq M\}_{i=1}^n$ and $\{\varepsilon_{ij} : 1\leq
j \leq m_i\}_{i=1}^n$. Then, we can express $\widetilde C(s,t) -
\mathbb{E}[\widetilde C(s,t)]$ as,
\begin{eqnarray}\label{eq:C_c_diff_offdiag}
&&\widetilde C(s,t) - \mathbb{E}[\widetilde C(s,t)] \nonumber\\
&=&
\sum_{1\leq k\neq k'\leq M} \sqrt{\lambda_k \lambda_{k'}}
\frac{1}{n} \sum_{i=1}^n \xi_{ki}\xi_{k'i} w(m_i)
B_{1i,k}(s)B_{1i,k'}(t)\nonumber\\
&& + \sum_{k=1}^M \lambda_k  \frac{1}{n} \sum_{i=1}^n
(\xi_{ki}^2-1) w(m_i) B_{1i,k}(s)B_{1i,k}(t) \nonumber\\
&& + \sum_{k=1}^M \lambda_k \frac{1}{n}\sum_{i=1}^n w(m_i)
(B_{1i,k}(s)B_{1i,k}(t)-\mathbb{E}(B_{1i,k}(s)B_{1i,k}(t)))\nonumber\\
&& + \sigma \sum_{k=1}^M \sqrt{\lambda_k} \frac{1}{n} \sum_{i=1}^n
w(m_i) \frac{1}{m_i}\sum_{j=1}^{m_i} \xi_{ik} \varepsilon_{ij}
(B_{1i,k}(s)B_i(t,T_{ij})+ B_{1i,k}(t)B_i(s,T_{ij})) \nonumber\\
&& +   \sigma^2 \frac{1}{n}\sum_{i=1}^n \frac{w(m_i)}{m_i^2}
\sum_{j\neq j'}^{m_i} \varepsilon_{ij}\varepsilon_{ij'}
B_i(s,T_{ij}) B_i(t,T_{ij'})\nonumber\\
&& +  \sigma^2 \frac{1}{n}\sum_{i=1}^n \frac{w(m_i)}{m_i^2}
\sum_{j=1}^{m_i} (\varepsilon_{ij}^2 - 1)
B_i(s,T_{ij})B_i(t,T_{ij})\nonumber\\
&& +   \sigma^2 \frac{1}{n}\sum_{i=1}^n
\frac{w(m_i)}{m_i^2}\sum_{j=1}^{m_i}
(B_i(s,T_{ij})B_i(t,T_{ij})-\mathbb{E}(B_i(s,T_{ij})B_i(t,T_{ij}))).
\end{eqnarray}
The last two terms in the above expression vanish since $|s-t| >
4(B_K+C_Q)h$. Note that, $\max_{1\leq i \leq n} w(m_i)$ is
bounded. By (\ref{eq:B_12_i_bound}), $|B_{1i,k}(s)B_{1i,k}(t)|
\leq C_4^2 h^{-2}$ are bounded for $k=1,\ldots,M$, and for all
$k,k'$,
\begin{equation}\label{eq:var_B_s_B_t}
\max_{1\leq i \leq n} \mbox{Var}(B_{1i,k}(s)B_{1i,k'}(t)) \leq C_6
\max\{(\underline{m}h)^{-2},(\underline{m}h)^{-1}\}
~~\mbox{for}~~C_6=C_6(A) > 0,
\end{equation}
(see Appendix F). Thus by \textit{Bernstein's inequality}, and
using the condition that $\overline{m}^2 = o(nh^2/\log n)$, given
$\eta
> 0$, there exists $c_{1,\eta} > 0$ such that for sufficiently large
$n$ (so that the bound in (\ref{eq:var_B_s_B_t}) is
$O((\underline{m}h)^{-2})$),
\begin{equation}
\mathbb{P}_g\left(\max_{k=1,\ldots,M}\left|\frac{1}{n}\sum_{i=1}^n
w(m_i)(B_{1i,k}(s)B_{1i,k}(t)-\mathbb{E}(B_{1i,k}(s)B_{1i,k}(t)))\right|
> c_{1,\eta}\sqrt{\frac{\log n}{nh^2\underline{m}^2}}\right) \leq
n^{-\eta}.
\end{equation}
Next, let ${\cal A}$ be the set of indices $i$ such that
$B_{1i,k}(s)B_{1i,k'}(t)\neq 0$ for some $k,k'$. And let $N_n =
|{\cal A}|$. Since for any $k,k'$, $P(B_{1i,k}(s)B_{1i,k'}(t)\neq 0)
\leq C_5 \overline{m}^2 h^2$, it follows by another application of
Bernstein's inequality that there exists a set $D_n$ (in the sigma
field generated by $\{T_{ij}\}$) and a constant $c_{2,\eta}
> 0$ such that
\begin{equation*}
D_n = \{ N_n \leq c_{2,\eta} n \overline{m}^2 h^2\}
~~~\mbox{and}~~\mathbb{P}(D_n) \geq 1 - n^{-\eta}.
\end{equation*}
Therefore we can restrict our attention to the set $D_n$, and
conditioning on $\mathbf{T}$  We can express
$\boldsymbol{\xi}_{{\cal A},k} = (\xi_{ik})_{i\in {\cal A}}$ as
$\boldsymbol{\xi}_{{\cal A},k} := (\mathbf{R}_{{\cal A}{\cal
A}})^{1/2}\overline{\boldsymbol{\xi}}_{{\cal A},k}$, where the
random vectors $\overline{\boldsymbol{\xi}}_{{\cal A},k}$ have
$N_{N_n}(0,I)$ distribution and are independent for different $k$'s.
Then we can write (conditionally on $\mathbf{T}$)
$$
\sum_{i=1}^n \xi_{ki}\xi_{k'i} w(m_i) B_{1i,k}(s)B_{1i,k'}(t) =
\overline{\boldsymbol{\xi}}_{{\cal A},k}^T
\Phi(\mathbf{T})\overline{\boldsymbol{\xi}}_{{\cal A},k'},
$$
where $\Phi(\mathbf{T}) = (\mathbf{R}_{{\cal A}{\cal
A}})^{1/2}~diag(w(m_i)B_{1i,k}(s)B_{1i,k'}(t))_{i\in {\cal A}}
(\mathbf{R}_{{\cal A}{\cal A}})^{1/2}$. Observe that by
(\ref{eq:B_12_i_bound}) and condition {\bf C3}, we have $\parallel
\Phi(\mathbf{T})\parallel \leq C_4 \kappa_n h^{-2}$. Therefore, by
an application of Lemma \ref{lemmaA.2}, we have, for some
$c_{3,\eta}
> 0$,
\begin{equation*}
\mathbb{P}(|\frac{1}{n}\sum_{i=1}^n \xi_{ki}\xi_{k'i} w(m_i)
B_{1i,k}(s)B_{1i,k'}(t)| > c_{3,\eta}\overline{m} \kappa_n
\sqrt{\frac{\log n}{nh^2}}, D_n) \leq n^{-\eta}.
\end{equation*}
Very similar arguments can be used to obtain bounds of order
$\overline{m} \kappa_n \sqrt{\frac{\log n}{nh^2}}$ (that hold with
probability at least $1-O(n^{-\eta})$, for any given $\eta > 0$) for
the second, fourth and fifth terms on the RHS of
(\ref{eq:C_c_diff_offdiag}). Thus, by conditions on $\kappa_n$ and
$h_n$, we have, for some constant $c_{4,\eta} > 0$,
\begin{equation}\label{eq:C_c_offdiag_prob_bound}
\mathbb{P}(|\overline{W}_{Ah}(s,t)(\widehat C_c(s,t) -
\mathbb{E}(\widehat C_c(s,t))| > c_{4,\eta} \overline{m} \kappa_n
\sqrt{\frac{\log n}{nh^2}}) \leq n^{-\eta}.
\end{equation}
\item [(ii)] $|s-t| \leq \frac{Ah}{2}$: In this case, we have
$\widehat C_c(s,t) - \mathbb{E}[\widehat C_c(s,t)] =
W_{Ah}(s,t)(\widehat C_*(\frac{s+t}{2}) - \mathbb{E}[\widehat
C_*(\frac{s+t}{2})])$ (ignoring the maximum over $h_n^2$ in the
definition). Then similar (but somewhat simpler) arguments, now
involving Lemma \ref{lemmaA.3}, show that for some $c_{5,\eta} >
0$,
\begin{equation}\label{eq:C_c_diag_prob_bound}
\mathbb{P}(|W_{Ah}(s,t)(\widehat C_*(\frac{s+t}{2}) -
\mathbb{E}[\widehat C_*(\frac{s+t}{2})])| > c_{5,\eta} \overline{m}
\kappa_n \sqrt{\frac{\log n}{nh^2}}) \leq n^{-\eta}.
\end{equation}
\end{itemize}
Combining (\ref{eq:C_c_offdiag_prob_bound}) and
(\ref{eq:C_c_diag_prob_bound}) we obtain the result.

\subsection*{Appendix F}

\subsubsection*{Details of computation of
$G_{j}^Q(\cdot)$}\label{subsec:spline_condition}

We want to give explicit functional form for $G_{j}^Q(y)$, for
$j=0,1$ and for any $y \in \mathbb{R}$. Let
\begin{eqnarray*}
B_1(x) &=& x^3/6\\
B_2(x) &=& (-3x^3 + 3x^2 + 3x+1)/6\\
B_3(x) &=& (3x^3 - 6x^2 +4)/6\\
B_4(x) &=& (1-x)^3/6.
\end{eqnarray*}
Then the centered version of the cubic $B$-spline $\overline{Q}$ has
the form
\begin{equation*}
\overline{Q}(x) =
\begin{cases}
B_1(x+2) & ~\mbox{for}~ -2 \leq x \leq -1\\
B_2(x+1) & ~\mbox{for}~ -1 \leq x \leq 0\\
B_3(x) & ~\mbox{for}~ 0 \leq x \leq 1\\
B_4(x-1) & ~\mbox{for}~ 1 \leq x \leq 2\\
0 & ~\mbox{otherwise}
\end{cases}
=
\begin{cases}
\frac{1}{6}(2+x)^3& ~\mbox{for}~ -2 \leq x \leq -1\\
\frac{1}{6}(-3x^3-6x^2 + 4) & ~\mbox{for}~ -1 \leq x \leq 0\\
\frac{1}{6}(3x^3-6x^2 + 4) & ~\mbox{for}~ 0 \leq x \leq 1\\
\frac{1}{6}(2-x)^3 & ~\mbox{for}~ 1 \leq x \leq 2\\
0 & ~\mbox{otherwise}.
\end{cases}
\end{equation*}
Note that $G_{j}^Q(y)$ can then be computed by utilizing the fact
that, for $j=0,1$,
\begin{equation*}
G_{j}^Q(y)=\int_{-2}^{(y+\frac{A}{2})\wedge 2} x^j \overline{Q}(x)dx
- \int_{-2}^{(y-\frac{A}{2})\wedge 2} x^j \overline{Q}(x)dx,
\end{equation*}
where the integrals on the right hand side are defined to be zero if
the corresponding upper limits are less than $-2$. The integrals on
the RHS of above equation can be computed from the representation of
$\overline{Q}(\cdot)$ as follows:
\begin{eqnarray*}\label{eq:int_Q_bar}
\int_{-2}^b \overline{Q}(x) dx &=& \frac{1}{24} (2+b)^4,
~~\mbox{for}~-2 \leq b \leq -1 \nonumber\\
\int_{-1}^b \overline{Q}(x) dx &=& \frac{1}{24}(-3b^4 - 8b^3 + 16b
+ 11),~~\mbox{for}~-1 \leq b \leq 0\nonumber\\
\int_0^b \overline{Q}(x) dx &=& \frac{1}{24}(3b^4 - 8b^3 + 16b),
~~\mbox{for}~~0 \leq b \leq 1 \nonumber\\
\int_1^b \overline{Q}(x) dx &=& \frac{1}{24}(1-(2-b)^4),
~~\mbox{for}~~1 \leq b \leq 2,
\end{eqnarray*}
\begin{eqnarray*}\label{eq:int_x_Q_bar}
\int_{-2}^b x \overline{Q}(x) dx &=& \frac{1}{30}(2+b)^5 -
\frac{1}{12}(2+b)^4, ~~\mbox{for}~-2 \leq b \leq -1 \nonumber\\
\int_{-1}^b x \overline{Q}(x) dx &=& \frac{1}{60}(-6b^5 - 15 b^4 +
20b^2 - 11),~~\mbox{for}~-1 \leq b \leq 0\nonumber\\
\int_0^b x \overline{Q}(x) dx &=& \frac{1}{60}(6b^5 - 15 b^4 +
20b^2),
~~\mbox{for}~~0 \leq b \leq 1 \nonumber\\
\int_1^b x \overline{Q}(x) dx &=& \frac{1}{30}(2-b)^5
-\frac{1}{12}(2-b)^4 + \frac{1}{20}, ~~\mbox{for}~~1 \leq b \leq 2.
\end{eqnarray*}

\subsubsection*{Details of the calculation of pointwise
bias}\label{subsec:asymp_bias_calc}

Performing a Taylor series expansion around $(s,t)$ we get,
\begin{eqnarray}\label{eq:g_expansion}
g(s_l + x h) &=& g(s) + h(\frac{s_l - s}{h} + x) g'(s) +
\frac{h^2}{2} (\frac{s_l - s}{h} + x)^2 g''(s) \nonumber\\
&& ~~~~+ O((|\frac{s - s_l}{h}|^{2+\alpha} + |x|^{2+\alpha})
h^{2+\alpha})\nonumber\\
g(s_{l'} + y h) &=& g(t) + h(\frac{s_{l'} - t}{h} + y) g'(t) +
\frac{h^2}{2} (\frac{s_{l'} - t}{h} + y)^2 g''(t) \nonumber\\
&&~~~~ + O((|\frac{t - s_{l'}}{h}|^{2+\alpha} + |y|^{2+\alpha})
h^{2+\alpha}),
\end{eqnarray}
and
\begin{eqnarray}\label{eq:C_expansion}
C(s_l + xh, s_{l'} + yh)  &=& C(s,t) + h(\frac{s_l - s}{h} +
x,\frac{s_{l'} - t}{h} + y)
\begin{bmatrix} C_s(s,t)\\ C_t(s,t) \end{bmatrix} \nonumber\\
&& + \frac{h^2}{2}(\frac{s_l - s}{h} + x,\frac{s_{l'} - t}{h} + y)
\begin{bmatrix} C_{ss}  & C_{st} \\ C_{ts} & C_{tt} \end{bmatrix}
\begin{bmatrix} \frac{s-s_l}{h} + x \\ \frac{t-s_{l'}}{h} + y
\end{bmatrix} \nonumber\\
&&~~+ O\left(\left(|\frac{s - s_l}{h}|^{2+\alpha} + |\frac{t -
s_{l'}}{h}|^{2+\alpha} + |x|^{2+\alpha} + |y|^{2+\alpha}\right)
h^{2+\alpha}\right).
\end{eqnarray}
First we consider the off-diagonal terms, i.e., compute
$\mathbb{E}\widetilde C(s,t)$, for $|s-t| > 2Ah$.
\begin{itemize}
\item {\bf $h^0$ terms :} Since $\int K(x) dx = 1$ and $\int K'(x)
dx = 0$,
\begin{equation}\label{eq:h_0_term}
\int \int C(s,t) (K(x) + \frac{s-s_l}{h} K'(-x))(K(y) +
\frac{t-s_{l'}}{h} K'(-y)) dx dy = C(s,t).
\end{equation}

\item {\bf $h^1$ terms :} Since $\int x K'(-x) dx = 1$,
$\int x K(x) dx = 0$, and $\int K(x) dx = 1$,
\begin{eqnarray}\label{eq:h_1_term1}
~~~~~~~~~ \int \int h\left[(\frac{s_l-s}{h} + x) C_s +
(\frac{s_{l'}-t}{h}+y)
C_t\right] ~~~~~~~~~~~~~~~~~~~~&& \nonumber\\
~~\cdot ~ (K(x) + \frac{s-s_l}{h} K'(-x))(K(y) + \frac{t-s_{l'}}{h}
K'(-y)) dx dy &=& 0,
\end{eqnarray}
and
\begin{eqnarray}\label{eq:h_1_term2}
~~~~~~~~~ \int\int h C(s,t)\left[ g(s)g'(t)(\frac{s_{l'} - t}{h} +
y)
+ g'(s)g(t) (\frac{s_l - s}{h} + x) \right] && \nonumber\\
\cdot~ (K(x) + \frac{s-s_l}{h} K'(-x))(K(y) + \frac{t-s_{l'}}{h}
K'(-y)) dx dy &=& 0.
\end{eqnarray}

\item {\bf $h^2$ terms :} Since
$\int x^2 K'(-x) dx = 0$, $\int x K'(-x) dx = 1$, $\int x K(x) dx =
0$, and $\int K(x) dx = 1$,
\begin{eqnarray*}
&& \frac{h^2}{2} C(s,t) \int \int \left[ g''(t)g(s) (\frac{s_{l'} -
t}{h}
+ y)^2 + g''(s)g(t) (\frac{s_l - s}{h} + x)^2 \right] \\
&& ~~\cdot~(K(x) + \frac{s-s_l}{h} K'(-x))(K(y) + \frac{t-s_{l'}}{h}
K'(-y)) dx dy \nonumber\\
&=& \frac{h^2}{2} C(s,t) \left[ g''(t)g(s) (K_2 - (\frac{s_{l'} -
t}{h})^2) + g''(s)g(t) (K_2 - (\frac{s_l - s}{h})^2)\right];
\nonumber
\end{eqnarray*}
\begin{eqnarray*}
~~~~~~~~ h^2 C(s,t) \int \int [(\frac{s_l - s}{h} + x)(\frac{s_{l'}
- t}{h} +
y) g'(s) g'(t)] ~~~~~~~&& \\
~~\cdot~ (K(x) + \frac{s-s_l}{h} K'(-x)) (K(y) + \frac{t-s_{l'}}{h}
K'(-y)) dx dy &=& 0; \nonumber
\end{eqnarray*}
\begin{eqnarray*}
&& h^2 \int \int \left[(\frac{s_l - s}{h} + x) C_s + (\frac{s_{l'} -
t}{h} + y)C_t\right] \\
&& ~~~~~\cdot~ \left[g(s)g'(t)(\frac{s_{l'} - t}{h} + y) +
g'(s)g(t)(\frac{s_l - s}{h}+ x)\right] \nonumber\\
&& ~~~~~\cdot~ (K(x) + \frac{s-s_l}{h} K'(-x)) (K(y) +
\frac{t-s_{l'}}{h} K'(-y)) dx dy \nonumber\\
&=& h^2 \left[C_s g'(s)g(t) (K_2 - (\frac{s_l - s}{h})^2) + C_t
g(s)g'(t) (K_2 - (\frac{s_{l'} - t}{h})^2 )\right]; \nonumber
\end{eqnarray*}
\begin{eqnarray*}
&& \frac{h^2}{2} \int\int \left[(\frac{s_l - s}{h} + x)^2 C_{ss} + 2
(\frac{s_l - s}{h} + x) (\frac{s_{l'} - t}{h} + y) C_{st} +
(\frac{s_{l'} - t}{h} + y)^2 C_{tt} \right] \\
&& ~~\cdot ~ (K(x) + \frac{s - s_l}{h} K'(-x))(K(y) +
\frac{t-s_{l'}}{h} K'(-y)) dx dy \nonumber\\
&=& \frac{h^2}{2} \left[C_{ss} (K_2 - (\frac{s_l - s}{h})^2) +
C_{tt} (K_2 - (\frac{s_{l'} - t}{h})^2) \right]. \nonumber
\end{eqnarray*}
In summary, the \textit{$h^2$ term} in the expansion is,
\begin{eqnarray}\label{eq:h_2_term}
&& h^2 \left(\frac{1}{2}g''(s)g(t) C + g'(s)g(t) C_s + \frac{1}{2}
C_{ss}\right)(K_2 - (\frac{s_l - s}{h})^2) \nonumber\\
&&~~ +~ h^2 \left(\frac{1}{2}g(s)g''(t) C + g(s)g'(t) C_t +
\frac{1}{2} C_{tt}\right)(K_2 - (\frac{s_{l'} - t}{h})^2).
\end{eqnarray}

\end{itemize}

\vskip.1in\noindent{\bf Proof of Lemma \ref{lemmaC.1} :} Combining
(\ref{eq:h_0_term}), (\ref{eq:h_1_term1}), (\ref{eq:h_1_term2}) and
(\ref{eq:h_2_term}), and using (\ref{eq:g_expansion}),
(\ref{eq:C_expansion}) and the fact that $\sum_{l=1}^{L_n}
|\frac{s-s_l}{h}|^\beta \overline{Q}_h(s-s_l) < \infty$, after some
algebra, we obtain (\ref{eq:expect_C_tilde}).

\subsubsection*{Combined bound on $\mathbb{E}\parallel H_\nu
\widetilde C_c \psi_\nu\parallel_2^2$}

We put the different pieces derived in Appendix D  together to
obtain a bound on $\mathbb{E}\parallel H_\nu \widetilde C_c
\psi_\nu\parallel_2^2$. For ease of notation, we denote  by ${\cal
H}_\nu \equiv {\cal H}_\nu(x,s_1,s_2,t_1,t_2)$ the \textit{integral
operator} with kernel
$H_\nu(x,s_1)H_\nu(x,s_2)\psi_\nu(t_1)\psi_\nu(t_2)$. Then, with
$r_1$, $r_2$ taking values 0 or 1,
\begin{eqnarray}
\int\int\int\int\int {\cal H}_\nu(x,s_1,s_2,t_1,t_2)
(C(s_1,t_1))^{r_1}(C(s_2,t_2))^{r_2} ds_1ds_2dt_1dt_2 dx &=& 0. \label{eq:cal_H_oper_1}\\
\int\int\int\int\int {\cal H}_\nu(x,s_1,s_2,t_1,t_2)
(C(s_1,t_2))^{r_1}(C(s_2,t_1))^{r_2}ds_1ds_2dt_1dt_2 dx &=& 0.
\label{eq:cal_H_oper_2}
\end{eqnarray}
\begin{eqnarray}\label{eq:cal_H_oper_3}
&& \int\int\int\int\int {\cal H}_\nu(x,s_1,s_2,t_1,t_2)
(C(s_1,s_2))^{r_1}(C(t_1,t_2))^{r_2} ds_1ds_2dt_1dt_2 dx \nonumber\\
&=& \lambda_\nu^{r_2}\left[\sum_{1\leq k\neq \nu\leq M}
\frac{\lambda_k}{(\lambda_k-\lambda_\nu)^2}\right]^{r_1}.
\end{eqnarray}
Implicitly using (\ref{eq:delta_delta_psi}) -
(\ref{eq:delta_delta_WW_psi}), we also have the bound
\begin{equation}\label{eq:cal_H_oper_4}
|\int\int\int\int\int {\cal
H}_\nu(x,s_1,s_2,t_1,t_2)R(s_1,s_2,t_1,t_2)
 ds_1ds_2dt_1dt_2 dx| =
O(\parallel R\parallel_\infty).
\end{equation}

From Proposition \ref{propC.1}, the total contribution in
(\ref{eq:H_C_psi_offdiag}) of
the first terms on the RHS of (\ref{eq:sum_E_U_ii}) and
(\ref{eq:sum_E_U_i1i2}) becomes
\begin{eqnarray}\label{eq:H_C_psi_offdiag_main}
&&
\left(\frac{1}{n}(1-\frac{1}{n}\sum_{i=1}^n\frac{4m_i-6}{m_i(m_i-1)})
+\frac{n-1}{n}\right) \nonumber\\
&& \cdot \int\left[\int\int H_\nu(x,s)\overline{W}_{\widetilde
h_n}(s,t)[C(s,t)+O(h_n^2)]\psi_\nu(t)dsdt\right]^2dx\nonumber\\
&&+
\left(\frac{1}{n}(1-\frac{1}{n}\sum_{i=1}^n\frac{4m_i-6}{m_i(m_i-1)})+\frac{1}{n^2}
\sum_{i_1\neq i_2}\rho_{i_1i_2}^2\right)\nonumber\\
&& \cdot \int \int\int\int\int
H_\nu(x,s_1)H_\nu(x,s_2)\overline{W}_{\widetilde
h_n}(s_1,t_1)\overline{W}_{\widetilde h_n}(s_2,t_2)\nonumber\\
&&~~~ (C(s_1,s_2)+O(h_n^2))(C(t_1,t_2)+O(h_n^2))
\psi_\nu(t_1)\psi_\nu(t_2)
ds_1ds_2dt_1dt_2 dx \nonumber\\
&& +
\left(\frac{1}{n}(1-\frac{1}{n}\sum_{i=1}^n\frac{4m_i-6}{m_i(m_i-1)})+\frac{1}{n^2}
\sum_{i_1\neq i_2}\rho_{i_1i_2}^2\right)\nonumber\\
&& \cdot \int \int\int\int\int
H_\nu(x,s_1)H_\nu(x,s_2)\overline{W}_{\widetilde
h_n}(s_1,t_1)\overline{W}_{\widetilde h_n}(s_2,t_2)\nonumber\\
&&~~~ (C(s_1,t_2)+O(h_n^2))(C(s_2,t_1)+O(h_n^2))
\psi_\nu(t_1)\psi_\nu(t_2) ds_1ds_2dt_1dt_2 dx.
\end{eqnarray}
Since $H_\nu C\psi_\nu \equiv 0$, it can be checked that the first
integral in (\ref{eq:H_C_psi_offdiag_main}) is $O(h_n^2)$.
On the other hand, from the definition of $\overline{W}_{\widetilde
h_n}(s,t)$ and the fact that $H_\nu C\psi_\nu \equiv 0$, it follows
that the last integral term is $O(h_n)$.

Next, apply ${\cal H}_\nu$ to the following functions~: ~~
$W_{\widetilde h_n}(s_1,t_1)W_{\widetilde
h_n}(s_2,t_2)D_2(s_1,s_2,t_1,t_2)$ ~~and \\
$2\overline{W}_{\widetilde
h_n}(s_1,t_1)W_{\widetilde h_n}(s_2,t_2)D_3(s_1,s_2,t_1,t_2)$, where
$D_2(s_1,s_2,t_1,t_2)$ and $D_3(s_1,s_2,t_1,t_2)$ are the terms
given by
the sum of the first three terms on the RHS of
(\ref{eq:sum_E_V_i1i2}) (including the isolated $O(h_n^2)$ term),
and the sum of the first three terms on the RHS of
(\ref{eq:sum_E_U_i1_V_i2}), respectively. Then, adding these terms
to (\ref{eq:H_C_psi_offdiag_main}), we have, by
(\ref{eq:cal_H_oper_1}) - (\ref{eq:cal_H_oper_4}),
(\ref{eq:delta_delta_WW_psi}) (for dealing with the isolated
$O(h_n^2)$ term in
(\ref{eq:sum_E_V_i1i2})), and the comment following
(\ref{eq:H_C_psi_offdiag_main}), that this sum equals
\begin{eqnarray}\label{eq:R_1_order}
R_1  &=& \frac{1}{n}\left(\sum_{1\leq k\neq \nu\leq M}
\frac{\lambda_k\lambda_\nu}{(\lambda_k-\lambda_\nu)^2}\right) +
\left(\frac{1}{n^2}\sum_{i_1\neq i_2}^{n}\rho_{i_1i_2}^2\right)
\left(\sum_{1\leq k\neq \nu\leq M}
\frac{\lambda_k\lambda_\nu}{(\lambda_k-\lambda_\nu)^2} +
O(h_n)\right)\nonumber\\
&& ~~~~ + O(h_n^4)  + O\left(\frac{1}{n\underline{m}}\right) +
O\left(\frac{h_n}{n}\right).
\end{eqnarray}
Next, for notational convenience, express the integral operator
${\cal H}_\nu$ applied to $Z_j$ (where $Z_j$ are as in Propositions
\ref{propC.1} - \ref{propC.2}) times $\overline{W}_{\widetilde
h_n}(s_1,t_1)\overline{W}_{\widetilde h_n}(s_2,t_2)W_{\widetilde
h_n}(s_1,t_2)$ by ${\cal H}_\nu
\overline{W}^{s_1,t_1}\overline{W}^{s_2,t_2}W^{s_1,t_2} Z_j$, etc.
Using (\ref{eq:delta_delta_psi}) - (\ref{eq:delta_delta_WtsW_psi}),
and the bounds in Proposition \ref{propC.1} for $Z_j$,
$j=1,\ldots,4$, we have,
\begin{eqnarray*}
R_2 &:=& {\cal H}_\nu \overline{W}^{s_1,t_1}\overline{W}^{s_2,t_2}
Z_1 = {\cal H}_\nu
\overline{W}^{s_1,t_1}\overline{W}^{s_2,t_2}W^{s_1,s_2} Z_1 =
O\left(\frac{1}{nh_n\underline{m}}\right),\label{eq:R_2_order}\\
R_3 &:=& {\cal H}_\nu \overline{W}^{s_1,t_1}\overline{W}^{s_2,t_2}
Z_2 = {\cal H}_\nu
\overline{W}^{s_1,t_1}\overline{W}^{s_2,t_2}W^{s_1,t_2} Z_2 =
O\left(\frac{1}{n\underline{m}}\right),\label{eq:R_3_order}\\
R_4 &:=& {\cal H}_\nu \overline{W}^{s_1,t_1}\overline{W}^{s_2,t_2}
Z_3 = {\cal H}_\nu
\overline{W}^{s_1,t_1}\overline{W}^{s_2,t_2}W^{s_2,t_1} Z_3 =
O\left(\frac{1}{n\underline{m}}\right),\label{eq:R_4_order}\\
R_5 &:=& {\cal H}_\nu \overline{W}^{s_1,t_1}\overline{W}^{s_2,t_2}
Z_4 = {\cal H}_\nu
\overline{W}^{s_1,t_1}\overline{W}^{s_2,t_2}W^{t_1,t_2} Z_4 =
O\left(\frac{1}{n\underline{m}}\right).\label{eq:R_5_order}
\end{eqnarray*}
Using analogous reasoning, from Propositions \ref{propC.1} and
\ref{propC.2} we also have
\begin{eqnarray*}
R_6 &:=& {\cal H}_\nu \overline{W}^{s_1,t_1}\overline{W}^{s_2,t_2}
Z_5 = O\left(\frac{1}{nh_n\underline{m}^2}\right),\label{eq:R_6_order}\\
R_7 &:=& {\cal H}_\nu \overline{W}^{s_1,t_1}\overline{W}^{s_2,t_2}
Z_6 = O\left(\frac{1}{n\underline{m}^2}\right)\label{eq:R_7_order}\\
R_8 &:=& {\cal H}_\nu W^{s_1,t_1}W^{s_2,t_2}
Z_7 = O\left(\frac{h_n}{n\underline{m}}\right),\label{eq:R_8_order}\\
R_9 &:=& {\cal H}_\nu \overline{W}^{s_1,t_1}W^{s_2,t_2}
Z_8 = O\left(\frac{1}{nh_n\underline{m}}\right),\label{eq:R_9_order}\\
R_{10} &:=& {\cal H}_\nu \overline{W}^{s_1,t_1}W^{s_2,t_2} Z_9 =
O\left(\frac{1}{n\underline{m}}\right).\label{eq:R_10_order}
\end{eqnarray*}
Hence, combining (\ref{eq:R_1_order}) with the bounds for $R_2$ to
$R_{10}$, using the definitions of $\mathbb{E}[\widetilde
C(s_1,t_1)\widetilde C(s_2,t_2)]$,
$\mathbb{E}[\widehat{\overline{C}}(\frac{s_1+t_1}{2})
\widehat{\overline{C}}(\frac{s_2+t_2}{2})]$, and
$\mathbb{E}[\widetilde
C(s_1,t_1)\widehat{\overline{C}}(\frac{s_2+t_2}{2})]$, and plugging
everything back into (\ref{eq:H_C_psi_L_2}), we complete the proof
of Proposition \ref{prop4}. The details of the key steps in this
derivation are given below.

\subsubsection*{Proof details for Proposition \ref{prop4}}

\noindent{\bf Proof of Proposition \ref{propC.1} :} We need to deal
with terms of the form
\begin{eqnarray*}\label{eq:general_sq_term}
&& \widetilde
E_{i_1i_2;j_1j_1'j_2j_2'}(s_1,t_1,s_2,t_2;l_1,l_1',l_2,l_2')
\nonumber\\
&:=& \mathbb{E}[Y_{i_1 j_1}Y_{i_1 j_1'} Y_{i_2j_2}Y_{i_2j_2'}
\widetilde K_{s_1,l_1}(T_{i_1j_1})\widetilde
K_{t_1,l_1'}(T_{i_1j_1'})\widetilde
K_{s_2,l_2}(T_{i_2j_2})\widetilde K_{t_2,l_2'}(T_{i_2j_2'})],
\end{eqnarray*}
for $1\leq j_1,j_1'\leq m_{i_1}$, $1\leq j_2,j_2'\leq m_{i_2}$,
$1\leq i_1,i_2 \leq n$. For computational convenience, we also
define,
\begin{eqnarray}\label{eq:E_i1i2_def}
&& E_{i_1i_2;j_1j_1'j_2j_2'}(s_1,t_1,s_2,t_2;l_1,l_1',l_2,l_2')
\nonumber\\
&=& \widetilde
E_{i_1i_2;j_1j_1'j_2j_2'}(s_1,t_1,s_2,t_2;l_1,l_1',l_2,l_2')
\overline{Q}_h(s_1-s_{l_1})\overline{Q}_h(t_1-s_{l_1'})
\overline{Q}_h(s_2-s_{l_2})\overline{Q}_h(t_2-s_{l_2'}).
\end{eqnarray}
First, consider the case $i_1 = i_2 = i$, say. Then, using to
$\star$ to denote
$E_{ii;j_1j_1'j_2j_2'}(s_1,t_1,s_2,t_2;l_1,l_1',l_2,l_2')$, we have
\begin{eqnarray}\label{eq:U_ii_expect_1}
&& \mathbb{E}[U_i(s_1,t_1)U_i(s_2,t_2)] \nonumber\\
&=& \frac{1}{m_i^4} \sum_{j_1\neq j_1'\neq
j_2 \neq j_2'}^{m_i} \sum_{l_1,l_1'=1}^{L_n}\sum_{l_2,l_2'=1}^{L_n} \star \nonumber\\
&&  + \frac{1}{m_i^4} \left[\sum_{j_1=j_1'\neq j_2 \neq j_2'}^{m_i}
+ \sum_{j_1 = j_2\neq j_1' \neq j_2'}^{m_i} +
\sum_{j_1 = j_2'\neq j_1'\neq j_2}^{m_i} \right.\nonumber\\
&& ~~~~ \left. + \sum_{j_1\neq j_1'\neq j_2=j_2'}^{m_i} + \sum_{j_1
\neq j_1' = j_2 \neq j_2'}^{m_i} + \sum_{j_1 \neq j_2\neq
j_1'=j_2'}^{m_i}
\right] \sum_{l_1,l_1'=1}^{L_n}\sum_{l_2,l_2'=1}^{L_n} \star \nonumber\\
&& + \frac{1}{m_i^4} \left[\sum_{j_1=j_1'\neq j_2 = j_2'}^{m_i} +
\sum_{j_1 = j_2\neq j_1' = j_2'}^{m_i} + \sum_{j_1 = j_2' \neq j_1'
= j_2}^{m_i} \right]
\sum_{l_1,l_1'=1}^{L_n}\sum_{l_2,l_2'=1}^{L_n} \star \nonumber\\
&& + \frac{1}{m_i^4} \left[\sum_{j_1=j_1'= j_2 \neq j_2'}^{m_i} +
\sum_{j_1 = j_1' = j_2' \neq j_2}^{m_i} + \sum_{j_1 = j_2 = j_2'
\neq j_1'}^{m_i} + \sum_{j_1\neq j_1'= j_2=j_2'}^{m_i} \right]
\sum_{l_1,l_1'=1}^{L_n}\sum_{l_2,l_2'=1}^{L_n} \star \nonumber\\
&& + \frac{1}{m_i^4} \sum_{j_1 = j_1' = j_2 = j_2'}^{m_i}
\sum_{l_1,l_1'=1}^{L_n}\sum_{l_2,l_2'=1}^{L_n} \star ~.
\end{eqnarray}
Next, consider the case $i_1 \neq i_2$. Then, with $\star$ denoting
$E_{i_1i_2;j_1j_1'j_2j_2'}(s_1,t_1,s_2,t_2;l_1,l_1',l_2,l_2')$,
\begin{eqnarray}\label{eq:U_i1i2_expect_1}
&& \mathbb{E}[U_{i_1}(s_1,t_1)U_{i_2}(s_2,t_2)] \nonumber\\
&=& \frac{1}{m_{i_1}^2m_{i_2}^2} \sum_{j_1\neq
j_1'}^{m_{i_1}}\sum_{j_2 \neq j_2'}^{m_{i_2}}
\sum_{l_1,l_1'=1}^{L_n}\sum_{l_2,l_2'=1}^{L_n} \star \nonumber\\
&& + \frac{1}{m_{i_1}^2m_{i_2}^2}
\left[\sum_{j_1=j_1'}^{m_{i_1}}\sum_{j_2 \neq j_2'}^{m_{i_2}} +
\sum_{j_1\neq j_1'}^{m_{i_1}}\sum_{j_2=j_2'}^{m_{i_2}} \right]
\sum_{l_1,l_1'=1}^{L_n}\sum_{l_2,l_2'=1}^{L_n} \star  +
\frac{1}{m_{i_1}^2m_{i_2}^2} \sum_{j_1=j_1'}^{m_{i_1}}\sum_{j_2 =
j_2'}^{m_{i_2}} \sum_{l_1,l_1'=1}^{L_n}\sum_{l_2,l_2'=1}^{L_n} \star
~.
\end{eqnarray}
Note that, for all $i_1$, $i_2$, if either $j_1 = j_1'$ or $j_2 =
j_2'$, then
\begin{equation*}
\sum_{l_1,l_1'=1}^{L_n}\sum_{l_2,l_2'=1}^{L_n}
E_{i_1i_2;j_1j_1'j_2j_2'}(s_1,t_1,s_2,t_2;l_1,l_1',l_2,l_2') = 0,
\end{equation*}
unless $|s_1 - t_1|\leq \frac{A}{2}h$, or $|s_2 - t_2|\leq
\frac{A}{2}h$, respectively, for $A$ satisfying $A\geq 4(B_K + C_Q)$
and $\widetilde h_n = A h_n$. This can be verified by using the
definition of
$E_{i_1i_2;j_1j_1'j_2j_2'}(s_1,t_1,s_2,t_2;l_1,l_1',l_2,l_2')$,
equations (\ref{eq:Y_four_terms_cond2}),
(\ref{eq:Y_four_terms_cond3}), (\ref{eq:Y_four_terms_cond5}) -
(\ref{eq:Y_four_terms_cond8}), and arguing as in the analysis of the
term (\ref{eq:Y_ij_square_expect}). Therefore, since
$\mathbf{1}_{|s_k-t_k|\leq \frac{A}{2}h}\overline{W}_{\widetilde
h_n}(s_k,t_k) =0$, for $k=1,2$, the sums corresponding to either
$j_1=j_1'$ or $j_2=j_2'$ in (\ref{eq:U_ii_expect_1}) and
(\ref{eq:U_i1i2_expect_1}) do not contribute anything to
(\ref{eq:H_C_psi_offdiag}). Thus, when $i_1 \neq i_2$, the only sum
that contributes to (\ref{eq:H_C_psi_offdiag}) corresponds to
$j_1\neq j_1'$, $j_2\neq j_2'$. When $i_1=i_2=i$, the sums that
contribute to (\ref{eq:H_C_psi_offdiag}) are the ones corresponding
to  $j_1\neq j_1'\neq j_2\neq j_2'$, $j_1 = j_2\neq j_1'\neq j_2'$,
$j_1= j_2'\neq j_1'\neq j_2'$, $j_1\neq j_1' = j_2\neq j_2'$,
$j_1\neq j_2\neq j_1'= j_2'$, $j_1=j_2\neq j_1'=j_2'$, and
$j_1=j_2'\neq j_1'=j_2$. We consider these cases one by one.

\begin{lemma}\label{lemmaD.1}
If $i_1=i_2$, $j_1\neq j_1'\neq j_2 \neq
j_2'$; or $i_1\neq i_2$ $j_1\neq j_1'$, $j_2\neq j_2'$, then
\begin{eqnarray}\label{eq:E_sum_i1i2_j_different}
&& \sum_{l_1,l_1'=1}^{L_n}\sum_{l_2,l_2'=1}^{L_n}
\frac{E_{i_1i_2;j_1j_1'j_2j_2'}(s_1,t_1,s_2,t_2;l_1,l_1',l_2,l_2')}
{g(s_1)g(t_1)g(s_2)g(t_2)} \nonumber\\
&=&  (C(s_1,t_1)+O(h^2))(C(s_2,t_2)+O(h^2))\nonumber\\
&&  + \rho_{i_1i_2}^2\left[(C(s_1,s_2)+O(h^2))(C(t_1,t_2)+O(h^2)) +
(C(s_1,t_2)+O(h^2))(C(s_2,t_1)+O(h^2))\right],
\end{eqnarray}
where the $O(h^2)$ terms are uniform in $s_1,t_1,s_2,t_2\in [0,1]$.
\end{lemma}

\vskip.1in The following lemma gives an expression and the
corresponding bound for the term $Z_1$.

\begin{lemma}\label{lemmaD.2}
If $i_1=i_2=i$, $j_1=j_2\neq j_1'\neq j_2'$, then
\begin{eqnarray}\label{eq:E_sum_ii_j1j2_same}
&&\frac{1}{n^2}\sum_{i=1}^n\frac{1}{m_i^4}\sum_{j_1 = j_2\neq j_1'
\neq j_2'}^{m_i}\sum_{l_1,l_1'=1}^{L_n}\sum_{l_2,l_2'=1}^{L_n}
\frac{E_{ii;j_1j_1'j_2j_2'}(s_1,t_1,s_2,t_2;l_1,l_1',l_2,l_2')}
{g(s_1)g(t_1)g(s_2)g(t_2)} \nonumber\\
&=& \begin{cases} O(\frac{1}{nh\underline{m}}) & ~\mbox{if}~
|s_1-s_2|\leq
\frac{Ah}{2}\\
0 & ~\mbox{otherwise}
\end{cases}
\end{eqnarray}
\end{lemma}

\vskip.1in The following lemma gives expressions and the
corresponding bounds for the term $Z_2$, $Z_3$ and $Z_4$.

\begin{lemma}\label{lemmaD.3}
If $i_1=i_2=i$ and $j_1 = j_2'\neq j_1'
\neq j_2$; $j_1\neq j_1'=j_2\neq j_2'$; $j_1\neq j_2\neq j_1'=j_2'$,
then
\begin{eqnarray}
&&\frac{1}{n^2}\sum_{i=1}^n\frac{1}{m_i^4}\sum_{j_1 = j_2'\neq j_1'
\neq j_2}^{m_i}\sum_{l_1,l_1'=1}^{L_n}\sum_{l_2,l_2'=1}^{L_n}
\frac{E_{ii;j_1j_1'j_2j_2'}(s_1,t_1,s_2,t_2;l_1,l_1',l_2,l_2')}
{g(s_1)g(t_1)g(s_2)g(t_2)} \nonumber \\
&=& \begin{cases} O(\frac{1}{nh\underline{m}}) & ~\mbox{if}~
|s_1-t_2|\leq
\frac{Ah}{2}\\
0 & ~\mbox{otherwise}
\end{cases} \label{eq:E_sum_ii_j1j2prime_same}\\
&& \frac{1}{n^2}\sum_{i=1}^n\frac{1}{m_i^4}\sum_{j_1\neq
j_1'=j_2\neq
j_2'}^{m_i}\sum_{l_1,l_1'=1}^{L_n}\sum_{l_2,l_2'=1}^{L_n}
\frac{E_{ii;j_1j_1'j_2j_2'}(s_1,t_1,s_2,t_2;l_1,l_1',l_2,l_2')}
{g(s_1)g(t_1)g(s_2)g(t_2)} \nonumber\\
&=& \begin{cases} O(\frac{1}{nh\underline{m}}) & ~\mbox{if}~
|t_1-s_2|\leq
\frac{Ah}{2}\\
0 & ~\mbox{otherwise}
\end{cases} \label{eq:E_sum_ii_j1primej2_same}\\
&& \frac{1}{n^2}\sum_{i=1}^n\frac{1}{m_i^4}\sum_{j_1\neq j_2\neq
j_1'=j_2'}^{m_i}\sum_{l_1,l_1'=1}^{L_n}\sum_{l_2,l_2'=1}^{L_n}
\frac{E_{ii;j_1j_1'j_2j_2'}(s_1,t_1,s_2,t_2;l_1,l_1',l_2,l_2')}
{g(s_1)g(t_1)g(s_2)g(t_2)} \nonumber\\
&=& \begin{cases} O(\frac{1}{nh\underline{m}}) & ~\mbox{if}~
|t_1-t_2|\leq
\frac{Ah}{2}\\
0 & ~\mbox{otherwise}
\end{cases} \label{eq:E_sum_ii_j1primej2prime_same}
\end{eqnarray}
\end{lemma}

\vskip.1in The following lemma gives expressions and the
corresponding bounds for the terms $Z_5$ and $Z_6$.

\begin{lemma}\label{lemmaD.4}
If $i_1=i_2=i$ and $j_1=j_2\neq j_1'=j_2'$, $j_1=j_2'\neq
j_1'=j_2$, then
\begin{eqnarray}\label{eq:E_sum_ii_j1j2_same_j1primej2prime}
&&\frac{1}{n^2}\sum_{i=1}^n\frac{1}{m_i^4}\sum_{j_1 = j_2\neq j_1' =
j_2'}^{m_i}\sum_{l_1,l_1'=1}^{L_n}\sum_{l_2,l_2'=1}^{L_n}
\frac{E_{ii;j_1j_1'j_2j_2'}(s_1,t_1,s_2,t_2;l_1,l_1',l_2,l_2')}
{g(s_1)g(t_1)g(s_2)g(t_2)}\nonumber\\
&=& \begin{cases} O(\frac{1}{nh^2\underline{m}^2}) & ~\mbox{if}~
\max\{|s_1-s_2|,|t_1-t_2|\}\leq\frac{Ah}{2}\\
O(\frac{1}{nh\underline{m}^2}) & ~\mbox{if}~|s_1-s_2|
\leq\frac{Ah}{2} ~\mbox{and}~ |t_1-t_2| > \frac{Ah}{2}\\
O(\frac{1}{nh\underline{m}^2}) & ~\mbox{if}~|s_1-s_2|
> \frac{Ah}{2}~\mbox{and}~ |t_1-t_2| \leq \frac{Ah}{2}\\
0 & ~\mbox{otherwise};
\end{cases}
\end{eqnarray}
\begin{eqnarray}\label{eq:E_sum_ii_j1j2prime_same_j1primej2}
&&\frac{1}{n^2}\sum_{i=1}^n\frac{1}{m_i^4}\sum_{j_1 = j_2'\neq j_1'
= j_2}^{m_i}\sum_{l_1,l_1'=1}^{L_n}\sum_{l_2,l_2'=1}^{L_n}
\frac{E_{ii;j_1j_1'j_2j_2'}(s_1,t_1,s_2,t_2;l_1,l_1',l_2,l_2')}
{g(s_1)g(t_1)g(s_2)g(t_2)} \nonumber\\
&=& \begin{cases} O(\frac{1}{nh^2\underline{m}^2}) & ~\mbox{if}~
\max\{|s_1-t_2|,|s_2-t_1|\}\leq\frac{Ah}{2}\\
O(\frac{1}{nh\underline{m}^2}) & ~\mbox{if}~|s_1-t_2|
\leq\frac{Ah}{2} ~\mbox{and}~ |s_2-t_1| > \frac{Ah}{2}\\
O(\frac{1}{nh\underline{m}^2}) & ~\mbox{if}~|s_1-t_2|
> \frac{Ah}{2} ~\mbox{and}~ |s_2-t_1| \leq \frac{Ah}{2}\\
0 & ~\mbox{otherwise}.
\end{cases}
\end{eqnarray}
\end{lemma}

\vskip.1in\noindent{\bf Proof of Proposition \ref{propC.2} :}
Define,
\begin{eqnarray}\label{eq:F_i1i2_def}
&& F_{i_1i_2;j_1,j_2}(s_1,t_1,s_2,t_2;l_1,l_2)\nonumber \\
&:=& \mathbb{E}[Y_{i_1j_1}^2Y_{i_2j_2}^2\widetilde
K_{(s_1+t_1)/2,l_1}(T_{i_1j_1})\widetilde
K_{(s_2+t_2)/2,l_2}(T_{i_2j_2})]
\overline{Q}_h(\frac{s_1+t_1}{2}-s_{l_1})
\overline{Q}_h(\frac{s_1+t_1}{2}-s_{l_2})
\end{eqnarray}
and
\begin{eqnarray}\label{eq:G_i1i2_def}
G_{i_1i_2;j_1,j_1',j_2}(s_1,t_1,s_2,t_2;l_1,l_1',l_2) &:=&
\mathbb{E}[Y_{i_1j_1}Y_{i_1j_1'}Y_{i_2j_2}^2\widetilde
K_{s_1,l_1}(T_{i_1j_1})\widetilde K_{t_1,l_1'}(T_{i_1j_1'})
\widetilde K_{(s_2+t_2)/2,l_2}(T_{i_2j_2})] \nonumber\\
&& ~~~~~\cdot~
\overline{Q}_h(s_1-s_{l_1})\overline{Q}_h(t_1-s_{l_1'})
\overline{Q}_h((s_2+t_2)/2-s_{l_2}).
\end{eqnarray}
First, if $i_1=i_2=i$ then, with $\star$ denoting
$F_{ii;j_1j_2}(s_1,t_1,s_2,t_2;l_1,l_2)$,
\begin{eqnarray}\label{eq:V_ii_expect_1}
\mathbb{E}[V_{i}(\frac{s_1+t_1}{2})V_{i}(\frac{s_2+t_2}{2})] &=&
\frac{1}{m_i^2} \sum_{j_1\neq j_2}^{m_i}
\sum_{l_1=1}^{L_n}\sum_{l_2=1}^{L_n} \star ~ + \frac{1}{m_i^2}
\sum_{j_1=j_2}^{m_i} \sum_{l_1=1}^{L_n}\sum_{l_2=1}^{L_n} \star~.
\end{eqnarray}
Next, if $i_1\neq i_2$ then, with $\star$ denoting
$F_{i_1i_2;j_1j_2}(s_1,t_1,s_2,t_2;l_1,l_2)$,
\begin{eqnarray}\label{eq:V_i1i2_expect_1}
\mathbb{E}[V_{i_1}(\frac{s_1+t_1}{2})V_{i_2}(\frac{s_2+t_2}{2})]
 &=& \frac{1}{m_{i_1}m_{i_2}} \sum_{j_1=1}^{m_{i_1}}\sum_{j_2
=1}^{m_{i_2}} \sum_{l_1=1}^{L_n}\sum_{l_2=1}^{L_n} \star ~.
\end{eqnarray}
Next, if $i_1=i_2=i$ then, with $\star$ denoting
$G_{ii;j_1,j_1',j_2}(s_1,t_1,s_2,t_2;l_1,l_1',l_2)$,
\begin{eqnarray}\label{eq:UV_ii_expect_1}
\mathbb{E}[U_{i}(s_1,t_1)V_{i}(\frac{s_2+t_2}{2})]  &=&
\frac{1}{m_i^3} \sum_{j_1\neq j_1'\neq j_2}^{m_i}
\sum_{l_1,l_1'=1}^{L_n}\sum_{l_2=1}^{L_n} \star ~~+ \frac{1}{m_i^3}
\sum_{j_1=j_1'\neq j_2}^{m_i}
\sum_{l_1,l_1'=1}^{L_n}\sum_{l_2=1}^{L_n} \star \nonumber\\
&& \frac{1}{m_i^3} \sum_{j_1'\neq j_1 = j_2}^{m_i}
\sum_{l_1,l_1'=1}^{L_n}\sum_{l_2=1}^{L_n} \star ~+ \frac{1}{m_i^3}
\sum_{j_1 \neq j_1' = j_2}^{m_i}
\sum_{l_1,l_1'=1}^{L_n}\sum_{l_2=1}^{L_n} \star \nonumber\\
&& ~~~~~~~+ \frac{1}{m_i^3} \sum_{j_1 = j_1' = j_2}^{m_i}
\sum_{l_1,l_1'=1}^{L_n}\sum_{l_2=1}^{L_n} \star ~.
\end{eqnarray}
Finally, if $i_1\neq i_2$, then, with $\star$ denoting
$G_{i_1i_2;j_1,j_1',j_2}(s_1,t_1,s_2,t_2;l_1,l_1',l_2)$,
\begin{eqnarray}\label{eq:UV_i1i2_expect_1}
&& \mathbb{E}[U_{i_1}(s_1,t_1)V_{i_2}(\frac{s_2+t_2}{2})] \nonumber\\
&=& \frac{1}{m_{i_1}^2m_{i_2}} \sum_{j_1\neq j_1'}^{m_{i_1}}
\sum_{j_2=1}^{m_{i_2}} \sum_{l_1,l_1'=1}^{L_n}\sum_{l_2=1}^{L_n}
\star ~~+ \frac{1}{m_{i_1}^2m_{i_2}} \sum_{j_1=j_1'}^{m_{i_1}}
\sum_{j_2=1}^{m_{i_2}} \sum_{l_1,l_1'=1}^{L_n}\sum_{l_2=1}^{L_n}
\star ~.
\end{eqnarray}
Arguments similar to those employed earlier show that the sums
corresponding to $j_1=j_1'$ in (\ref{eq:UV_ii_expect_1}) and
(\ref{eq:UV_i1i2_expect_1}) do not contribute anything to
$\mathbb{E}\parallel H_\nu \widehat C_c \psi_\nu\parallel^2$.

\vskip.1in We first consider $\mathbb{E}(V_{i_1}(z_1)V_{i_2}(z_2))$.
Then Lemmas \ref{lemmaD.5} and \ref{lemmaD.6}, stated below, give
expressions for the leading term and the term $Z_7$ (and
corresponding bound), respectively, in (\ref{eq:sum_E_V_i1i2}).

\begin{lemma}\label{lemmaD.5}
If $i_1\neq i_2$ or $i_1=i_2$, and $j_1\neq j_2$ then for
$|s_k-t_k|\leq \frac{Ah}{2}$, $k=1,2$, with $A \geq 4(B_K+C_Q)$,
\begin{eqnarray}\label{eq:F_sum_i1i2_j_different}
&& \frac{1}{n^2} \sum_{i=1}^n \frac{1}{m_i^2} \sum_{j_1\neq
j_2}^{m_i} \sum_{l_1=1}^{L_n}\sum_{l_2=1}^{L_n}
\frac{F_{ii;j_1,j_2}(s_1,t_1,s_2,t_2;l_1,l_2)}{g(z_1)g(z_2)} \nonumber\\
&& + \frac{1}{n^2}\sum_{i_1\neq i_2}^n \frac{1}{m_{i_1}m_{i_2}}
\sum_{j_1=1}^{m_{i_1}} \sum_{j_2=1}^{m_{i_2}}
\sum_{l_1=1}^{L_n}\sum_{l_2=1}^{L_n}
\frac{F_{i_1i_2;j_1,j_2}(s_1,t_1,s_2,t_2;l_1,l_2)}{g(z_1)g(z_2)}
\nonumber\\
&& - \sigma^2[\mathbb{E}(\widehat C_*(z_1))+\mathbb{E}(\widehat
C_*(z_2))] + \sigma^4
\nonumber\\
&=& C(s_1,t_1)C(s_2,t_2)- \left(\frac{1}{n^2}\sum_{i=1}^n
\frac{1}{m_i}\right) (C(s_1,t_1)+\sigma^2)(C(s_2,t_2)+\sigma^2)
+ O(h^2)\nonumber\\
&& \hskip-.2in + \left(\frac{1}{n}(1-\frac{1}{n}\sum_{i=1}^n
\frac{1}{m_i}) + \frac{1}{n^2}\sum_{i_1\neq i_2}^n
\rho_{i_1i_2}^2\right) (C(s_1,s_2)C(t_1,t_2) + C(s_1,t_2)C(s_2,t_1)
+ O(h)).
\end{eqnarray}
\end{lemma}

\begin{lemma}\label{lemmaD.6}
If $i_1=i_2=i$, $j_1=j_2$, then
\begin{eqnarray}\label{eq:F_sum_ii_j_same}
\frac{1}{n^2} \sum_{i=1}^n \frac{1}{m_i^2} \sum_{j_1=1}^{m_i}
\sum_{l_1=1}^{L_n}\sum_{l_2=1}^{L_n}
\frac{F_{ii;j_1,j_1}(s_1,t_1,s_2,t_2;l_1,l_2)}{g(z_1)g(z_2)}  &=&
\begin{cases}
O(\frac{1}{nh\underline{m}}) & ~\mbox{if}~|z_1-z_2| \leq \frac{Ah}{2} \\
0 & ~\mbox{otherwise.}
\end{cases}
\end{eqnarray}
\end{lemma}

\vskip.1in Finally, consider the term
$\mathbb{E}(U_{i_1}(s_1,t_1)V_{i_2}(z_2))$. Lemma \ref{lemmaD.7}
gives an expression for the leading term in
(\ref{eq:sum_E_U_i1_V_i2}), Lemma \ref{lemmaD.8} gives expressions
and the corresponding bounds for the terms $Z_8$ and $Z_9$.

\begin{lemma}\label{lemmaD.7}
If $i_1\neq i_2$, $j_1\neq j_1'$; $i_1=i_2$, $j_1\neq j_1'\neq j_2$,
then for $|s_1-t_1| > \frac{Ah}{2}$, and $|s_2-t_2|\leq
\frac{Ah}{2}$, with $A \geq 4(B_K+C_Q)$,
\begin{eqnarray}\label{eq:G_sum_i1i2_j_different}
&& \frac{1}{n^2} \sum_{i=1}^n w(m_i)\frac{1}{m_i^3} \sum_{j_1\neq
j_1'\neq j_2}^{m_i} \sum_{l_1,l_1'=1}^{L_n}\sum_{l_2=1}^{L_n}
\frac{G_{ii;j_1,j_1',j_2}(s_1,t_1,s_2,t_2;l_1,l_1',l_2)}{g(s_1)g(t_1)g(z_2)} \nonumber\\
&& + \frac{1}{n^2}\sum_{i_1\neq i_2}^n w(m_{i_1}) \frac{1}{m_{i_1}^2
m_{i_2}} \sum_{j_1\neq j_1'}^{m_{i_1}} \sum_{j_2=1}^{m_{i_2}}
\sum_{l_1,l_1'=1}^{L_n}\sum_{l_2=1}^{L_n}
\frac{G_{i_1i_2;j_1,j_1',j_2}(s_1,t_1,s_2,t_2;l_1,l_1',l_2)}{g(s_1)g(t_1)g(z_2)}
- \sigma^2 \mathbb{E}\widetilde C(s_1,t_1)
\nonumber\\
&=& (C(s_1,t_1) + O(h^2))(C(s_2,t_2) + O(h^2))\nonumber\\
&&  - \left(\frac{1}{n^2}\sum_{i=1}^n
\frac{2}{m_i}\right)(C(s_1,t_1)
+ O(h^2))(C(s_2,t_2) + \sigma^2 + O(h^2))\nonumber\\
&& + \left(\frac{1}{n}(1-\frac{1}{n}\sum_{i=1}^n \frac{2}{m_i}) +
\frac{1}{n^2}\sum_{i_1\neq i_2}^n \rho_{i_1i_2}^2\right)
(C(s_1,s_2)C(t_1,t_2) + C(s_1,t_2)C(s_2,t_1) + O(h)).
\end{eqnarray}
\end{lemma}

\begin{lemma}\label{lemmaD.8}
If $i_1=i_2=i$ and
$j_1'\neq j_1=j_2$, $j_1\neq j_1'=j_2$, then
\begin{eqnarray}\label{eq:G_sum_ii_j1j2_same}
&& \frac{1}{n^2} \sum_{i=1}^n w(m_i)\frac{1}{m_i^3} \sum_{j_1'\neq
j_1 = j_2}^{m_i} \sum_{l_1,l_1'=1}^{L_n}\sum_{l_2=1}^{L_n}
\frac{G_{ii;j_1,j_1',j_2}(s_1,t_1,s_2,t_2;l_1,l_1',l_2)}{g(s_1)g(t_1)g(z_2)} \nonumber\\
&=& \begin{cases} O(\frac{1}{nh^2\underline{m}}) &~\mbox{if}~ |s_1-s_2|\leq \frac{Ah}{2}\\
0 & ~\mbox{otherwise}.
\end{cases}
\end{eqnarray}
\begin{eqnarray}\label{eq:G_sum_ii_j1primej2_same}
&& \frac{1}{n^2} \sum_{i=1}^n w(m_i)\frac{1}{m_i^3} \sum_{j_1\neq
j_1' = j_2}^{m_i} \sum_{l_1,l_1'=1}^{L_n}\sum_{l_2=1}^{L_n}
\frac{G_{ii;j_1,j_1',j_2}(s_1,t_1,s_2,t_2;l_1,l_1',l_2)}{g(s_1)g(t_1)g(z_2)} \nonumber\\
&=& \begin{cases} O(\frac{1}{nh^2\underline{m}}) &~\mbox{if}~ |t_1-s_2|\leq \frac{Ah}{2}\\
0 & ~\mbox{otherwise}.
\end{cases}
\end{eqnarray}
\end{lemma}

\subsubsection*{Details of the calculation of pointwise
variance (\ref{eq:C_hat_variance})}\label{subsec:asymp_var_calc}

\noindent{\bf Proof of Proposition \ref{propC.3} :} Consider first
\begin{equation*}
\overline{W}_{\widetilde h_n}(s,t)\mbox{Var}(\widetilde C(s,t)) =
\overline{W}_{\widetilde h_n}(s,t) \mathbb{E}(\widetilde C(s,t))^2 -
(\mathbb{E}(\overline{W}_{\widetilde h_n}(s,t) \widetilde
C(s,t)))^2.
\end{equation*}
Using (\ref{eq:U_i_def}), (\ref{eq:E_i1i2_def}) and the arguments
leading to (\ref{eq:E_sum_i1i2_j_different}), we have
\begin{eqnarray}\label{eq:C_tilde_var dominant}
&& \overline{W}_{\widetilde h_n}(s,t) \frac{1}{n^2} \sum_{i_1\neq
i_2}^n \frac{w(m_{i_1})w(m_{i_2})}{m_{i_1}^2m_{i_2}^2} \sum_{j_1\neq
j_1'}^{m_{i_1}} \sum_{j_2\neq j_2'}^{m_{i_2}}
\sum_{l_1,l_1'=1}^{L_n}\sum_{l_2,l_2'=1}^{L_n}
\frac{E_{i_1i_2;j_1j_1'j_2j_2'}(s,t,s,t;l_1,l_1',l_2,l_2')}
{(g(s)g(t))^2}\nonumber\\
&& ~~~~~~~~- (\mathbb{E}(\overline{W}_{\widetilde h_n}(s,t)
\widetilde
C(s,t)))^2\nonumber\\
&=& \overline{W}_{\widetilde h_n}(s,t) \frac{1}{n^2} \sum_{i_1\neq
i_2}^n \frac{w(m_{i_1})w(m_{i_2})}{m_{i_1}^2m_{i_2}^2} \sum_{j_1\neq
j_1'}^{m_{i_1}} \sum_{j_2\neq j_2'}^{m_{i_2}}
\sum_{l_1,l_1'=1}^{L_n}\sum_{l_2,l_2'=1}^{L_n}\nonumber\\
&&  \frac{\mathbb{E}[C(T_{i_1j_1},T_{i_1j_1'})\widetilde
K_{s,l_1}(T_{i_1j_1})\widetilde K_{t,l_1'}(T_{i_1j_1'})]}{g(s)g(t)}
\frac{\mathbb{E}[C(T_{i_2j_2},T_{i_2j_2'})\widetilde
K_{s,l_2}(T_{i_2j_2})\widetilde
K_{t,l_2'}(T_{i_2j_2'})]}{g(s)g(t)}\nonumber\\
&& - \overline{W}_{\widetilde h_n}(s,t) \frac{1}{n^2}
\left[\sum_{i=1}^n \frac{w(m_i)}{m_i^2} \sum_{j_1\neq j_1'}^{m_i}
\sum_{l_1,l_1'=1}^{L_n}\frac{\mathbb{E}[C(T_{ij_1},T_{ij_1'})\widetilde
K_{s,l_1}(T_{ij_1'})\widetilde
K_{t,l_1'}(T_{ij_1'})]}{g(s)g(t)}\right]^2\nonumber\\
&& + \overline{W}_{\widetilde h_n}(s,t)  \frac{1}{n^2} \sum_{i_1\neq
i_2}^n \rho_{i_1i_2}^2 \left(
(C(s,s)+O(h^2))(C(t,t)+O(h^2))+(C(s,t)+O(h^2))^2\right)\nonumber\\
&=& - \overline{W}_{\widetilde h_n}(s,t) \frac{1}{n}
(C(s,t)+O(h^2))^2\nonumber\\
&& + \overline{W}_{\widetilde h_n}(s,t)  \frac{1}{n^2} \sum_{i_1\neq
i_2}^n \rho_{i_1i_2}^2 \left(
(C(s,s)+O(h^2))(C(t,t)+O(h^2)) + (C(s,t)+O(h^2))^2\right)\nonumber\\
&=& O\left(\frac{1}{n}\right) + \left(\frac{1}{n^2} \sum_{i_1\neq
i_2}^n \rho_{i_1i_2}^2\right)  O(1).
\end{eqnarray}
Combining (\ref{eq:C_tilde_var dominant}) with
(\ref{eq:E_sum_i1i2_j_different}) and (\ref{eq:E_sum_ii_j1j2_same})
- (\ref{eq:E_sum_ii_j1j2prime_same_j1primej2}), we obtain
(\ref{eq:var_C_tilde_offdiag}).

\vskip.1in\noindent{\bf Proof of Proposition \ref{propC.4} :} Write
\begin{equation*}
W_{\widetilde h_n}(s,t)\mbox{Var}(\widehat C_*(\frac{s+t}{2})) =
W_{\widetilde h_n}(s,t)\mathbb{E}(\widehat C_*(\frac{s+t}{2}))^2  -
(\mathbb{E}(W_{\widetilde h_n}(s,t)\widehat C_*(\frac{s+t}{2})))^2,
\end{equation*}
and observe that, by (\ref{eq:V_i_def}), (\ref{eq:F_i1i2_def}) and
(\ref{eq:F_sum_i1i2_j_different}), and following steps very similar
to those leading to (\ref{eq:C_tilde_var dominant}), we have
\begin{eqnarray}\label{eq:C_hat_star_var_dominant}
&& W_{\widetilde h_n}(s,t) \frac{1}{n^2}\sum_{i_1\neq i_2}^n
\frac{1}{m_{i_1}m_{i_2}} \sum_{j_1=1}^{m_{i_1}}
\sum_{j_2=1}^{m_{i_2}} \sum_{l_1=1}^{L_n}\sum_{l_2=1}^{L_n}
\frac{F_{i_1i_2;j_1,j_2}(s,t,s,t;l_1,l_2)}{g(\frac{s+t}{2})^2} -
(\mathbb{E}(W_{\widetilde h_n}(s,t)\widehat
C_*(\frac{s+t}{2})))^2\nonumber\\
&=&  - W_{\widetilde h_n}(s,t) \frac{1}{n} (C(s,t)+\sigma^2 +
O(h^2))^2 + W_{\widetilde h_n}(s,t) \left( \frac{1}{n^2}
\sum_{i_1\neq i_2}^n
\rho_{i_1i_2}^2 \right) (2(C(s,t))^2 + O(h))\nonumber\\
&=& O\left(\frac{1}{n}\right) + \left(\frac{1}{n^2} \sum_{i_1\neq
i_2}^n \rho_{i_1i_2}^2\right)  O(1).
\end{eqnarray}
Combining (\ref{eq:C_hat_star_var_dominant}) with the steps leading
to (\ref{eq:F_sum_i1i2_j_different}) and (\ref{eq:F_sum_ii_j_same}),
we obtain (\ref{eq:var_C_star_diag}).

\subsubsection*{Proofs of Lemmas \ref{lemmaD.1} - \ref{lemmaD.8}}

\noindent{\bf Proof of Lemma \ref{lemmaD.1} :} Since $\rho_{ii} =1$,
from expressions (\ref{eq:Y_four_terms_cond1}) and
(\ref{eq:Y_four_terms_cond4}) we can treat the terms corresponding
to $i_1=i_2=i$ and $i_1 \neq i_2$ in a unified way. From
(\ref{eq:fourth_uncond_moment}) and
(\ref{eq:fourth_uncond_factorize}), the expression (\ref{eq:C_uv_2})
and the calculations leading to (\ref{eq:expect_C_tilde}),
(\ref{eq:E_sum_i1i2_j_different}) follows.

\vskip.1in\noindent{\bf Proof of Lemma \ref{lemmaD.2} :} It follows
from (\ref{eq:Y_four_terms_cond5}), (\ref{eq:C_uu_2}) and
(\ref{eq:C_uv_uw_4}) (taking $s=s_1$, $s'=s_2$, $t=t_1$ and $t'=t_2$
in the latter).

\vskip.1in\noindent{\bf Proof of Lemma \ref{lemmaD.3} :} Follows by
arguments analogous to those for deriving
(\ref{eq:E_sum_ii_j1j2_same}).

\vskip.1in\noindent{\bf Proof of Lemma \ref{lemmaD.4} :}  Follows
from (\ref{eq:Y_four_terms_cond7}), (\ref{eq:C_uu_2}) and
(\ref{eq:C_uv_sq_4}).

\vskip.1in\noindent{\bf Proof of Lemma \ref{lemmaD.5} :} By
(\ref{eq:Y_four_terms_cond3}) and (\ref{eq:Y_four_terms_cond7}),
\begin{eqnarray}
&&
\mathbb{E}[Y_{i_1j_1}^2Y_{i_2j_2}^2|\mathbf{T}_{i_1},\mathbf{T}_{i_2}]
\nonumber\\
&=&
(C(T_{i_1j_1},T_{i_1j_1})+\sigma^2)(C(T_{i_2j_2},T_{i_2j_2})+\sigma^2)
+ 2\rho_{i_1i_2}^2(C(T_{i_1j_1},T_{i_2j_2}))^2.
\end{eqnarray}
The expression for
$\mathbb{E}[(C(T_{i_1j_1},T_{i_2j_2}))^2\widetilde
K_{z_1,l_1}(T_{i_1j_1})\widetilde K_{z_2,l_2}(T_{i_2j_2})]$ is given
by
\begin{equation*}
\int\int (C(u,v))^2 g(u)g(v)\widetilde K_{z_1,l_1}(u) \widetilde
K_{z_2,l_2}(v) du dv,
\end{equation*}
and it can be shown that when we sum over $l_1,l_2=1,\ldots,L_n$,
the sum equals $(C(z_1,z_2))^2g(z_1)g(z_2) + O(h^2)$. From this, and
the calculations leading to (\ref{eq:expect_C_star}), we have, for
$|s_k-t_k|\leq \frac{Ah}{2}$, $k=1,2$, with $A \geq 4(B_K+C_Q)$,
\begin{eqnarray*}
&& \frac{1}{n^2} \sum_{i=1}^n \frac{1}{m_i^2} \sum_{j_1\neq
j_2}^{m_i} \sum_{l_1=1}^{L_n}\sum_{l_2=1}^{L_n}
\frac{F_{ii;j_1,j_2}(s_1,t_1,s_2,t_2;l_1,l_2)}{g(z_1)g(z_2)}
\nonumber\\
&& + \frac{1}{n^2}\sum_{i_1\neq i_2}^n \frac{1}{m_{i_1}m_{i_2}}
\sum_{j_1=1}^{m_{i_1}} \sum_{j_2=1}^{m_{i_2}}
\sum_{l_1=1}^{L_n}\sum_{l_2=1}^{L_n}
\frac{F_{i_1i_2;j_1,j_2}(s_1,t_1,s_2,t_2;l_1,l_2)}{g(z_1)g(z_2)}
\nonumber\\
&& - \sigma^2[\mathbb{E}(\widehat C_*(z_1))+\mathbb{E}(\widehat
C_*(z_2))] + \sigma^4
\nonumber\\
&=& \left(\frac{1}{n}(1-\frac{1}{n}\sum_{i=1}^n \frac{1}{m_i}) +
\frac{n-1}{n}\right)(C(z_1,z_1)+\sigma^2 +
O(h^2))(C(z_2,z_2)+\sigma^2 + O(h^2))\nonumber\\
&& \hskip-.2in + \left(\frac{1}{n}(1-\frac{1}{n}\sum_{i=1}^n
\frac{1}{m_i}) + \frac{1}{n^2}\sum_{i_1\neq i_2}^n
\rho_{i_1i_2}^2\right)(2(C(z_1,z_2))^2 + O(h^2))\nonumber\\
&& - \sigma^2(C(z_1,z_1) + C(z_2,z_2)+ 2\sigma^2 + O(h^2)) +
\sigma^4  \nonumber\\
&=& \left(1-\frac{1}{n^2}\sum_{i=1}^n
\frac{1}{m_i}\right)(C(s_1,t_1)+\sigma^2 + O(h^2))(C(s_2,t_2)+\sigma^2+O(h^2))\nonumber\\
&& \hskip-.2in + \left(\frac{1}{n}(1-\frac{1}{n}\sum_{i=1}^n
\frac{1}{m_i}) + \frac{1}{n^2}\sum_{i_1\neq i_2}^n
\rho_{i_1i_2}^2\right)(C(s_1,s_2)C(t_1,t_2) + C(s_1,t_2)C(s_2,t_1) +
O(h)),\nonumber\\
&& - \sigma^2(C(s_1,t_1) + C(s_2,t_2)) - \sigma^4 +
O(h^2)\nonumber\\
&=& C(s_1,t_1)C(s_2,t_2)- \left(\frac{1}{n^2}\sum_{i=1}^n
\frac{1}{m_i}\right) (C(s_1,t_1)+\sigma^2)(C(s_2,t_2)+\sigma^2)  + O(h^2)\nonumber\\
&& \hskip-.2in + \left(\frac{1}{n}(1-\frac{1}{n}\sum_{i=1}^n
\frac{1}{m_i}) + \frac{1}{n^2}\sum_{i_1\neq i_2}^n
\rho_{i_1i_2}^2\right)(C(s_1,s_2)C(t_1,t_2) + C(s_1,t_2)C(s_2,t_1) +
O(h)).
\end{eqnarray*}

\vskip.1in\noindent{\bf Proof of Lemma \ref{lemmaD.6} :} Note that,
$\mathbb{E}(Y_{ij_1}^4|\mathbf{T}_i) =
3(C(T_{ij_1},T_{ij_1})+\sigma^2)^2$. Thus, from
(\ref{eq:C_uu_sq_2}), we have
\begin{eqnarray*}
&&
\sum_{l_1,l_2=1}^{L_n}\mathbb{E}[\mathbb{E}(Y_{ij_1}^4|\mathbf{T}_i)\widetilde
K_{z_1,l_1}(T_{ij_1})\widetilde
K_{z_2,l_2}(T_{ij_1})]\overline{Q}_h(z_1-s_{l_1})\overline{Q}_h(z_2-s_{l_2})
\\
&=&
\begin{cases}
O(h^{-1}) & ~\mbox{if}~|z_1-z_2| \leq \frac{Ah}{2} \\
0 & ~\mbox{otherwise}
\end{cases}
\end{eqnarray*}
uniformly in $s_1,t_1,s_2,t_2 \in [0,1]$ and $1\leq j_1\leq m_i$,
$1\leq i \leq n$. Therefore (\ref{eq:F_sum_ii_j_same}) follows.

\vskip.1in\noindent{\bf Proof of Lemma \ref{lemmaD.7} :} By
(\ref{eq:Y_four_terms_cond2}) and (\ref{eq:Y_four_terms_cond5}),
\begin{equation*}
\mathbb{E}[Y_{i_1j_1}Y_{i_1j_1'}Y_{i_2j_2}^2|
\mathbf{T}_{i_1},\mathbf{T}_{i_2}] =
C(T_{i_1j_1},T_{i_1j_1'})(C(T_{i_2j_2},T_{i_2j_2})+\sigma^2) +
2\rho_{i_1i_2}^2 C(T_{i_1j_1},T_{i_2j_2})C(T_{i_1j_1'},T_{i_2j_2}).
\end{equation*}
The expression for
$\mathbb{E}[C(T_{i_1j_1},T_{i_2j_2})C(T_{i_1j_1'},T_{i_2j_2})\widetilde
K_{s_1,l_1}(T_{i_1j_1})\widetilde
K_{t_1,l_1'}(T_{i_1j_1'})\widetilde K_{z_2,l_2}(T_{i_2j_2})]$ is
given by
\begin{equation*} 
\int\int\int C(u,w)C(v,w) g(u)g(v)g(w)\widetilde K_{s,l}(u)
\widetilde K_{t,l'}(v) \widetilde K_{z,m}(w) du dv dw,
\end{equation*}
and it can be shown that when we sum this over
$l_1,l_1',l_2=1,\ldots,L_n$, the sum equals
\begin{equation*}
C(s_1,z_2) C(t_1,z_2)g(s_1)g(t_1)g(z_2) + O(h^2).
\end{equation*}
From this, and similar arguments as before, we have, for $|s_1-t_1|
> \frac{Ah}{2}$, and $|s_2-t_2|\leq \frac{Ah}{2}$, with $A \geq
4(B_K+C_Q)$,
\begin{eqnarray*}
&& \frac{1}{n^2} \sum_{i=1}^n w(m_i)\frac{1}{m_i^3} \sum_{j_1\neq
j_1'\neq j_2}^{m_i} \sum_{l_1,l_1'=1}^{L_n}\sum_{l_2=1}^{L_n}
\frac{G_{ii;j_1,j_1',j_2}(s_1,t_1,s_2,t_2;l_1,l_1',l_2)}{g(s_1)g(t_1)g(z_2)}
\nonumber\\
&& + \frac{1}{n^2}\sum_{i_1\neq i_2}^n w(m_{i_1}) \frac{1}{m_{i_1}^2
m_{i_2}} \sum_{j_1\neq j_1'}^{m_{i_1}} \sum_{j_2=1}^{m_{i_2}}
\sum_{l_1,l_1'=1}^{L_n}\sum_{l_2=1}^{L_n}
\frac{G_{i_1i_2;j_1,j_1',j_2}(s_1,t_1,s_2,t_2;l_1,l_1',l_2)}{g(s_1)g(t_1)g(z_2)}
\nonumber\\
&& - \sigma^2 \mathbb{E}\widetilde C(s_1,t_1)
\nonumber\\
&=& \frac{1}{n}(1-\frac{1}{n}\sum_{i=1}^n\frac{2}{m_i}) (C(s_1,t_1)
+ O(h^2))(C(z_2,z_2) + \sigma^2 + O(h^2)) \nonumber\\
&& + \frac{1}{n^2} \sum_{i_1\neq i_2}^n
w(m_{i_1})\frac{1}{m_{i_1}^2} \sum_{j_1\neq j_1'}^{m_{i_1}}
\sum_{l_1,l_1'=1}^{L_n}
\frac{\mathbb{E}[C(T_{i_1j_1},T_{i_1j_1'})\widetilde
K_{s_1,l_1}(T_{i_1j_1})\widetilde
K_{t_1,l_1'}(T_{i_1j_1'})]}{g(s_1)g(t_1)}\nonumber\\
&& ~~~~~~\cdot~
\frac{1}{m_{i_2}}\sum_{j_2=1}^{m_{i_2}}\sum_{l_2=1}^{L_n}
\frac{\mathbb{E}[(C(T_{i_2j_2},T_{i_2j_2})+\sigma^2)\widetilde
K_{z_2,l_2}(T_{i_2j_2})]}{g(z_2)}
\nonumber\\
&& - \sigma^2 \frac{1}{n}\sum_{i=1}^{n}w(m_i)
\frac{1}{m_i^2}\sum_{j_1\neq j_1'}^{m_i}\sum_{l_1,l_1'=1}^{L_n}
\frac{\mathbb{E}[C(T_{i_1j_1},T_{i_1j_1'})\widetilde
K_{s_1,l_1}(T_{i_1j_1})\widetilde
K_{t_1,l_1'}(T_{i_1j_1'})]}{g(s_1)g(t_1)} \nonumber\\
&&  + \left(\frac{1}{n}(1-\frac{1}{n}\sum_{i=1}^n \frac{2}{m_i}) +
\frac{1}{n^2}\sum_{i_1\neq i_2}^n
\rho_{i_1i_2}^2\right)(2C(s_1,z_2)C(t_1,z_2) + O(h^2))\nonumber\\
&=& \left(\frac{1}{n}(1-\frac{1}{n}\sum_{i=1}^n \frac{2}{m_i}) +
\frac{n-1}{n}\right)(C(s_1,t_1) + O(h^2))(C(z_2,z_2) + \sigma^2 +
O(h^2))\nonumber\\
&& - \sigma^2 (C(s_1,t_1) + O(h^2)) \nonumber\\
&& + \left(\frac{1}{n}(1-\frac{1}{n}\sum_{i=1}^n \frac{2}{m_i}) +
\frac{1}{n^2}\sum_{i_1\neq i_2}^n
\rho_{i_1i_2}^2\right)(2C(s_1,z_2)C(t_1,z_2) + O(h^2))\nonumber\\
&=& (C(s_1,t_1) + O(h^2))(C(s_2,t_2) + O(h^2))\nonumber\\
&& - \left(\frac{1}{n^2}\sum_{i=1}^n \frac{2}{m_i}\right)(C(s_1,t_1)
+ O(h^2))(C(s_2,t_2) + \sigma^2 + O(h^2))\nonumber\\
&&  + \left(\frac{1}{n}(1-\frac{1}{n}\sum_{i=1}^n \frac{2}{m_i}) +
\frac{1}{n^2}\sum_{i_1\neq i_2}^n \rho_{i_1i_2}^2\right)
(C(s_1,s_2)C(t_1,t_2) + C(s_1,t_2)C(s_2,t_1) + O(h)).
\end{eqnarray*}
The last equality follows from the fact that the terms $C(s_1,t_1) +
O(h^2))$  appearing lines four, nine and ten are the same.

\vskip.1in\noindent{\bf Proof of Lemma \ref{lemmaD.8} :} Follows
from (\ref{eq:Y_four_terms_cond6}) and (\ref{eq:C_uu_uv_4}).

\subsubsection*{Proof of (\ref{eq:var_B_s_B_t})}

Define $W_{ij} = \psi_k(T_{ij})B_i(s,T_{ij})$ and $\bar W_{ij} =
\psi_{k'}(T_{ij})B_i(t,T_{ij})$. Since $|s-t| > Ah/2$, it follows
that for all $i$, $W_{ij}^k \bar W_{ij}^l = 0$, for all $k,l \geq
1$, for all $j=1,\ldots,m_i$. Thus, if $|s-t| > Ah/2$, then
\begin{eqnarray*}
m_i^4 \mbox{Var}(B_{1i,k}(s)B_{1i,k'}(t)) &=& \mathbb{E}[\sum_{j\neq
j'}^{m_i} (W_{ij} \bar W_{ij'} -
\mathbb{E} (W_{ij} \bar W_{ij'}))]^2\\
&=& \sum_{j\neq j'}^{m_i} [\mathbb{E}(W_{ij} \bar W_{ij'})^2 -
(\mathbb{E} (W_{ij} \bar W_{ij'}))^2]\\
&& + \sum_{j_1 = j_2'\neq j_1'=j_2}^{m_i} [\mathbb{E}(W_{ij_1} \bar
W_{ij_1'}W_{ij_1'}\bar W_{ij_1}) - \mathbb{E} (W_{ij_1} \bar
W_{ij_1'}) \mathbb{E} (W_{ij_1'} \bar W_{ij_1})]\\
&& + \sum_{j_1 = j_2 \neq j_1' \neq j_2'}^{m_i}
[\mathbb{E}(W_{ij_1}^2 \bar W_{ij_1'} \bar W_{ij_2'}) - \mathbb{E}
(W_{ij_1} \bar W_{ij_1'}) \mathbb{E} (W_{ij_1} \bar W_{ij_2'})]\\
&& + \sum_{j_1 = j_2' \neq j_1' \neq j_2}^{m_i} [\mathbb{E}(W_{ij_1}
\bar W_{ij_1'}W_{ij_2}\bar W_{ij_1}) - \mathbb{E} (W_{ij_1} \bar
W_{ij_1'}) \mathbb{E} (W_{ij_2} \bar W_{ij_1})]\\
&& + \sum_{j_1' = j_2 \neq j_1 \neq j_2'}^{m_i} [\mathbb{E}(W_{ij_1}
\bar W_{ij_1'}W_{ij_1'}\bar W_{ij_2'}) - \mathbb{E} (W_{ij_1} \bar
W_{ij_1'}) \mathbb{E} (W_{ij_1'} \bar W_{ij_2'})]\\
&& + \sum_{j_1' = j_2' \neq j_1 \neq j_2}^{m_i} [\mathbb{E}(W_{ij_1}
\bar W_{ij_1'}^2 W_{ij_2}) - \mathbb{E}(W_{ij_1} \bar W_{ij_1'})
\mathbb{E} (W_{ij_2} \bar W_{ij_1})],
\end{eqnarray*}
since the term corresponding to $j_1\neq j_1'\neq j_2 \neq j_2'$
vanishes.

Now by using the fact that $T_{ij}$'s are i.i.d., we can simplify
each sum on the RHS.
\begin{eqnarray*}
1\mbox{st term} &=& m_i(m_i-1) [\mathbb{E}(W_{i1}^2) \mathbb{E}(\bar
W_{i1}^2) - (\mathbb{E}(W_{i1}))^2(\mathbb{E}(\bar W_{i1}))^2] \\
2\mbox{nd term} &=& m_i(m_i-1)[0 - (\mathbb{E}(W_{i1}))^2
(\mathbb{E}(\bar W_{i1}))^2]\\
3\mbox{rd term} &=& m_i(m_i-1)(m_i-2)
[\mathbb{E}(W_{i1}^2)(\mathbb{E}(\bar W_{i1}))^2 -
(\mathbb{E}(W_{i1}))^2(\mathbb{E}(\bar W_{i1}))^2]\\
4\mbox{th term} &=& m_i(m_i-1)(m_i-2) [0 - (\mathbb{E}(W_{i1}))^2
(\mathbb{E}(\bar W_{i1}))^2]\\
5\mbox{th term} &=& m_i(m_i-1)(m_i-2) [0 - (\mathbb{E}(W_{i1}))^2
(\mathbb{E}(\bar W_{i1}))^2]\\
6\mbox{th term} &=& m_i(m_i-1)(m_i-2)
[(\mathbb{E}(W_{i1}))^2\mathbb{E}(\bar W_{i1}^2) -
(\mathbb{E}(W_{i1}))^2(\mathbb{E}(\bar W_{i1}))^2]
\end{eqnarray*}
Thus,
\begin{eqnarray*}
&& m_i^4 \mbox{Var}(B_{1i,k}(s)B_{1i,k'}(t))\nonumber\\
&=& m_i(m_i-1) [\mathbb{E}(W_{i1}^2) \mathbb{E}(\bar W_{i1}^2) +
(m_i-2) \mathbb{E}(W_{i1}^2)(\mathbb{E}(\bar W_{i1}))^2 + (m_i-2)
(\mathbb{E}(W_{i1}))^2\mathbb{E}(\bar W_{i1}^2)]\\
&& ~~~- m_i(m_i-1)(4m_i - 6)(\mathbb{E}(W_{i1}))^2 (\mathbb{E}(\bar
W_{i1}))^2
\end{eqnarray*}
Now, using the facts that $\mathbb{E}(W_{i1}^2) = O(h^{-1}) =
\mathbb{E}(\bar W_{i1}^2)$ and $|\mathbb{E}(W_{i1})| = O(1) =
|\mathbb{E}(W_{i1})|$, we conclude (\ref{eq:var_B_s_B_t}).


\subsubsection*{Computation of conditional mixed moments}

The computation of the moments is done by using the \textit{Wick
formula} (Lemma \ref{lemmaA.4}). We consider all the different
generic cases below:
\begin{itemize}
\item{\bf Case : $i_1 \neq i_2$, $j_1 \neq j_1'$, $j_2\neq j_2'$ :}
In this case,
\begin{eqnarray}\label{eq:Y_four_terms_cond1}
&& \mathbb{E}(Y_{i_1 j_1} Y_{i_1j_1'}Y_{i_2j_2}Y_{i_2j_2'}
|\mathbf{T}_{i_1},\mathbf{T}_{i_2}) \nonumber\\
&=& C(T_{i_1j_1},T_{i_1j_1'}) C(T_{i_2j_2},T_{i_2j_2'}) \nonumber\\
&& + \rho_{i_1i_2}^2
\left[C(T_{i_1j_1},T_{i_2j_2})C(T_{i_1j_1'},T_{i_2j_2'})  +
C(T_{i_1j_1},T_{i_2j_2'})C(T_{i_1j_1'},T_{i_2j_2'})\right].
\end{eqnarray}

\item
{\bf Case : $i_1 \neq i_2$, $j_1 = j_1'$, $j_2 \neq j_2'$
(equivalent to $i_1\neq i_2$, $j_1 \neq j_1'$, $j_2 = j_2'$):} In
this case,
\begin{eqnarray*}
\mathbb{E}(Y_{i_1 j_1} Y_{i_1j_1'}Y_{i_2j_2}Y_{i_2j_2'}
|\mathbf{T}_{i_1},\mathbf{T}_{i_2})  &=& \mathbb{E}(X_{i_1j_1}^2
X_{i_2j_2}X_{i_2j_2'} |\mathbf{T}_{i_1},\mathbf{T}_{i_2}) + \sigma^2
\mathbb{E}(X_{i_2j_2}X_{i_2j_2'}
|\mathbf{T}_{i_1},\mathbf{T}_{i_2}).
\end{eqnarray*}
Therefore, by (\ref{eq:Wick}),
\begin{equation*}
\mathbb{E}(X_{i_1j_1}^2 X_{i_2j_2}X_{i_2j_2'}
|\mathbf{T}_{i_1},\mathbf{T}_{i_2})
= C(T_{i_1j_1},T_{i_1j_1})C(T_{i_2j_2},T_{i_2j_2'}) +
2\rho_{i_1i_2}^2 C(T_{i_1j_1},T_{i_2j_2})C(T_{i_1j_1},T_{i_2j_2'}).
\end{equation*}
Combining, we have
\begin{eqnarray}\label{eq:Y_four_terms_cond2}
&\hskip-.2in& \mathbb{E}(Y_{i_1 j_1}
Y_{i_1j_1'}Y_{i_2j_2}Y_{i_2j_2'}
|\mathbf{T}_{i_1},\mathbf{T}_{i_2}) \nonumber\\
&\hskip-.5in=& \hskip-.2in
C(T_{i_1j_1},T_{i_1j_1})C(T_{i_2j_2},T_{i_2j_2'}) + 2\rho_{i_1i_2}^2
C(T_{i_1j_1},T_{i_2j_2})C(T_{i_1j_1},T_{i_2j_2'}) + \sigma^2
C(T_{i_2j_2},T_{i_2j_2'}).
\end{eqnarray}


\item
{\bf Case : $i_1 \neq i_2$, $j_1 = j_1'$, $j_2 = j_2'$ :} In this
case,
\begin{eqnarray}\label{eq:Y_four_terms_cond3}
&& \mathbb{E}(Y_{i_1 j_1} Y_{i_1j_1'}Y_{i_2j_2}Y_{i_2j_2'}
|\mathbf{T}_{i_1},\mathbf{T}_{i_2}) \nonumber\\
&=&
(C(T_{i_1j_1},T_{i_1j_1})+\sigma^2)(C(T_{i_2j_2},T_{i_2j_2})+\sigma^2)
+ 2\rho_{i_1i_2}^2 (C(T_{i_1j_1},T_{i_2j_2}))^2.
\end{eqnarray}



\item
{\bf Case :  $i_1 = i_2$, $j_1 \neq j_1' \neq j_2 \neq j_2'$ :} In
this case
\begin{eqnarray}\label{eq:Y_four_terms_cond4}
&& \mathbb{E}(Y_{i_1 j_1} Y_{i_1j_1'}Y_{i_2j_2}Y_{i_2j_2'}
|\mathbf{T}_{i_1},\mathbf{T}_{i_2}) \nonumber\\
&=& C(T_{i_1j_1},T_{i_1j_1'})C(T_{i_1j_2},T_{i_1j_2'}) +
C(T_{i_1j_1},T_{i_1j_2})C(T_{i_1j_1'},T_{i_1j_2'}) \nonumber\\
&& ~~~~~+ C(T_{i_1j_1},T_{i_1j_2'})C(T_{i_1j_1'},T_{i_1j_2}).
\end{eqnarray}

\item
{\bf Case : $i_1 = i_2$, $j_1 = j_1' \neq j_2 \neq j_2'$ (equivalent
to $i_1 = i_2$, $j_1 = j_2 \neq j_1' \neq j_2'$; $i_1 = i_2$, $j_1 =
j_2' \neq j_1' \neq j_2'$; $i_1 = i_2$, $j_1 \neq j_1' = j_2 \neq
j_2'$; $i_1 = i_2$, $j_1 \neq j_1'= j_2' \neq j_2$; and $i_1 = i_2$,
$j_1 \neq j_1' \neq j_2 = j_2'$):} In this case
\begin{eqnarray}\label{eq:Y_four_terms_cond5}
&&\mathbb{E}(Y_{i_1 j_1} Y_{i_1j_1'}Y_{i_2j_2}Y_{i_2j_2'}
|\mathbf{T}_{i_1},\mathbf{T}_{i_2}) \nonumber\\
&=&  (C(T_{i_1j_1},T_{i_1j_1})+\sigma^2)C(T_{i_1j_2},T_{i_1j_2'}) +
2C(T_{i_1j_1},T_{i_1j_2})C(T_{i_1j_1},T_{i_1j_2'}).
\end{eqnarray}

\item
{\bf Case : $i_1 = i_2$, $j_1 = j_1' = j_2 \neq j_2'$ (equivalent to
$i_1 = i_2$, $j_1 = j_1' = j_2' \neq j_2$; $i_1 = i_2$, $j_1 = j_2 =
j_2' \neq j_1'$; and $i_1 = i_2$, $j_1 \neq j_1' = j_2 = j_2'$):} In
this case
\begin{equation}\label{eq:Y_four_terms_cond6}
\mathbb{E}(Y_{i_1 j_1} Y_{i_1j_1'}Y_{i_2j_2}Y_{i_2j_2'}
|\mathbf{T}_{i_1},\mathbf{T}_{i_2})
= 3C(T_{i_1j_1},T_{i_1j_1})C(T_{i_1j_1},T_{i_1j_2'}) + 3\sigma^2
C(T_{i_1j_1},T_{i_1j_2'}).
\end{equation}

\item
{\bf Case : $i_1 = i_2$, $j_1 = j_1' \neq j_2 = j_2'$ (equivalent to
$i_1 = i_2$, $j_1 = j_2 \neq j_1' = j_2'$; and $i_1 = i_2$, $j_1 =
j_2' \neq j_1' = j_2$):} This this case
\begin{eqnarray}\label{eq:Y_four_terms_cond7}
&&\mathbb{E}(Y_{i_1 j_1} Y_{i_1j_1'}Y_{i_2j_2}Y_{i_2j_2'}
|\mathbf{T}_{i_1},\mathbf{T}_{i_2}) \nonumber\\
&=&
(C(T_{i_1j_1},T_{i_1j_1})+\sigma^2)(C(T_{i_1j_2},T_{i_1j_2})+\sigma^2)
+2(C(T_{i_1j_1},T_{i_1j_2}))^2 .
\end{eqnarray}

\item
{\bf Case : $i_1 = i_2$, $j_1 = j_1' = j_2 = j_2'$ :}  In this case
\begin{eqnarray}\label{eq:Y_four_terms_cond8}
\mathbb{E}(Y_{i_1 j_1} Y_{i_1j_1'}Y_{i_2j_2}Y_{i_2j_2'}
|\mathbf{T}_{i_1},\mathbf{T}_{i_2})
&=& 3(C(T_{i_1j_1},T_{i_1j_1})+\sigma^2)^2.
\end{eqnarray}

\end{itemize}

\subsubsection*{Computation of unconditional mixed moments
(off-diagonal part)}

Here, we obtain simplified forms the certain expectations that are
used in the proof of Propositions \ref{propC.3} and \ref{propC.4}.
Observe that, based on our calculations in
Appendix A, we only need to compute the expectations of the form
\begin{equation}\label{eq:fourth_uncond_moment}
\mathbb{E}[C(T_{i_1j_1},T_{i_1'j_1'})C(T_{i_2j_2},T_{i_2'j_2'})
\widetilde K_{s_1,l_1}(T_{i_1j_1}) \widetilde
K_{t_1,l_1'}(T_{i_1'j_1'}) \widetilde K_{s_2,l_2}(T_{i_2j_2})
\widetilde K_{t_2,l_2'}(T_{i_2'j_2'})].
\end{equation}
Notice that, when the pairs $(T_{i_1j_1},T_{i_1'j_1'})$ and
$(T_{i_2j_2}, T_{i_2'j_2'})$ are independent, the expectation in
(\ref{eq:fourth_uncond_moment}) factorizes as
\begin{equation}\label{eq:fourth_uncond_factorize}
\mathbb{E}[C(T_{i_1j_1},T_{i_1'j_1'})\widetilde
K_{s_1,l_1}(T_{i_1j_1}) \widetilde K_{t_1,l_1'}(T_{i_1'j_1'})] ~
\mathbb{E}[C(T_{i_2j_2},T_{i_2'j_2'})\widetilde
K_{s_2,l_2}(T_{i_2j_2}) \widetilde K_{t_2,l_2'}(T_{i_2'j_2'})].
\end{equation}
Each individual term is exactly of the same form that we encountered
while calculating the bias of our estimate. The expectations
appearing above are of the form
\begin{equation}\label{eq:C_uv_2}
\int\int C(u,v) g(u)g(v) \widetilde K_{s,l}(u) \widetilde
K_{s',l'}(v) du.
\end{equation}
For other terms we need to evaluate or approximate various other
integrals. The general forms of these integrals are given below, for
$1\leq l,l',m,m'\leq L_n$ and $s,s',t,t'\in [0,1]$.
\begin{eqnarray}
&& \int (C(u,u))^r g(u) \widetilde K_{s,l}(u) \widetilde
K_{s',l'}(u) du  \nonumber\\
&=& \begin{cases} O(h^{-1}) &~\mbox{if}~
\max\{|s-s_l|,|s'-s_{l'}|\}\leq 2B_K h\\
0 & ~\mbox{otherwise}
\end{cases}; ~~~~\mbox{for}~~~r=0,1,2;
\label{eq:C_uu_2}\\
&& \int C(u,u)g(u)\widetilde K_{s,l}(u) \widetilde K_{s',l'}(u)
\widetilde K_{t,m}(u) \widetilde K_{t',m'}(u) du ; \label{eq:C_uu_4}\\
&& \int (C(u,u))^2g(u)\widetilde K_{s,l}(u) \widetilde K_{s',l'}(u)
\widetilde K_{t,m}(u) \widetilde K_{t',m'}(u) du.
\label{eq:C_uu_sq_4}
\end{eqnarray}
\begin{eqnarray}
&& \int\int (C(u,v))^{r} g(u)g(v)\widetilde K_{s,l}(u) \widetilde
K_{s',l'}(u) \widetilde K_{t,m}(v) \widetilde K_{t',m'}(v)
dudv  \nonumber\\
&=&
\begin{cases} O(h^{-2}) & ~\mbox{if}~
\max\{|s-s_l|,|s'-s_{l'}|,|t-s_m|,|t'-s_{m'}|\} \leq 2B_K h\\
0 & ~\mbox{otherwise}.
\end{cases}; \nonumber\\
&& ~~~~~~~~~\mbox{for}~r=0,1,2 \label{eq:C_uv_sq_4}\\
&&\int\int (C(u,u))^rC(u,v) g(u)g(v)\widetilde K_{s,l}(u) \widetilde
K_{s',l'}(u) \widetilde K_{t,m}(u) \widetilde K_{t',m'}(v) dudv
\nonumber\\
&=&
\begin{cases} O(h^{-2}) &
~\mbox{if}~\max\{|s-s_l|,|s'-s_{l'}|,|t-m|\} \leq 2B_Kh\\
0 & ~\mbox{otherwise}
\end{cases};~~\mbox{for}~~r=0,1. \label{eq:C_uu_uv_4}
\end{eqnarray}
\begin{eqnarray}\label{eq:C_uv_uw_4}
&&\int\int\int C(u,v)C(u,w) g(u)g(v)g(w)\widetilde K_{s,l}(u)
\widetilde K_{s',l'}(u) \widetilde K_{t,m}(v) \widetilde
K_{t',m'}(w) dudvdw \nonumber\\
&=&
\begin{cases} O(h^{-1}) & ~\mbox{if}~
\max\{|s-s_l|,|s'-s_{l'}|\} \leq 2B_K h\\
0 & ~\mbox{otherwise}.
\end{cases}
\end{eqnarray}

\subsubsection*{Computation of unconditional mixed moments (diagonal and mixed part)}


We have the following bound:
\begin{eqnarray} \label{eq:C_uu_sq_2}
&& \int (C(u,u))^r g(u) \widetilde K_{z_1,l_1}(u) \widetilde
K_{z_2,l_2}(u)du \nonumber\\
&=& \begin{cases}
O(h^{-1}) & \mbox{if}~|z_k-s_{l_k}|\leq 2B_K h, ~k=1,2\\
0 & \mbox{otherwise}
\end{cases}
~~~\mbox{for}~~r=0,1,2.
\end{eqnarray}

\subsubsection*{Some error bounds involving Dirac-$\delta$}

Here, we provide some key estimates that are crucial to obtaining
the overall risk bound. They all involve the operator $H_\nu$. Due
to the decomposition (\ref{eq:H_nu_repr}) we can reduce the
computations of these bounds to integrals involving
$\{\psi_k(\cdot)\}_{k=1}^M$ and $\delta(\cdot,\cdot)$. Throughout we
assume that $R(s_1,s_2,t_1,t_2)$ is a ``nice'' function satisfying
certain (boundedness) conditions. Then,
\begin{eqnarray}\label{eq:delta_delta_psi}
&&|\int\int\int\int
\delta(x,s_1)\delta(x,s_2)R(s_1,s_2,t_1,t_2)\psi_\nu(t_1)\psi_\nu(t_2)ds_1ds_2dt_1dt_2|
\nonumber\\
&=& |\int\int R(x,x,t_1,t_2)\psi_\nu(t_1)\psi_\nu(t_2)dt_1dt_2|
~\leq~ \parallel R\parallel_\infty \parallel
\psi_\nu\parallel_\infty^2.
\end{eqnarray}
\begin{eqnarray}\label{eq:delta_delta_W_psi}
&& | \int\int\int\int \delta(x,s_1)\delta(x,s_2)W_{Ah}(s_1,t_1)
R(s_1,s_2,t_1,t_2)\psi_\nu(t_1)\psi_\nu(t_2)ds_1ds_2dt_1dt_2|
\nonumber\\
&=& |\int\int\int\int \delta(x,s_1)\delta(x,s_2)
\int_{(s_1-\frac{Ah}{2})\vee 0}^{(s_1+\frac{Ah}{2})\wedge 1}
R(s_1,s_2,t_1,t_2)\psi_\nu(t_1)\psi_\nu(t_2)dt_1dt_2ds_1ds_2| \nonumber\\
&=& |\int \int_{(x-\frac{Ah}{2})\vee 0}^{(x+\frac{Ah}{2})\wedge 1}
R(x,x,t_1,t_2)\psi_\nu(t_1)\psi_\nu(t_2) dt_1dt_2| ~\leq~ Ah
\parallel R\parallel_\infty \parallel \psi_\nu\parallel_\infty^2.
\end{eqnarray}
\begin{eqnarray}\label{eq:delta_delta_WW_psi}
&& |\int\int\int\int
\delta(x,s_1)\delta(x,s_2)W_{Ah}(s_1,t_1)W_{Ah}(s_2,t_2)
R(s_1,s_2,t_1,t_2)
\psi_\nu(t_1)\psi_\nu(t_2)ds_1ds_2dt_1dt_2 |\nonumber\\
&=& |\int\int\int\int \delta(x,s_1)\delta(x,s_2)
\int_{(s_2-\frac{Ah}{2})\vee 0}^{(s_2+\frac{Ah}{2})\wedge 1}
\int_{(s_1-\frac{Ah}{2})\vee 0}^{(s_1+\frac{Ah}{2})\wedge 1}
R(s_1,s_2,t_1,t_2)\psi_\nu(t_1)\psi_\nu(t_2)dt_1dt_2ds_1ds_2| \nonumber\\
&=& |\int_{(x-\frac{Ah}{2})\vee 0}^{(x+\frac{Ah}{2})\wedge 1}
\int_{(x-\frac{Ah}{2})\vee 0}^{(x+\frac{Ah}{2})\wedge 1}
R(x,x,t_1,t_2)\psi_\nu(t_1)\psi_\nu(t_2) dt_1dt_2| \nonumber\\
&\leq& (Ah)^2
\parallel R\parallel_\infty \parallel \psi_\nu\parallel_\infty^2.
\end{eqnarray}
\begin{eqnarray}\label{eq:delta_delta_Wt_psi}
&& |\int\int\int\int \delta(x,s_1)\delta(x,s_2)W_{Ah}(t_1,t_2)
R(s_1,s_2,t_1,t_2)
\psi_\nu(t_1)\psi_\nu(t_2)ds_1ds_2dt_1dt_2|\nonumber\\
&=& |\int\int\int \delta(x,s_1)\delta(x,s_2)
\int_{(t_2-\frac{Ah}{2})\vee 0}^{(t_2+\frac{Ah}{2})\wedge 1}
R(s_1,s_2,t_1,t_2)\psi_\nu(t_1)\psi_\nu(t_2)dt_1dt_2 ds_1ds_2| \nonumber\\
&=& |\int \int_{(t_2-\frac{Ah}{2})\vee 0}^{(t_2+\frac{Ah}{2})\wedge
1} R(x,x,t_1,t_2)\psi_\nu(t_1)\psi_\nu(t_2) dt_1dt_2|
\nonumber\\
&\leq& Ah
\parallel R\parallel_\infty \parallel \psi_\nu\parallel_\infty^2.
\end{eqnarray}
\begin{eqnarray}\label{eq:delta_delta_WtW_psi}
&&|\int\int\int\int
\delta(x,s_1)\delta(x,s_2)W_{Ah}(t_1,t_2)W_{Ah}(s_2,t_2)
R(s_1,s_2,t_1,t_2)
\psi_\nu(t_1)\psi_\nu(t_2)ds_1ds_2dt_1dt_2| \nonumber\\
&=& |\int\int \delta(x,s_1)\delta(x,s_2)
\int_{(s_2-\frac{Ah}{2})\vee 0}^{(s_2+\frac{Ah}{2})\wedge 1}
\int_{(t_2-\frac{Ah}{2})\vee 0}^{(t_2+\frac{Ah}{2})\wedge 1}
R(s_1,s_2,t_1,t_2)\psi_\nu(t_1)\psi_\nu(t_2)dt_1dt_2 ds_1ds_2| \nonumber\\
&=& |\int_{(x-\frac{Ah}{2})\vee 0}^{(x+\frac{Ah}{2})\wedge 1}
\int_{(t_2-\frac{Ah}{2})\vee 0}^{(t_2+\frac{Ah}{2})\wedge 1}
R(x,x,t_1,t_2)\psi_\nu(t_1)\psi_\nu(t_2) dt_1dt_2| \nonumber\\
&\leq& (Ah)^2
\parallel R\parallel_\infty \parallel \psi_\nu\parallel_\infty^2.
\end{eqnarray}
\begin{eqnarray}\label{eq:delta_delta_WtsW_psi}
&&|\int\int\int\int
\delta(x,s_1)\delta(x,s_2)W_{Ah}(t_1,s_2)W_{Ah}(s_2,t_2)
R(s_1,s_2,t_1,t_2)
\psi_\nu(t_1)\psi_\nu(t_2)ds_1ds_2dt_1dt_2| \nonumber\\
&=& |\int\int \delta(x,s_1)\delta(x,s_2)
\int_{(s_2-\frac{Ah}{2})\vee 0}^{(s_2+\frac{Ah}{2})\wedge 1}
\int_{(s_2-\frac{Ah}{2})\vee 0}^{(s_2+\frac{Ah}{2})\wedge 1}
R(s_1,s_2,t_1,t_2)\psi_\nu(t_1)\psi_\nu(t_2)dt_1dt_2 ds_1ds_2| \nonumber\\
&=& |\int_{(x-\frac{Ah}{2})\vee 0}^{(x+\frac{Ah}{2})\wedge 1}
\int_{(x-\frac{Ah}{2})\vee 0}^{(x+\frac{Ah}{2})\wedge 1}
R(x,x,t_1,t_2)\psi_\nu(t_1)\psi_\nu(t_2) dt_1dt_2| \nonumber\\
&\leq& (Ah)^2
\parallel R\parallel_\infty \parallel \psi_\nu\parallel_\infty^2.
\end{eqnarray}

\subsection*{Appendix G : Proof of Theorem \ref{thm3}}

In order to prove this result, we use a strategy very similar to the
one used in the proof of Corollary 1 in Paul and Peng (2007). 
In view of the statement of the theorem, it suffices
to consider a submodel consisting of kernels $\overline{\Sigma}$ of
rank 1. Let
\begin{equation*}
\Sigma^{(0)}(s,t) = \overline{\lambda}
\overline{\psi}(s)\overline{\psi}(t), ~~~s,t \in [0,1]
\end{equation*}
for $\overline{\lambda} \geq C_1$, where
$\overline{\psi}(\cdot) \equiv 1$. Then $\overline{\psi}$ is the
first (and only) eigenfunction of $\overline{\Sigma}^{(0)}$ with
corresponding eigenvalue $\overline{\lambda}$. Let us suppose that
the design $D$ satisfies $\underline{m} = \overline{m} = m \geq 4$.
Finally, choose $g$ to be the uniform density on $[0,1]$. Let $M_*
\sim (nm)^{1/5}$, and let $\{\gamma_l\}_{l=1}^{M_*}$ be orthonormal
functions such that (i) $\gamma_l$'s are twice continuously
differentiable, and $\max_l
\parallel\gamma_l^{(j)}\parallel_\infty = O(M_*^{1/2+j})$, for
$j=0,1,2$; (ii) $\int_0^1\gamma_l(s) ds = 0$ for all $l$, and (iii)
$\gamma_l$ is centered around $l/M_*$ with length of support
$O(M_*^{-1})$ uniformly over $l$. Note that, condition (iii) implies
that $\{\gamma_l\}$ are orthogonal to $\overline{\psi}$. Let $M_0 =
[\frac{2M_*}{9}]$. Let ${\cal F}_0$ be an index set satisfying $\log
|{\cal F}_0| \asymp M_*$, and $\{z_l^{(j)}:l=1,\ldots,M_*\}_{j \in
{\cal F}_0}$ be a collection with $z_l^{(j)}$ taking values in
$\{-M_0^{-1/2},0,M_0^{-1/2}\}$, such that with $\mathbf{z}^{(j)}$
denoting the vector $(z_l^{(j)})_{l=1}^{M_*}$, we have $\parallel
\mathbf{z}^{(j)}
\parallel_2 = 1$ and $\parallel \mathbf{z}^{(j)} -
\mathbf{z}^{(j')}\parallel_2 \geq 1$ for $j \neq j' \in {\cal F}_0$.
The construction is by a ``sphere packing'' argument as in Paul and
Johnstone (2007). 
Let $\delta \asymp (nm)^{-2/5} \asymp M_*^{-2}$ Then, define
\begin{equation*}
\psi^{(j)}(s) = \sqrt{1-\delta^2} \overline{\psi}(s) + \delta
\sum_{l=1}^{M_*} z_l^{(j)} \gamma_l(s),~~~j \in {\cal F}_0.
\end{equation*}
Note that by construction, (i') $\parallel \psi^{(j)}\parallel_2 =
1$; (ii') $\psi^{(j)}$ are twice differentiable, with second
derivative bounded; (iii') $\parallel \psi^{(j)}  -
\psi^{(j')}\parallel_2 \geq \delta$ for $j\neq j' \in {\cal F}_0$;
(iv') $\parallel \overline{\psi} - \psi^{(j)} \parallel_\infty =
O(\delta)$ uniformly over $j \in {\cal F}_0$. Property (iv') will be
crucial for much of our analysis later on.

In order to prove Theorem \ref{thm3}, we need to show the following:
\begin{equation}\label{eq:KL_rate}
\sum_{i=1}^n
\mathbb{E}K(\overline{\Sigma}_i^{(j)},\overline{\Sigma}_i^{(0)})
~\asymp~ nm\delta^2, ~~~\mbox{uniformly in}~~j \in {\cal F}_0,
\end{equation}
where $\overline{\Sigma}_i^{(j)}$ denotes the covariance of the
observation $i$ given $\{T_{il}\}_{l=1}^{m}$ under the model
parameterized by $\overline{\Sigma}_0^{(j)}$, and $\mathbb{E}$
denotes expectation with respect to the design points $\mathbf{T}$.

\subsubsection*{Proof of (\ref{eq:KL_rate})}

From now onwards, we shall fix $j \in {\cal F}_0$, and drop the
superscript $(j)$ for convenience. Denote the $m\times 1$ vectors
$(\overline{\psi}(T_{ij})_{j=1}^m$ and $(\psi(T_{ij})_{j=1}^m$ by
$\overline{\bs{\psi}}_i$ and $\bs{\psi}_i$, respectively. Of course,
$\overline{\bs{\psi}}_i$ is the nonrandom vector with all the
entries equal to 1. Next, observe that,
\begin{eqnarray}\label{eq:Sigma_diff_bound_1}
\parallel \overline{\Sigma}_i^{(0)} - \overline{\Sigma}_i
\parallel_F &=& \overline{\lambda} \parallel \overline{\bs{\psi}}_i
(\overline{\bs{\psi}}_i - \bs{\psi}_i)^T + (\overline{\bs{\psi}}_i -
\bs{\psi}_i)\bs{\psi}_i^T \parallel_F \nonumber\\
&\leq& \overline{\lambda} (\parallel
\overline{\bs{\psi}}_i\parallel_2 + \parallel
\bs{\psi}_i\parallel_2)
\parallel \overline{\bs{\psi}}_i -
\bs{\psi}_i \parallel_2.
\end{eqnarray}
Since $\parallel \overline{\bs{\psi}}_i - \bs{\psi}_i \parallel_2
\leq \sqrt{m} \parallel \overline{\psi} - \psi \parallel_\infty =
O(\sqrt{m} \delta)$ (by property (iv')), and $\parallel
\overline{\bs{\psi}}_i \parallel_2 = \sqrt{m}$, from
(\ref{eq:Sigma_diff_bound_1}) it follows that,
\begin{equation}\label{eq:Sigma_diff_bound_2}
\max_{1\leq i \leq n} \parallel \overline{\Sigma}_i^{(0)} -
\overline{\Sigma}_i \parallel_F^2 = O(m^2 \delta^2).
\end{equation}
Since $m \delta \asymp m (nm)^{-2/5}$ and $m = o(n^{2/3})$, the RHS
of (\ref{eq:Sigma_diff_bound_2}) is $o(1)$ (\textit{a nonrandom
bound}) uniformly over ${\cal F}_0$, and hence, using arguments as
in the proof of Proposition 2 in Paul and Peng (2007), 
we have
\begin{equation*}
\sum_{i=1}^n K(\overline{\Sigma}_i,\overline{\Sigma}_i^{(0)})
~\asymp~ \sum_{i=1}^n \parallel (\overline{\Sigma}_i^{(0)})^{-1/2}
(\overline{\Sigma}_i^{(0)} - \overline{\Sigma}_i)
(\overline{\Sigma}_i^{(0)})^{-1/2}\parallel_F^2, ~~\mbox{uniformly
over}~~{\cal F}_0.
\end{equation*}
Thus, (\ref{eq:KL_rate}) will follow once we prove:

\begin{prop}\label{propE.1}
Uniformly over ${\cal F}_0$,
\begin{equation}\label{eq:mean_diff_sigma}
\mathbb{E}\parallel (\overline{\Sigma}_1^{(0)})^{-1/2}
(\overline{\Sigma}_1^{(0)} - \overline{\Sigma}_1)
(\overline{\Sigma}_1^{(0)})^{-1/2}\parallel_F^2 ~\asymp~ m\delta^2.
\end{equation}
\end{prop}

\vskip.1in\noindent{\bf Proof of Proposition \ref{propE.1} :} First,
note that, $\overline{\theta} = \overline{\bs{\psi}}_1/\sqrt{m}$ is
a vector of $l_2$ norm 1, and hence, by using a standard matrix
inversion formula,
\begin{equation*}
(\overline{\Sigma}_1^{(0)})^{-1} = (I + \overline{\lambda} m
\overline{\theta}\overline{\theta}^T)^{-1} = I - \kappa
\overline{\theta}\overline{\theta}^T,~~~\mbox{where}~~\kappa =
\frac{\overline{\lambda} m}{1 + \overline{\lambda} m}~.
\end{equation*}
Let $\Delta = \overline{\Sigma}_1 - \overline{\Sigma}_1^{(0)} =
\overline{\lambda}( \bs{\psi}_1\bs{\psi}_1^T - m
\overline{\theta}\overline{\theta}^T)$. Then,
\begin{eqnarray}\label{eq:diff_norm_expand}
&& \parallel (\overline{\Sigma}_1^{(0)})^{-1/2}
(\overline{\Sigma}_1^{(0)} - \overline{\Sigma}_1)
(\overline{\Sigma}_1^{(0)})^{-1/2}\parallel_F^2 \nonumber\\
&=& \tr[(I - \kappa \overline{\theta}\overline{\theta}^T) \Delta (I
- \kappa \overline{\theta}\overline{\theta}^T) \Delta]\nonumber\\
&=& \tr[(I - \overline{\theta}\overline{\theta}^T)\Delta (I -
\overline{\theta}\overline{\theta}^T)\Delta] + 2(1-\kappa)
\overline{\theta}^T \Delta (I -
\overline{\theta}\overline{\theta}^T)\Delta \overline{\theta} +
(1-\kappa)^2 (\overline{\theta}^T \Delta
\overline{\theta})^2 \nonumber\\
&=& \overline{\lambda}^2 \left[\parallel (I -
\overline{\theta}\overline{\theta}^T)\bs{\psi}_1 \parallel_2^4 + 2
(1-\kappa) (\overline{\theta}^T \bs{\psi}_1)^2
\parallel (I - \overline{\theta}\overline{\theta}^T)\bs{\psi}_1
\parallel_2^2 + (1-\kappa)^2 (m - (\overline{\theta}^T
\bs{\psi}_1)^2)^2\right]\nonumber\\
&=& \overline{\lambda}^2 \left[(\parallel \bs{\psi}_1 \parallel_2^2
- (\overline{\theta}^T \bs{\psi}_1)^2)^2 +  2 (1-\kappa)
(\overline{\theta}^T \bs{\psi}_1)^2 (\parallel \bs{\psi}_1
\parallel_2^2 - (\overline{\theta}^T \bs{\psi}_1)^2) +
(1-\kappa)^2 (m - (\overline{\theta}^T \bs{\psi}_1)^2)^2\right]
\nonumber\\
&&
\end{eqnarray}
where the third and last steps follow from the fact that $(I -
\overline{\theta}\overline{\theta}^T) \overline{\theta} = 0$ and $(I
- \overline{\theta}\overline{\theta}^T)^2 = I -
\overline{\theta}\overline{\theta}^T$. From
(\ref{eq:diff_norm_expand}), the proof will follow once we establish
the following results.

\begin{lemma}\label{lemmaE.1}
With $T_{ij}$ i.i.d. from Uniform$[0,1]$, we have (uniformly over
${\cal F}_0$)
\begin{eqnarray*}
\mathbb{E}[m \parallel \bs{\psi}_1 \parallel_2^2 -
(\overline{\bs{\psi}}_1^T \bs{\psi}_1)^2] &=& m(m-1) \delta^2
(1+o(1)), \\
~~\mbox{and}~~\mbox{Var}[m
\parallel \bs{\psi}_1 \parallel_2^2 - (\overline{\bs{\psi}}_1^T
\bs{\psi}_1)^2] &=& O(m^3 \delta^4).
\end{eqnarray*}
\end{lemma}

\begin{lemma}\label{lemmaE.2}
With $T_{ij}$ i.i.d. from Uniform$[0,1]$, we have (uniformly over
${\cal F}_0$)
\begin{eqnarray*}
\mathbb{E}[m - \overline{\bs{\psi}}_1^T \bs{\psi}_1]^2 &=& m
\delta^2(1+o(1)),\\
\mathbb{E} \parallel \overline{\bs{\psi}}_1 -
\bs{\psi}_1\parallel_2^4 &=& O(m^2 \delta^4).
\end{eqnarray*}
\end{lemma}

\begin{lemma}\label{lemmaE.3}
With $T_{ij}$ i.i.d. from Uniform$[0,1]$, we have (uniformly over
${\cal F}_0$)
\begin{equation*}
\mathbb{E}[\parallel \bs{\psi}_1\parallel_2^2 (m \parallel
\bs{\psi}_1\parallel_2^2 - (\overline{\bs{\psi}}^T \bs{\psi}_1)^2)]
= m^2(m-1) \delta^2 (1+o(1)).
\end{equation*}
\end{lemma}

To see how (\ref{eq:mean_diff_sigma}) follows from Lemmas
\ref{lemmaE.1} - \ref{lemmaE.3}, note first that,
\begin{eqnarray}\label{eq:mean_diff_sigma_main}
&&\mathbb{E} (\overline{\bs{\psi}}_1^T \bs{\psi}_1)^2  (m \parallel
\bs{\psi}_1 \parallel_2^2 - (\overline{\bs{\psi}}_1^T
\bs{\psi}_1)^2) \nonumber\\
&=& m \mathbb{E}[ \parallel \bs{\psi}_1 \parallel_2^2 (m \parallel
\bs{\psi}_1 \parallel_2^2 -
(\overline{\bs{\psi}}_1^T\bs{\psi}_1)^2)] - \mathbb{E}(m \parallel
\bs{\psi}_1 \parallel_2^2 -
(\overline{\bs{\psi}}_1^T\bs{\psi}_1)^2)^2\nonumber\\
&=& m^3(m-1) \delta^2 (1+o(1)) - O(m^4 \delta^4) = m^3(m-1)\delta^2
(1+o(1))
\end{eqnarray}
by Lemmas \ref{lemmaE.1} and \ref{lemmaE.3}. Now, from
(\ref{eq:diff_norm_expand}) we obtain
\begin{eqnarray*}
\mathbb{E}\parallel (\overline{\Sigma}_1^{(0)})^{-1/2}
(\overline{\Sigma}_1^{(0)} - \overline{\Sigma}_1)
(\overline{\Sigma}_1^{(0)})^{-1/2}\parallel_F^2  &\geq&
2\overline{\lambda}^2 (1-\kappa)\frac{1}{m^2} \mathbb{E}
(\overline{\bs{\psi}}_1^T \bs{\psi}_1)^2  (m \parallel \bs{\psi}_1
\parallel_2^2 -
(\overline{\bs{\psi}}_1^T \bs{\psi}_1)^2)\nonumber\\
&=& \frac{2\overline{\lambda}^2 m(m-1)}{1+\overline{\lambda} m}
\delta^2 (1+o(1)),
\end{eqnarray*}
where the last step is by (\ref{eq:mean_diff_sigma_main}). This
establishes the lower bound in (\ref{eq:mean_diff_sigma}).

To establish the upper bound in (\ref{eq:mean_diff_sigma}), we also
need to consider the expectations of the other two terms on the RHS
of (\ref{eq:diff_norm_expand}). First, by Lemma \ref{lemmaE.1},
\begin{equation}\label{eq:mean_diff_sigma_first}
\overline{\lambda}^2 \mathbb{E}(\parallel \bs{\psi}_1
\parallel_2^2 - (\overline{\theta}^T \bs{\psi}_1)^2)^2
=  \frac{\overline{\lambda}^2}{m^2} \mathbb{E}(m \parallel
\bs{\psi}_1\parallel_2^2 - (\overline{\bs{\psi}}_1^T
\bs{\psi}_1)^2)^2 = O(m^2 \delta^4).
\end{equation}
Next, writing
\begin{equation*}
m  - (\overline{\theta}^T \bs{\psi}_1)^2 = m - \parallel \bs{\psi}_1
\parallel_2^2 + \frac{1}{m} [m \parallel \bs{\psi}_1
\parallel_2^2 - (\overline{\bs{\psi}}_1^T
\bs{\psi}_1)^2],
\end{equation*}
and then using the fact that for any $\epsilon
> 0$, and $a,b \in \mathbb{R}$, $(a+b)^2 \leq (1+\epsilon) a^2 + (1+\epsilon^{-1})b^2$, we
have, for arbitrary but fixed $\epsilon > 0$,
\begin{eqnarray}\label{eq:mean_diff_sigma_third}
&& \mathbb{E}(m  - (\overline{\theta}^T \bs{\psi}_1)^2)^2 \nonumber\\
&\leq& (1+\epsilon) \mathbb{E}[m -
\parallel \bs{\psi}_1\parallel_2^2]^2 + \frac{(1+\epsilon^{-1})}{m^2}\mathbb{E}[m
\parallel \bs{\psi}_1\parallel_2^2 - (\overline{\bs{\psi}}_1^T
\bs{\psi}_1)^2]^2 \nonumber\\
&=& (1+\epsilon) \mathbb{E}[\parallel \overline{\bs{\psi}}_1 -
\bs{\psi}_1
\parallel_2^2 - 2(m - \overline{\bs{\psi}}_1^T \bs{\psi}_1)]^2 +
O(m^2 \delta^4)~~(\mbox{by Lemma \ref{lemmaA.1}})\nonumber\\
&\leq& 4(1+\epsilon)^2 \mathbb{E}[m - \overline{\bs{\psi}}_1^T
\bs{\psi}_1]^2 + (1+\epsilon)(1+\epsilon^{-1}) \mathbb{E}\parallel
\overline{\bs{\psi}}_1 - \bs{\psi}_1\parallel_2^4 + O(m^2 \delta^4)
\nonumber\\
&=& 4(1+\epsilon)^2 m\delta^2 (1+o(1)) + O(m^2\delta^4),
\end{eqnarray}
where the last step follows from Lemma \ref{lemmaE.2}. Finally,
substituting (\ref{eq:mean_diff_sigma_first}),
(\ref{eq:mean_diff_sigma_main}) and (\ref{eq:mean_diff_sigma_third})
in (\ref{eq:diff_norm_expand}) we obtain an upper bound of the form
\begin{equation*}
\frac{2\overline{\lambda}^2 m(m-1)}{1+\overline{\lambda} m}
\delta^2(1+o(1)) + (1+\epsilon)^2 \frac{4\overline{\lambda}^2
m}{(1+\overline{\lambda} m)^2} \delta^2 (1+o(1)) + O(m^2 \delta^4) =
O(m\delta^2),
\end{equation*}
which concludes the proof.

\subsubsection*{Proof of Lemmas \ref{lemmaE.1} - \ref{lemmaE.3}}

In order to prove the lemmas, we define $\xi = \overline{\psi} -
\psi$ and notice the very important set of relations :
\begin{equation}\label{eq:xi_int_order}
\int \xi = \int (\overline{\psi} - \psi) = \int (\overline{\psi} -
\psi) \overline{\psi} = 1 - \int \overline{\psi}\psi =
\frac{1}{2}\int |\overline{\psi} - \psi|^2 = \frac{1}{2}\int \xi^2 =
\frac{1}{2} \delta^2 + O(\delta^4).
\end{equation}

\vskip.1in\noindent{\bf Proof of Lemma \ref{lemmaE.1} :} The
decomposition
\begin{equation}\label{eq:diff_psi_basic}
m \parallel \bs{\psi}_1\parallel_2^2 - (\overline{\bs{\psi}}_1^T
\bs{\psi}_1)^2 = (m-1) \sum_{k=1}^m \psi^2(T_{1k}) - \sum_{k \neq
k'}^m \psi(T_{1k}) \psi(T_{1k'})
\end{equation}
yields
\begin{eqnarray*}
&& \mathbb{E}[m \parallel \bs{\psi}_1\parallel_2^2 -
(\overline{\bs{\psi}}_1^T \bs{\psi}_1)^2] \\
&=& (m-1) m \int \psi^2 -
m(m-1) (\int \psi)^2 = m(m-1) [1 - (1 - \int \xi)^2]\\
&=& m(m-1) [2 \int \xi - (\int \xi)^2] = m(m-1)\delta^2
(1+O(\delta^2)),~~~~~(\mbox{by}~(\ref{eq:xi_int_order})).
\end{eqnarray*}
Define $\tau = \int \psi$. Then, using (\ref{eq:diff_psi_basic}),
\begin{eqnarray*}
&&\mbox{Var}[m \parallel \bs{\psi}_1\parallel_2^2 -
(\overline{\bs{\psi}}_1^T \bs{\psi}_1)^2] \\
&=& (m-1)^2\mathbb{E}[ \sum_{k=1}^m (\psi^2(T_{1k}) - 1)]^2 \\
&& ~~~ + \sum_{k_1 \neq k_1'} \sum_{k_2 \neq k_2'}
\mathbb{E}[(\psi(T_{1k_1})\psi(T_{1k_1'})-
\tau^2)(\psi(T_{1k_2})\psi(T_{1k_2'})- \tau^2)]\\
&& - 2(m-1)\sum_{k_1=1}^m \sum_{k_2 \neq
k_2'}\mathbb{E}[(\psi^2(T_{1k_1}) -
1)(\psi(T_{1k_2})\psi(T_{1k_2'})- \tau^2)] \\
&=& (m-1)^2 m \mathbb{E}(\psi^2(T_{11}) - 1)^2 + 2m(m-1)
\mathbb{E}(\psi(T_{11})\psi(T_{12})- \tau^2)^2 \\
&& + 4m(m-1)(m-2) \mathbb{E}[(\psi(T_{11})\psi(T_{12})- \tau^2)
(\psi(T_{11})\psi(T_{13})- \tau^2)] \\
&& - 4m(m-1)^2 \mathbb{E}[(\psi^2(T_{11}) -
1)(\psi(T_{11})\psi(T_{12})- \tau^2)]\\
&=& m(m-1)\left[(m-1) (\int \psi^4 - 1) + 2(\int \psi^2 \int \psi^2
- \tau^4) \right. \\
&& ~~ \left. + 4(m-2)(\int \psi^2\int \psi \int \psi -
\tau^4)\right] -4m(m-1)^2(\int \psi^3 \int \psi -
\tau^2)\\
&=& m(m-1)\left[(m-1)(\int (1-\xi)^4 -1) + 2(1-\tau^4) \right. \\
&& ~~~ \left.+ 4(m-2)\tau^2 (1 - \tau^2) - 4(m-1)\tau(\int (1-\xi)^3
- \tau)\right].
\end{eqnarray*}
Simplifying this expression, and using (\ref{eq:xi_int_order}),
first term within square bracket is $(m-1) (4\int \xi^2 - 4 \int
\xi^3 + \int \xi^4)$, and the last term within square bracket is
$-4(m-1)\tau(2\int \xi^2 -\int \xi^3)$. Collecting terms and using
the fact that $1-\tau^2 = 2(1-\tau) - (1-\tau)^2 = \xi^2 -
(\xi^2)^4$ (again by (\ref{eq:xi_int_order})), we can express the
sum as
\begin{eqnarray*}
&& m(m-1)\left[(4(m-1) + 4 + 4(m-2)-8(m-1))\int \xi^2 \right. \\
&& ~~~~~~~~~~~\left. - (4(m-1) -
4(m-1))\int \xi^3 + (m-1)\int \xi^4\right] \\
&&  + m(m-1)\left[ -4(1-\tau)^2 - 2(1-\tau^2)^2 - 4(m-2) ((1-\tau)^2
- (1-\tau^2)^2) \right.\\
&& ~~~~~~~~~ \left. + 4 (m-1) (1-\tau)(2 \int \xi^2 - \int \xi^3)\right] \\
&=& O(m^3\delta^4).
\end{eqnarray*}

\vskip.1in\noindent{\bf Proof of Lemma \ref{lemmaE.2} :} First
observe that,
\begin{eqnarray*}
&& \mathbb{E}[m - \overline{\bs{\psi}}_1^T \bs{\psi}_1]^2 \\
&=&
\mathbb{E}[\sum_{k=1}^m (1- \psi(T_{1k}))]^2 = \sum_{k=1}^m
\mathbb{E}(1- \psi(T_{1k}))^2 + \sum_{k\neq k'}^m \mathbb{E}[(1-
\psi(T_{1k})(1- \psi(T_{1k'})]\\
&=& m \int (\overline{\psi} - \psi)^2 + m(m-1)(\int (\overline{\psi}
- \psi))^2 \\
&=& m (\delta^2+O(\delta^4)) + \frac{m(m-1)}{4}(\delta^2 +
O(\delta^4))^2 = m\delta^2 (1+o(1)),
~~~(\mbox{by}~(\ref{eq:xi_int_order})).
\end{eqnarray*}
Next,
\begin{eqnarray*}
\mathbb{E}\parallel \overline{\bs{\psi}}_1 -
\bs{\psi}_1\parallel_2^4 &=& \mathbb{E}[\sum_{k=1}^m (1 -
\psi(T_{1k}))^2]^2 \\
&=& \sum_{k=1}^m \mathbb{E} (1 - \psi(T_{1k}))^4 + \sum_{k\neq k'}^m
\mathbb{E}[(1-
\psi(T_{1k})^2(1- \psi(T_{1k'})^2] \\
&=& m \int (\overline{\psi} -\psi)^4 + m (m-1) (\int
(\overline{\psi} -\psi)^2)^2 \\
&\leq& m \parallel \overline{\psi} -\psi \parallel_\infty^2 \int
(\overline{\psi} -\psi)^2 + m(m-1) (\int (\overline{\psi}
-\psi)^2)^2 \\
&=& O(m \delta^4) + m(m-1) \delta^4 (1+o(1)) = O(m^2 \delta^4),
\end{eqnarray*}
where in the last step we used (iv') and (\ref{eq:xi_int_order}).

\vskip.1in\noindent{\bf Proof of Lemma \ref{lemmaE.3} :} Use
(\ref{eq:diff_psi_basic}) to write the expectation as
\begin{eqnarray*}
&&(m-1) \mathbb{E}[\sum_{k=1}^m \psi^2(T_{1k})]^2 - \mathbb{E}
\left[(\sum_{k_1=1}^m \psi^2(T_{1k_1}))(\sum_{k_2 \neq k_2'}^m
\psi(T_{1k_2}) \psi(T_{1k_2'}))\right]\\
&=& (m-1)\left[\sum_{k=1}^m \mathbb{E}\psi^4(T_{1k}) + \sum_{k\neq
k'}^m \mathbb{E}[\psi^2(T_{1k})\psi^2(T_{1k'})\right] \\
&& - \left[\sum_{k_1 = k_2 \neq k_2'}
\mathbb{E}[\psi^3(T_{1k_1})\psi(T_{1k_2'})] + \sum_{k_1 = k_2' \neq
k_2} \mathbb{E}[\psi^3(T_{1k_1})\psi(T_{1k_2})] \right. \\
&& ~~~~~~~\left. + \sum_{k_1 \neq k_2 \neq k_2'}
\mathbb{E}[\psi^2(T_{1k_1})\psi(T_{1k_2})\psi(T_{1k_2'})]\right]\\
&=& (m-1)[m \int \psi^4 + m(m-1) (\int \psi^2)^2] \\
&& ~~~~~~~ - [2m(m-1) (\int
\psi^3)(\int \psi) + m(m-1)(m-2) (\int \psi^2) (\int \psi)^2]\\
&=& m(m-1)[\int (1-\xi)^4 + (m-1) - 2 (\int (1-\xi)^3)(\int (1-\xi))
- (m-2)(\int (1-\xi))^2]\\
&=& m(m-1)[m\int \xi^2 - (\int \xi^3)(2-\int \xi^2) -
\frac{1}{4}(m-2) (\int \xi^2)^2 + \int \xi^4]\\
&=& m^2 (m-1) \delta^2 (1+o(1)),
\end{eqnarray*}
where in the fourth and last steps we used (\ref{eq:xi_int_order})
and (iv').

\end{document}